\documentclass[dvipdfmx,a4paper,11pt]{article}
\usepackage{graphicx}
\usepackage{jheppub}
\usepackage[T1]{fontenc}
\usepackage{braket}	
\usepackage{dcolumn}

\newcommand{\be}{\begin{equation}}
\newcommand{\ee}{\end{equation}}

\newcommand{\LQ}{\Lambda_{\rm QCD}}

\newcommand{\lt}{\left}
\newcommand{\rt}{\right}
\newcommand{\bz}{\beta_0}
\newcommand{\non}{\nonumber \\}
\newcommand{\mf}{\mu_f}

\newcommand{\US}{\rm US}
\newcommand{\DV}{\Delta V}
\newcommand{\fn}{\footnote}
\newcommand{\RF}{\rm RF}
\newcommand{\MSb}{\overline{\rm MS}}
\newcommand{\latt}{\rm latt}
\newcommand{\cont}{\rm cont}
\newcommand{\LMS}{\Lambda_{\overline{\rm MS}}}

\title{Determination of $\alpha_s$ from static QCD potential:
OPE with renormalon subtraction and lattice QCD}

\author[a,1]{Hiromasa Takaura,\note{Corresponding author.}}
\author[b]{Takashi Kaneko,}
\author[c]{Yuichiro Kiyo,}
\author[d]{and Yukinari Sumino}

\affiliation[a]{Department of Physics, Kyushu University, Fukuoka, 819-0395 Japan}
\affiliation[b]{Theory Center, KEK, Tsukuba, Ibaraki, 305-0801 Japan}
\affiliation[c]{Department of Physics, Juntendo University, Inzai, 270-1695, Japan}
\affiliation[d]{Department of Physics, Tohoku University, Sendai, 980-8578 Japan}

\emailAdd{takaura@phys.kyushu-u.ac.jp}
\emailAdd{ykiyo@juntendo.ac.jp}
 \emailAdd{takashi.kaneko@kek.jp}
\emailAdd{sumino@tuhep.phys.tohoku.ac.jp}

\abstract{
We determine the strong coupling constant $\alpha_s$
from the static QCD potential by matching a theoretical calculation
with a lattice QCD computation.
We employ a new theoretical formulation based on the operator product expansion,
in which renormalons are subtracted from the leading Wilson coefficient.
We remove not only the leading renormalon uncertainty of $\mathcal{O}(\Lambda_{\rm QCD})$ 
but also the first $r$-dependent uncertainty of $\mathcal{O}(\Lambda_{\rm QCD}^3 r^2)$.
The theoretical prediction for the potential turns out to be valid
at the static color charge distance $\LMS r \lesssim 0.8$ ($r \lesssim 0.4$~fm),
which is significantly larger than ordinary perturbation theory.
With lattice data down to $\LMS r \sim 0.09$ ($r \sim 0.05$~fm),
we perform the matching in a wide region of $r$,
which has been difficult in previous determinations of $\alpha_s$
from the potential.
Our final result is 
$\alpha_s(M_Z^2) = 0.1179^{+0.0015}_{-0.0014}$ with 1.3\,\% accuracy.
The dominant uncertainty comes from higher order corrections
to the perturbative prediction
and can be straightforwardly reduced by simulating finer lattices.
}

\begin{document}

\preprint{KYUSHU--HET--187, KEK--CP--368, TU--1070}

%\makeatletter
%\patchcmd{\maketitle}{\@fpheader}{}{}{}
%\makeatother

\maketitle
\flushbottom

\section{Introduction}
Today, facing frontier experiments of particle physics, 
such as the ones at LHC and Super $B$ Factory (Belle II),
there exist increasing demands for more accurate theoretical predictions 
based on QCD on various phenomena of the strong interaction.
Precise determination of the strong coupling constant $\alpha_s$,
which is a fundamental parameter of QCD, sets a benchmark for such predictions.
In fact, many theoretical developments are required 
for improving accuracy of $\alpha_s$  determination, 
and once $\alpha_s$ is determined,
it serves as an input parameter for various predictions.
For instance, a precise value of $\alpha_s$ will play crucial roles 
in measurements of Higgs boson properties,
in searches for new physics, or in high-precision flavor physics.
It is also demanded in the context of precise determination
of the top quark mass,
predicting running of the Higgs quartic coupling, etc.

Let us quote the current value of $\alpha_s$,
given as the world-combined result by the Particle Data Group (PDG), 
$
\alpha_s(M_Z^2)=0.1181 \pm 0.0011 
$ \cite{37ce3e5843594be4beddf3c7540d08bc}.
Dominant contributions to this value come from
determinations by lattice QCD,
which have smaller errors than other determinations using 
more direct experimental inputs.
The Flavor Lattice Averaging Group (FLAG) 
reports an average of lattice determinations as
$\alpha_s(M_Z^2)=0.1182 \pm 0.0012$ \cite{Aoki:2016frl}
based on the studies in Refs.~\cite{Maltman:2008bx, Aoki:2009tf, McNeile:2010ji, Chakraborty:2014aca, Bazavov:2014soa}.
The relative accuracies of these current values are 0.9--1.0 per cent.

In determinations of $\alpha_s$ by lattice QCD,
we need to pay attention to  
the so-called  "window problem," 
as pointed out in the FLAG report  \cite{Aoki:2016frl}.
This is a problem that  
it is difficult to find a wide enough region where 
both lattice QCD and perturbative QCD predictions are accurate.
A lattice simulation is carried out with a finite lattice spacing $a$,
whose inverse plays the role of an ultraviolet (UV) cutoff scale.
Hence, the lattice results are accurate in the energy region $Q \ll a^{-1}$.
On the other hand, perturbative calculations are accurate at $Q \gtrsim 1 ~ {\rm GeV} (\gg \LQ \sim 300 ~ {\rm MeV})$.
Determinations of $\alpha_s$ are performed by matching of both results.
It turns out that, for currently available lattice cutoff scales, 
the energy window $1 ~ {\rm GeV} \lesssim Q \ll a^{-1}$ cannot be taken widely.

The method of finite volume scheme combined with step-scaling 
\cite{Luscher:1991wu, Luscher:1992an, Luscher:1993gh, Sint:1993un}
can resolve this problem even at currently available lattice cutoffs.
In this method, discretization and finite volume effects are kept under control by a finite volume scheme,
while lattice data after the step-scaling running can be matched with
perturbation theory at sufficiently high scale.
As a result, matching with perturbative prediction can be performed at 10--100 ${\rm GeV}$.
A recent determination based on this method gives $\alpha_s(M_Z^2)$ 
with 0.7 per cent relative accuracy \cite{Bruno:2017gxd}
(not yet included in the above average values). 

In this paper, we determine $\alpha_s$ by taking an alternative approach to the window problem:
We enlarge the validity range of a theoretical calculation 
to lower energy where lattice calculations are accurate due to $Q \ll a^{-1}$.
For this purpose, we use the operator product expansion (OPE) as a theoretical framework.
Its difference from perturbative calculations can be stated as follows.
Perturbative predictions have an inevitable uncertainty known as renormalon uncertainty,
which stems from a certain divergent behavior of perturbative series at large orders. 
(See Ref.~\cite{Beneke:1998ui} for a review of renormalon.)
For a dimensionless observable $R(Q)$ with typical energy scale $Q$,
a renormalon uncertainty is estimated as $\mathcal{O}((\LQ/Q)^n)$ with a positive integer $n$ 
(dependent on the observable).
 %It ultimately limits validity range of perturbative calculations on low energy side.
In the context of the OPE of the same observable, given by
\be
R(Q)=C_1(Q)+C_{\mathcal{O}_1}(Q) \frac{\braket{0|\mathcal{O}_1|0}}{Q^n}
+\dots \, , \label{OPE1}
\ee
the perturbative result is encoded in the leading Wilson coefficient $C_1$.
In fact, the renormalon uncertainty of $C_1$ generally has the same order of magnitude 
as the leading nonperturbative effect (the second term),
which corresponds to ${\rm dim} [\mathcal{O}_1]=n$ \cite{Mueller:1984vh}.
It is expected that the renormalon uncertainty in the leading Wilson coefficient 
gets canceled when the nonperturbative matrix element is added.
Hence, the OPE may realize a wider validity range due to the absence of the renormalon uncertainty, 
in particular at lower energy side.

However,
the OPE cannot be made a maximal use 
as long as we naively calculate $C_1$ in the ordinary perturbation theory.
This is because we do not know sufficiently about the nonperturbative matrix element.
It is not obvious how to practically 
eliminate the renormalon uncertainty of $C_1$ using the OPE.
In the case where the renormalon uncertainty remains in $C_1$,
one encounters a difficulty that 
the nonperturbative effect cannot be estimated using the OPE \eqref{OPE1}
since $C_1$ has an error comparable to this nonperturbative effect.
In other words, a renormalon uncertainty causes a mixing 
between $C_1$ and the nonperturbative effect.
Many studies considering the OPE in the literature are not free from such a difficulty.

In Refs.~\cite{Sumino:2005cq, Mishima:2016vna}, 
a method to cope with a renormalon uncertainty has been proposed.
This method enables us to divide $C_1$ into a renormalon uncertainty and a renormalon free part.
By this, we remove a renormalon uncertainty from $C_1$ 
before referring to the nonperturbetive matrix element.
In this method, we first define $C_1$ as a UV quantity \`a la Wilson
by introducing an IR cutoff scale $\mf$ 
(corresponding to a factorization scale of an effective field theory).
Then, we separate $C_1(Q^2;\mf)$ 
into its cutoff independent part and dependent part.
While a cutoff dependent part exhibits a connection to the IR physics,
a cutoff independent part is regarded as a genuine UV contribution.
This cutoff independent part corresponds to a renormalon free part, 
determined within perturbation theory.
Furthermore, by absorbing the cutoff dependent part into the leading nonperturbative matrix element,
the nonperturbative matrix element can also be defined as a renormalon free quantity.
Hence, we can define the leading Wilson coefficient and the leading nonperturbative effect
such that they are clearly separated.
This enables us to estimate the nonperturbative effect
without being affected significantly from the higher order uncertainty of $C_1$.

We apply this calculation method for the static QCD potential following Ref.~\cite{Sumino:2005cq}.
The typical energy scale of the static QCD potential is $r^{-1}$, 
which is the inverse of the distance between the static color charges.
The renormalons of the static QCD potential are located at half integers in the Borel $u$-plane.
The first renormalon at $u=1/2$ gives an $\mathcal{O}(\LQ)$ uncertainty.
This renormalon is known to be cancelled against twice the pole mass in the total energy 
once the pole mass is expressed in terms of a short-distance mass \cite{Beneke:1998rk, Hoang:1998nz}.
(At this stage, the OPE in $r$ is not necessary.)
The next-to-leading renormalon at $u=3/2$ gives the 
leading $r$-dependent uncertainty of $\mathcal{O}(\LQ^3 r^2)$. 
In the present work, we remove not only the $u=1/2$ renormalon but also
the $u=3/2$ renormalon using the above renormalon subtraction in the OPE framework.

The OPE of the static QCD potential in $r$ can be performed  
in the form of the multipole expansion within
the effective field theory (EFT), potential non-relativistic QCD (pNRQCD) \cite{Brambilla:1999xf}.
Thanks to this solid basis of pNRQCD, the $u=3/2$ renormalon cancellation has been convincingly shown \cite{Brambilla:1999xf},
which gives a solid basis to our OPE formula.
We construct a renormalon subtracted Wilson coefficient 
(which will be denoted by $V_S^{\RF}$ below)
based on the fixed order result, 
which is currently known up to the next-to-next-to-next-to-leading order (${\rm N^3LO}$), i.e., $\mathcal{O}(\alpha_s^4)$ 
\cite{Smirnov:2008pn, Anzai:2009tm, Smirnov:2009fh, Lee:2016cgz}.  
As mentioned, a unique feature of our renormalon subtraction is that 
not only the leading renormalon (at $u=1/2$) but also
the next-to-leading renormalon (at $u=3/2$) is removed from $V_S^{\RF}$.
In the OPE, the leading term is given by $V_S^{\RF} \sim \mathcal{O}(1/r)$.
The NLO term represents the leading nonperturbative effect and is $\mathcal{O}(r^2)$.
We include the NLO term with an unknown coefficient
which is to be determined by a fit.
We will explicitly show consistency with the OPE by comparing $V_S^{\RF}$ with a lattice result:
the difference between them can be fitted  by an $\mathcal{O}(r^2)$-term.
Our OPE prediction turns out to be valid up to $\LQ r \lesssim 0.8$,
corresponding to $r^{-1} \gtrsim 0.5~{\rm GeV}$. 
This shows that our theoretical prediction indeed has a wider validity range than
the ordinary perturbation theory, which is valid at $r^{-1} \gtrsim 1~{\rm GeV}$.

We determine $\alpha_s$ from the static QCD potential by matching 
a lattice result with the above OPE where the renormalon uncertainty is subtracted. 
The lattice results that we use are obtained by the JLQCD collaboration at large cutoffs up to 4.5 GeV \cite{Kaneko:2013jla, JLQCD:future}, 
which facilitate the matching between lattice and the OPE calculations.

Determinations of $\alpha_s$ using the static QCD potential 
have been performed in Ref.~\cite{Bazavov:2014soa} with 3-flavor lattice simulation
and in Ref.~\cite{Karbstein:2018mzo} with 2-flavor lattice simulation.
In these studies,  perturbative calculations are matched with lattice results
in the perturbative regime $\LQ r \lesssim  0.2\text{--}0.3$.
Our determination is carried out with the OPE calculation,
and the matching range is taken as
$\LQ r \lesssim 0.6\text{--}0.8$.
We have briefly reported our analysis in Ref.~\cite{Takaura:2018lpw}.

This paper is organized as follows. 
In Sec.~\ref{sec:2}, we present our theoretical formula 
to subtract renormalons in the OPE (partially supplemented in Appendix~\ref{app:VSRF}).
In Sec.~\ref{sec:alphas}, we determine $\alpha_s$ by matching 
the theoretical calculation with a lattice result.
Lattice results and the way to determine $\alpha_s$ are also explained therein.
Conclusions and discussion are given in Sec.~\ref{sec:4}.
Some referential materials and supplementary arguments are given in Appendices.

\section{Theoretical framework}
\label{sec:2}
Our renormalon subtraction formula \cite{Sumino:2005cq} 
is constructed based on the EFT,
potential non-relativistic QCD (pNRQCD) \cite{Brambilla:1999xf}.
This EFT factorizes two typical scales of the static QCD potential.
One of the scales is the soft scale $\sim 1/r$, 
which is the inverse of the distance between the static color charges.
The other is the ultrasoft (US) scale $\sim \DV (\ll 1/r)$, 
which is the energy difference between the octet and singlet bound states. 
pNRQCD enables us to investigate the US scale physics 
in a systematic expansion in $r \DV \ll 1$. 
The Lagrangian of this EFT consists of the singlet and octet matter fields and the US gluon fields,
while the soft scale contributions are integrated out and
encoded in the Wilson coefficients.
Our formula is based on the multipole expansion,
which expands the static QCD potential in $r$ using the hierarchy $r \DV  \ll 1$.

In Sec.~\ref{sec:BF}, we present our formula to subtract renormalons 
after a brief review of the multipole expansion, on which the formula is based. 
In Sec.~\ref{sec:US}, we explain our treatment of the IR divergence in the three-loop result of the static QCD potential,
which is related to the US scale dynamics.
In Sec~\ref{sec:HO}, we estimate the higher order perturbative uncertainty of the theoretical calculation,
which is required in estimation of systematic errors in $\alpha_s$ determination.

\subsection{Formula to subtract renormalons}
\label{sec:BF}
Our theoretical calculation is based on the multipole expansion,
which gives an expansion of the static QCD potential in $r$ \cite{Brambilla:1999xf}:
\be
V_{\rm QCD}(r)=V_S(r)+\delta E_{\US}(r)+\dots \, . \label{OPE}
\ee
The explicit $r$-dependence is given by $V_S \sim \mathcal{O}(1/r)$ and 
$\delta E_{\US} \sim \mathcal{O}(r^2)$. (The dots denote higher order terms in $r$.)
The singlet potential $V_S$ originates from the soft scale\fn{
In the pNRQCD terminology, the "soft scale" corresponds to the UV scale,
which has been integrated out.} $\sim 1/r \, (\gg \LQ)$
and can be evaluated in perturbation theory. 
In terms of the pNRQCD Lagrangian, $V_S$ is a Wilson coefficient.
Perturbative result in coordinate space is usually obtained through Fourier transform (FT)
of the perturbative evaluation of $\alpha_V(q^2)$,
\be
V_S(r)=- 4\pi C_F \int \frac{d^3 \vec{q}}{(2 \pi)^3} \, e^{i \vec{q} \cdot \vec{r}} \frac{\alpha_V(q^2)}{q^2} ~~~~~(q=|\vec{q}|) \, , \label{singlet}
\ee
where the perturbative result of $\alpha_V(q^2)$ is currently known up to three-loop order \cite{Anzai:2009tm, Smirnov:2009fh, Lee:2016cgz}:
\be
\alpha_V(q^2)=\alpha_s(\mu^2) \sum_{n=0}^3 [P_n(\log(\mu/q))+\delta P_n(\log(\mu/q))] \lt(\frac{\alpha_s(\mu^2)}{4 \pi} \rt)^n \, . \label{pertexp}
\ee
Here, $P_n$ is an $n$-th order polynomial and we denote its constant part by $a_n$:
\be
a_n=P_n(0)=P_n(\log(\mu/q))|_{\mu=q} \, . \label{pertexp2}
\ee
The logarithmic terms in $P_n$ can be calculated from the renormalization group (RG) equation
and are expressed by $a_k$ with $k<n$ and the coefficients of the beta function.
$\delta P_n$ represents the IR divergence and associated logarithmic dependence.
It is zero for $n \leq 2$, and non-zero for $n=3$; see Eq.~\eqref{deltaP3} for $\delta P_3$.
This IR divergence is different from renormalon uncertainties and its presence 
hardly affects the following renormalon subtraction formula.
We explain our prescription for regularizing this divergence in the next section \ref{sec:US}.
We collect the explicit expressions for $a_n$ in Appendix~\ref{app:coeff}.

The NLO term of Eq.~\eqref{OPE}, 
$\delta E_{\US}$, is dominantly determined by dynamics of the US scale $\sim \DV$.
It is given by a correlator of the US fields in pNRQCD:
\be
\delta E_{\US}(r) =-i \frac{2 \pi \alpha_s}{3}  \int_0^{\infty} dt \,  e^{-i \DV t} \braket{0|\vec{r} \cdot \vec{E}^a(t) \varphi_{\rm adj}(t;0)^{ab}  \vec{r} \cdot \vec{E}^b(0)|0} \, , \label{deltaE}
\ee 
where $\vec{E}^a$ is the US chromoelectric field; see Ref.~\cite{Brambilla:1999xf} for details.

Despite the fact that conceptually the singlet potential is a soft quantity,
the integration region is usually taken as $0\leq q < \infty$ as shown in Eq.~\eqref{singlet}. 
In particular, IR region of the integral is known to cause renormalon uncertainties in $V_S$,
and it brings about a mixing between $V_S$ and $\delta E_{\US}$.
To avoid this feature, we construct $V_S$ as a renormalon free quantity below, 
following Ref.~\cite{Sumino:2005cq}.

We first introduce a factorization (cutoff) scale $\mf$ to divide the energy region
as $\LQ  \ll \mf \ll  1/r$,
and define $V_S$ as a soft quantity in terms of this cutoff scale:
\be
V_S(r;\mf)=- 4 \pi C_F \int_{q>\mf} \frac{d^3 \vec{q}}{(2 \pi)^3} \, e^{i \vec{q} \cdot \vec{r}} \frac{\alpha_V(q^2)}{q^2}   \label{VS} \, .
\ee
Since all the known renormalons stem from the low energy region of the $\vec{q}$-integral, the above definition
renders $V_S$ free from renormalons.\fn{
More accurately, dominant renormalons which arise from the $\vec{q}$-integral are removed.
Renormalons contained in $\alpha_V(q^2)$ are regarded as subdominant and have not been studied,
to our knowledge.
We neglect them in our analysis.}
In $V_S(r;\mf)$, there is a cutoff dependent part by construction,
which is regarded as an IR sensitive part of $V_S$.
Such a dependence vanishes only when it is combined with the IR quantities such as $\delta E_{\US}$.
Hence, the mixing is induced through the factorization scale.
In this respect the cutoff dependent part can be regarded as a renormalon related part.
In contrast, the cutoff independent part is determined within perturbation theory independently of 
IR contributions, and hence it can be regarded as a genuine renormalon free part
and as a pure UV contribution.
%Ref.~\cite{Sumino:2005cq} has proposed a method to extract a cutoff independent part from Eq.~\eqref{VS}.

\begin{figure}[hbpt]
\begin{center} 
\includegraphics[width=10cm]{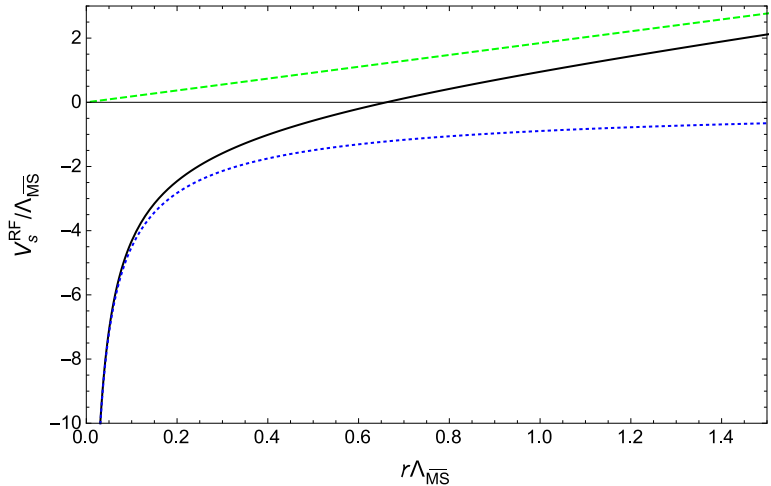}
\end{center}
\caption{Renormalon free singlet potential $V_S^{\RF}/\Lambda_{\MSb}$ as a function of $\Lambda_{\MSb} r$
(black solid line).
$V_C(r)/\Lambda_{\MSb}$ is shown by the blue dotted line, and the linear contribution $\mathcal{C}_1 r/\LMS$ 
[Eq.~\eqref{linear}] is shown by the green dashed line.
The results are obtained with regularization method I [Eq.~\eqref{regI}].}
\label{fig:VSRF}
\end{figure}

The renormalon free quantity, which we denote by $V_S^{\rm RF}(r)$ [= cutoff independent part of $V_S(r;\mu_f)$], 
can be constructed systematically as follows.
For $\alpha_V(q^2)$ in Eq.~\eqref{VS}, we adopt the next-to-next-to-next-to-leading log (${\rm N^3LL}$) result,
which is obtained by RG improvement of the ${\rm N^3LO}$ fixed order result: 
\be
\alpha_V(q^2)|_{\rm N^3 LL}=\alpha_s(q^2)
\lt[ a_0 +a_1 \frac{\alpha_s(q^2)}{4 \pi}+a_2 \lt(\frac{\alpha_s(q^2)}{4 \pi} \rt)^2
+a_3^{\rm Reg. I \, or \, II}(q) \lt(\frac{\alpha_s(q^2)}{4 \pi} \rt)^3 \rt] \, , \label{alphaVNNNLL}
\ee
where $\alpha_s(q^2)$ is the running coupling constant, namely, 
the solution to the RG equation at four-loop (for consistency):
\be
q^2 \frac{d}{d q^2} \alpha_s(q^2)
=\beta(\alpha_s(q^2))|_{\text{4-loop}}=-\alpha_s(q^2) \sum_{i=0}^3 \beta_i  \lt( \frac{\alpha_s(q^2)}{4 \pi} \rt)^{i+1} \, . \label{running}
\ee
We solve this equation numerically.\fn{
Although the running coupling constant $\alpha_s(\mu^2)$ can approximately  be expressed by 
series expansion in $1/\log{(\mu^2/\LMS^2)}$,
we do not use this approximation but solve the RG equation for $\alpha_s(\mu^2)$ numerically.
\label{fn:running}}
The three-loop coefficient $a_3$ is originally IR divergent as mentioned and 
we regularize it with the prescription explained below 
[$a_3^{\rm Reg. I \, or \, II}(q)$ is given by Eq.~\eqref{regI} or \eqref{regII} in Sec.~\ref{sec:US}].
Related to this divergence, the regularized $a_3$ has a $q$-dependence unlike the coefficients up to $a_2$. 
(As noted, this feature has nothing to do with renormalons.)
We set $n_f=3$ and the corresponding light quarks ($u,d,s$) are 
treated in the massless approximation in our main analysis.
(Finite mass effects are taken into account as a systematic error of our $\alpha_s$ determination in Sec.~\ref{sec:AnaII}.)
Up to here, the integrand of Eq.~\eqref{VS} is determined.
Then, by deforming the integration contour of Eq.~\eqref{VS} in the complex $q$-plane,
we can separate a cutoff independent part from a cutoff dependent part.
We explain this formulation explicitly in Appendix~\ref{app:VSRF},
which is a brief review of Refs.~\cite{Sumino:2005cq, Mishima:2016vna}.
After this procedure, we obtain the following expression:
 \be
 V_S(r;\mu_f) =V_S^{\rm RF}(r) + \mathcal{C}_0(\mu_f)+\mathcal{C}_2(\mu_f) r^2+\mathcal{O}(r^3) \, , \label{decom}
 \ee
where $V_S^{\rm RF}(r)$ is a $\mu_f$-independent and renormalon-free quantity.
The cutoff dependence of $V_S(r;\mu_f)$ is encoded in the $r^0$ and $r^2$ terms (and in further higher order terms),
which correspond to the $u=1/2$ and $u=3/2$ renormalons, respectively.
$V_S^{\rm RF}$ consists of a Coulomb-like part and a linear part:
\be
V_S^{\rm RF}(r)=V_C(r)+\mathcal{C}_1 r \label{VSRF} \, .
\ee
$V_C(r)$ is expressed by a one-dimensional integral, whose explicit form is given in Eq.~\eqref{VC}.
We evaluate this integral numerically. 
$V_C$ has a Coulomb-like form with logarithmic corrections at short distances.
The coefficient of the linear part, $\mathcal{C}_1$, 
is proportional to $\Lambda_{\MSb}^2$, 
and it can unambiguously be calculated as\fn{
The value is obtained in regularization I,
where $a_3$ is regularized as in Eq.~\eqref{regI}.}
\be
\mathcal{C}_1= 1.844 \Lambda_{\MSb}^2 ~~~~\text{for~ $n_f=3$~ at ${\rm N^3 LL}$} \, . \label{linear} 
\ee
One obtains $V_S^{\rm RF}/\LMS$ as a function of $\LMS r$
without free parameters,
where $\LMS$ is the only dimensionful parameter in massless QCD.
Here and hereafter, $\LMS$ is the $\Lambda$-parameter at four-loop in the $\MSb$ scheme
with $n_f=3$, unless otherwise stated: $\LMS={\LMS}^{\text{4-loop}}_{,n_f=3}$. 
(See Appendix~\ref{app:Lambda} for the definition of $\LMS$.)
We show $V_S^{\rm RF}(r)$ in Fig.~\ref{fig:VSRF}.

So far, we have concentrated on the perturbative part $V_S$. 
Now let us see how the result is combined with the multipole expansion \eqref{OPE}.
Since we define the soft quantity $V_S$ with the IR cutoff scale $\mf$, 
it is natural to define the US quantity $\delta E_{\US}$ with the UV cutoff scale $\mf$.
Accordingly, the multipole expansion is written as
\be
V_{\rm QCD}(r)=V_S(r;\mf)+\delta E_{\US}(r;\mf)+\dots \, .
\ee
It is confirmed in Ref.~\cite{Takaura:2017lwd} 
that the cutoff dependence of $V_S(r;\mu_f)$ of the $r^2$-term
gets canceled against the leading cutoff dependence of $\delta E_{\US}(r;\mf)$ at the LL level. 
This corresponds to the $u=3/2$ renormalon cancellation, 
which was first reported in Ref.~\cite{Brambilla:1999xf}.
Although the explicit confirmation at the ${\rm N^3 LL}$ level (which we consider) is still missing,
we assume a parallel scenario. 
Hence, by using Eq.~\eqref{decom},
we can perform the multipole expansion as
\be
V_{\rm QCD}(r)=V_S^{\RF}(r)+\delta E_{\US}^{\RF}(r)+\dots \, . \label{RFOPE}
\ee
Here, $\delta E_{\US}^{\RF}$ is the sum of $\delta E_{\US}(r;\mu_f)$ and $\mathcal{C}_2(\mf) r^2$;
hence it is $\mu_f$ independent and free from renormalons.
In this way, $V_S^{\RF}$ and $\delta E_{\US}^{\RF}$ are defined as genuine UV and IR quantities, respectively.
We omit the constant $\mathcal{C}_0(\mf)$,
which does not have a significant meaning 
in $\alpha_s$ determination; see Sec.~\ref{sec:alphas}.
Eq.~\eqref{RFOPE} is the central formula of our theoretical calculation.
The first term is given by Eq.~\eqref{VSRF} and shown in Fig.~\ref{fig:VSRF}.
In our analyses, we regard $\delta E_{\US}^{\RF}$, which is an US quantity, 
as a nonperturbative object (non-local gluon condensate), 
and assume $\delta E_{\US}^{\RF} \sim \LMS^3 r^2$.
This is because we focus on relatively long distances 
where $\DV \gg \LMS$ is in general {\it not} assured.
The perturbative result for $\delta E_{\US}$ (obtained within pNRQCD) is used for a limited purpose.
Hence, we treat the second term as $\delta E_{\US}=A_2 r^2$ 
where $A_2$ is a fitting parameter, showing the size of the (renormalon-free) nonperturbative effect.\fn{
The $r^2$ behavior may receive logarithmic corrections, for instance, 
from higher order computations of Wilson coefficients. 
We discuss their effects on $\alpha_s$ determination in Appendix~\ref{app:further}.}

Let us state the unique features of $V_S^{\rm RF}$,
which is a central object in Eq.~\eqref{RFOPE}.
First, let us clarify the difference from the usual RG improved predictions.
Usual ${\rm N}^k$LL predictions for the static QCD potential are reliable at short distances, 
but they have an unphysical singularity around $r \sim \Lambda_{\MSb}^{-1}$,\fn{
An N$^k$LL prediction is given by
\be
r V_{\rm QCD}(r)=d_0 \alpha_s(1/r^2)+d_1 \alpha_s(1/r^2)^2+\dots+d_k \alpha_s(1/r^2)^{k+1}   \, ,
\ee
where $\alpha_s(1/r^2)$ is the $(k+1)$-loop running coupling.
Due to the singularity of $\alpha_s(1/r^2)$ around $r^{-1} \sim \LMS$,
the prediction has an unphysical singularity.
\label{fn:unphysical}} 
which distorts the behavior around this region drastically.
In contrast, $V_S^{\RF}(r)$ does not have an unphysical singularity,
while ${\rm N^3LL}$ accuracy is held at short distances.
Therefore, reliable range of $V_S^{\RF}(r)$ on the low energy side is not limited a priori.\fn{
The naive expectation for the validity range of $V_S^{\RF}$
is the region up to where the $\Lambda_{\MSb}^3 r^2$ becomes non-negligible,
due to the structure of the OPE.}
%In addition to this, 
%once $V_S^{\RF}(r)$ is expanded in $\alpha_s(1/r^2)$,
%its coefficients are non-zero to all orders.
%Namely, $V_S^{\RF}(r)$ contains an estimation of higher order coefficients
%which are not known only from the RG equation.
%\fn{
%This is analogue to the large-$\bz$ approximation, 
%where RG independent coefficients are approximately obtained 
%by considering an effective charge before performing loop integrals.
%In the current case, the effective charge corresponds to RG improved $\alpha_V(q^2)$,
%and loop integral does to integration of FT.} 
Secondly, $V_S^{\RF}$ does not have any renormalons.
In particular, it is free not only from the $u=1/2$ renormalon but also from the 
$u=3/2$ renormalon, and thus,
it is free from the leading $r$-dependent renormalon uncertainty 
of $\mathcal{O}(\LMS^3 r^2)$. 
Thanks to these features, 
$V_S^{\RF}$ is reliable at short to relatively long distances $\LMS r \sim \mathcal{O}(1)$.
This allows our OPE prediction to have a wide validity range, 
as will be shown in Sec.~\ref{sec:CC}.

\subsection{Treatment of US scale}
\label{sec:US}

We explain our prescriptions for regularizing the IR divergence in the three-loop coefficient. 
The IR divergence was first discovered in Ref.~\cite{Appelquist:1977tw} and calculated in Ref.~\cite{Brambilla:1999qa}.
In dimensional regularization with $D=4 -2 \epsilon$, it reads
\be
\delta P_3=72 \pi^2 \lt( \frac{1}{\epsilon}+6 \log(\mu/q) \rt) \, . \label{deltaP3}
\ee
This IR divergence signals breakdown of naive perturbative expansion
and is attributed to the dynamics at the US scale.
The counterpart of the above divergence is provided from $\delta E_{\US}$ \eqref{deltaE}.
Namely, the FT of $\delta E_{\US}(r)$ at $\mathcal{O}(\alpha_s^4)$, defined by 
\be
\delta E_{\US}(r)|_{\mathcal{O}(\alpha_s^4)}=
-4 \pi C_F \int \frac{d^3 \vec{q}}{(2 \pi)^3} \, e^{i \vec{q} \cdot \vec{r}} \frac{{\widetilde{{\delta E}}_{\US}}(q)|_{\mathcal{O}(\alpha_s^4)}}{q^2}  \, ,
\ee
is evaluated as \cite{Anzai:2009tm}
\begin{align}
\widetilde{\delta E}_{\US}(q)|_{\mathcal{O}(\alpha_s^4)}
&=\alpha_s \lt(\frac{\alpha_s}{4 \pi} \rt)^3
\lt(P_3^{\US}+\delta P_3^{\US}(\log(\mu/q)) \rt)  \label{a3US}
\end{align}
with
\be
P_3^{\US}=72 \pi^2  
\lt(2 \log(C_A \alpha_s(\mu))+2 \gamma_E-\frac{5}{3} \rt)
\ee
\be
\delta P_3^{\US}(\log(\mu/q))=-72 \pi^2  
\lt(\frac{1}{\epsilon}+6 \log(\mu/q) \rt) \, .
\ee
Namely, $\delta E_{\US}$ has the UV divergence $\delta P_3^{\US}$, and $\delta P_3^{\US}=-\delta P_3$. 
Hence, the sum of the soft contribution ($V_S$) and US contribution ($\delta E_{\US}$)
at order $\alpha_s^4$ is finite.

In our analysis, it is appropriate to remove the IR divergence ($\delta P_3$) from the definition of $V_{S}^{\RF}$
as well as its IR renormalons since $V_{S}^{\RF}$ is defined as a pure UV quantity.
For this, we make the UV divergence of $\delta E_{\US}$ absorbed into $V_S$.
This is compatible with our formulation since we define $\delta E_{\US}^{\rm RF}$ as a pure IR contribution.
%it is appropriate to remove the UV divergence ($\delta P_3^{\US}$) from the definition of $\delta E_{\US}^{\RF}$
%(as well as its UV renormalons)
%since $\delta E_{\US}^{\rm RF}$ is defined as a pure IR quantity.
%Concurrently, the UV divergence should be absorbed into $V_S$ 
%to make $V_S^{\RF}$ free of IR divergence.
%This is compatible with our formulation, where we define $V_S^{\rm RF}$ as a pure UV contribution.

We consider the following two prescriptions for regularizing the three-loop perturbative coefficient of $V_S$.
In the first one,  we regard $P_3^{\US}+\delta P_3^{\US}$, the perturbative contribution of $\delta E_{\US}$,
as a UV contribution and make $V_S^{\rm RF}$ finite by including it as
\begin{align} 
&~ a_3^{\rm Reg. I}(q) \equiv (P_3 +\delta P_3) + (P_3^{\US}+\delta P_3^{\US} )|_{\mu=q}  \non
&=a_3+72 \pi^2 \lt(2 \log(C_A \alpha_s(\mu^2))+2 \gamma_E-\frac{5}{3} \rt)\biggr|_{\mu=q} \, . \label{regI}
\end{align}
When RG improvement is applied, we also replace $\alpha_s(\mu^2)$ inside logarithm of Eq.~\eqref{regI} by $\alpha_s(q^2)$,
in the same way as in Eq.~\eqref{alphaVNNNLL}. 
In the second prescription, we divide 
$P_3^{\US}+\delta P_3^{\US}$ into UV and IR contributions by a cutoff scale $\mu_{\US}$
which is taken above the US scale.
In this case we adopt
\be
a_3^{\rm Reg. II}(q)=a_3+144 \pi^2 \log(\mu_{\US}/q) \, . \label{regII}
\ee
This can be compared to Eq.~\eqref{regI} by regarding the regulator $C_A \alpha_s$ as $\DV(r) r \sim \mu_{\rm US}/q$.
We choose $\mu_{\rm US}= 3 \Lambda_{\MSb}$ or $4 \Lambda_{\MSb}$.

The motivation to consider the above two prescriptions in $\alpha_s$ determination
is to check sensitivity to the treatment of the US perturbative contribution.
We will see that it does not induce a significant effect.
We use the regularization I, given in Eq.~\eqref{regI}, in our main analysis.

\subsection{Higher order perturbative uncertainty}
\label{sec:HO}
\begin{figure}[h!]
\begin{center}
\includegraphics[width=11cm]{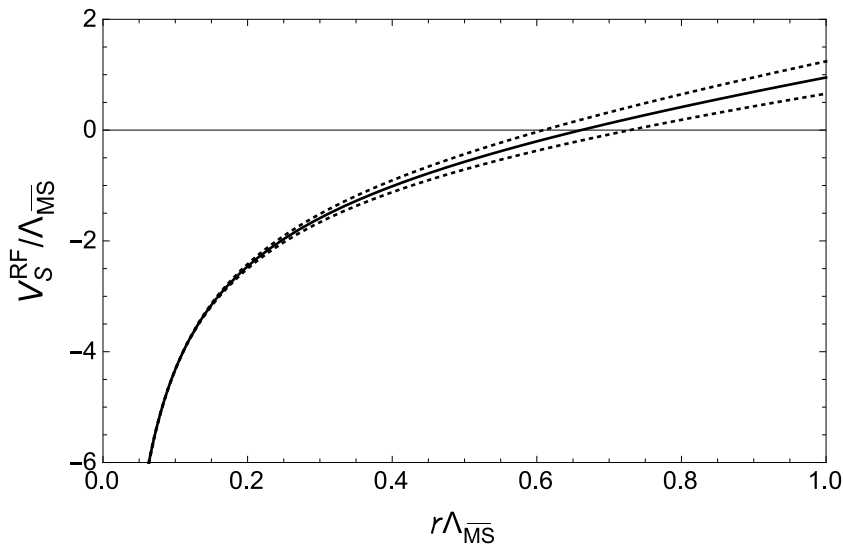}
\end{center}
\caption{$V_S^{\RF}$ at ${\rm N^3 LL}$ (solid) and the higher order uncertainty
given by the region between the dotted lines.
The $r$-independent constants are adjusted such that
three lines take the same value at $\LMS r=0.1$.}
\label{fig:theoerror}
\end{figure}

$V_S^{\RF}$, which we calculate at ${\rm N^3LL}$ accuracy,
receives higher order corrections.
We assume that $V_S^{\RF}$ can vary to 
\be
V_S^{\RF} \pm \delta V_S^{\RF} \, ,
\ee
due to higher order corrections. 
We take $\delta V_S^{\RF}$ conservatively as 
\be
\delta V_S^{\rm RF}=V_S^{\RF}|_{\rm N^3 LL}-V_S^{\RF}|_{\rm N^2 LL} \, . \label{deltaV}
\ee
In calculating $V_S^{\RF}|_{\rm N^2 LL}$, we use the 4-loop beta function\fn{
This is for simplicity of the analysis.
}
in evaluating $\alpha_V(q^2)$,
while the fixed order results are used up to $a_2$ rather than up to $a_3$. 
Hence, $\delta V_S^{\RF}$ reduces to 
the Coulomb+linear part originating from the $a_3$-term alone (at ${\rm N^3 LL}$).
We show the perturbative uncertainty of $V_S^{\RF}$ (i.e. $V_S^{\rm RF} \pm \delta V_S^{\rm RF}$) in Fig.~\ref{fig:theoerror}.

Higher order uncertainty $\delta V_S^{\RF}$
takes a Coulomb+linear form (with log corrections) similarly to $V_S$.
Hence, it is qualitatively different from 
the nonperturbative effect, whose form is quadratic in $r$.
This enables us to estimate the nonperturbative effect
while distinguishing it from the perturbative uncertainty.\fn{
In previous studies considering the OPE,
an estimation of nonperturbative effect suffers significantly from perturbative uncertainty
since renormalons are not subtracted from perturbative calculation.
See sec.~\ref{sec:CC}.}
%In practice, an estimation of the nonperturbative effect is not completely
%independent of the higher order uncertainty 
%since we use a limited number of data.

Furthermore, we note that the perturbative error is smaller than the one in usual perturbative calculation
thanks to renormalon subtraction.
We will revisit this point in Sec.~\ref{sec:CC}.
We also remark that the higher order uncertainty 
is expected to become smaller as the order grows.
Since such a property does not hold in the presence of renormalons,
this is another non-trivial merit of renormalon subtraction.

%Although $V_S$ and $\delta E_{\US}$ are independent at the first glance,
%they are actually mixed through a renormalon uncertainty.
%In perturbative evaluation of $V_S$, the renormalon uncertainties appear, 
%whose leading $r$-dependent one is $\mathcal{O}(\LQ^3 r^2)$, caused by the $u=3/2$ renormalon.\fn{
%The first renormalon at $u=1/2$ gives an uncertainty just for the constant part of the potential, and 
%can be omitted after finding a renormalon free part, as done below.}
%This is the same magnitude as $\delta E_{\US}$ when it is treated as the non-local gluon condensate. 
%Although this renormalon is known to get canceled against that of $\delta E_{\US}$,
%it is practically impossible to distinguish perturbative result of $V_S$ from
%a nonperturbative correction $\delta E_{\US}$
%because we do not know the whole structure, namely both ambiguous and unambiguous part of $\delta E_{\US}$.

\section{$\alpha_s$ determination}
\label{sec:alphas}

We determine the strong coupling constant at the $Z$ boson mass scale, $\alpha_s(M_Z^2)$.
We achieve this by matching the theoretical calculation 
[presented in the previous section in particular in Eq.~\eqref{RFOPE}]
with a lattice result. 
We perform two analyses.

The first analysis, which we call Analysis (I), consists of two steps.
In the first step, we take the continuum limit of the lattice result 
(without referring to the theoretical prediction above).
In the second step, we compare it with the theoretical prediction to extract $\alpha_s(M_Z^2)$.
We proceed while checking (a) if the lattice result can smoothly be extrapolated to the continuum limit,
and (b) if $V_S^{\RF}$ can explain the lattice result up to $\mathcal{O}(r^2)$ difference,
consistently with the OPE structure of Eq.~\eqref{RFOPE}.
After confirming these features, we determine $\alpha_s(M_Z^2)$. 

In the second analysis, Analysis (II), 
we perform a global fit to determine $\alpha_s(M_Z^2)$.
The two tasks, i.e., an extrapolation to the continuum limit of the lattice data and
a determination of $\alpha_s$ by comparison with the theoretical prediction,
are carried out at once. 

Our final result will be adopted from Analysis (II), 
where we achieve a smaller error than Analysis (I).
Analysis (II) is a first-principle analysis, which avoids model interpolating function, 
required in Analysis (I) for continuum extrapolation.
Nevertheless, Analysis (II) is performed without revealing detailed profiles at intermediate steps.
To fill the gap, Analysis (I) is performed, 
where the intermediate steps of the analysis are examined and exhibited explicitly. 
%Hence, it serves to strengthen the validity of Analysis (II).

We start with an explanation of lattice simulations in Sec.~\ref{sec:LS}.
Subsequently we present Analysis (I) in Sec.~\ref{sec:AnaI} and Analysis (II) in Sec.~\ref{sec:AnaII}. 
Necessary formulas for the analyses are given in Appendix~\ref{app:corrmat}.

\subsection{Lattice simulations}
\label{sec:LS}

Our analysis is performed by using lattice QCD data $V_{\rm latt}(r)$  
obtained by the JLQCD collaboration~\cite{Kaneko:2013jla, JLQCD:future}.
Their numerical simulations are carried out in three-flavor QCD in the isospin limit
by employing the the Symanzik gauge~\cite{Weisz:1982zw}
and M\"obius domain-wall quark actions~\cite{Brower:2012vk}.
A careful choice of the detailed structure of the quark action
reduces the computational cost to simulate fine lattices remarkably
while preserving chiral symmetry to good precision~\cite{Kaneko:2013jla}.
Lattice data of $V_{\rm latt}$ are available
at three lattice cutoffs,
which are determined as $a^{-1}\!=\!2.453(4)$, 3.610(9) and 4.496(9)~GeV
from the Wilson-flow scale \cite{Luscher:2010iy}.
In the following, we denote the three lattice spacings 
by $a_1$, $a_2$ and $a_3$ ($a_1\!>\!a_2\!>a_3$).
Discretization errors of $V_{\rm latt}$ start at $O(a^2)$,
since chiral symmetry forbids $O(a)$ errors.

%// volume, mass 

The lattice sizes are $N_s^3 \times T=32^3\!\times\!64$, $48^3\!\times\!96$
and $64^3\!\times\!128$ at $a_1$, $a_2$ and $a_3$, respectively.
In order to control finite volume effects,
their physical sizes are roughly kept constant $L=N_s a \!\approx\! 2.6$~fm,
and sufficiently larger than the short distance region $r\!\lesssim\!0.5$~fm,
where we perform the matching with the OPE.
At each cutoff, we take lattice data $V_{\rm latt}(r)$
at a single combination of the degenerate up and down quark mass $m_{ud}$
and the strange quark mass $m_s$.
While $m_s$ is close to its physical value,
$m_{ud}$ corresponds to unphysically heavy pion mass $M_\pi\!\approx\!300$~MeV.
The correction to $V_{\rm latt}(r)$ due to the unphysical quark masses
is taken into account in Analysis (II),
but turns out to be small (see Sec.~\ref{sec:AnaII}).
Gauge ensembles are generated by using the Hybrid Monte Carlo algorithm.
The statistics are 5,000 Molecular Dynamics (MD) time at each simulation point.
Simulation parameters are summarized in table~\ref{tab:param}.

\begin{table}
\small
\begin{center}
\begin{tabular}{llllllll}
\hline
$a^{-1}$ [GeV] &  size & $m_{ud}$ & $m_s$ & $M_\pi$ [MeV] & $M_K$ [MeV] &
  \# bin 
\\ \hline
$a_1^{-1}\!=\!2.453(4)$ & $32\!\times\!64$ &
  0.0070 & 0.0400 & 309(1) & 547(1) & $N_1\!=\!200$ 
\\
$a_2^{-1}\!=\!3.610(9)$ &  $48\!\times\!96$ & 
  0.0042 & 0.0250 & 300(1) & 547(2) & $N_2\!=\!100$ 
\\
$a_3^{-1}\!=\!4.496(9)$ &  $64\!\times\!128$ & 
  0.0030 & 0.0150 & 284(1) & 486(1) & $N_3\!=\!100$ 
\\ \hline
\end{tabular}
\end{center}
\caption{
  Lattice simulation parameters.
  For the quark masses $m_{ud}$ and $m_s$, we list bare values in lattice units. 
  The renormalization factor to the $\overline{\rm MS}$ scheme is available
  in Ref.~\cite{Tomii:2016xiv}.
}
\label{tab:param}
\end{table}

% measurement

The potential $V_{\rm latt}(r)$ is extracted
from the asymptotic behavior of the rectangular Wilson loop 
\begin{align}
  W(r,t) &= C(r)\, \exp\left[ -V_{\rm latt}(r)\,t \right]
  \hspace{3mm}
  (t\!\to\!\infty),
\label{WtoV}
\end{align}
where $r$ and $t$ represent its spatial and temporal sizes, respectively.
A gauge link smearing~\cite{Bali:1992ab} is applied to
the spatial Wilson lines to suppress excited state contaminations
at reasonably small $t$.
The spatial Wilson lines and, hence, the quark pair separation $\vec{r}$ 
are chosen to be parallel to the spatial directions
$(1,0,0)$ and $(1,1,0)$, which we call direction $d=1$ and 2 in the following.
Throughout this study,
we estimate the statistical error by the jackknife method.
The bin size is chosen as 25 ($a_1$)
or 50 MD time ($a_2$ and $a_3$)
by inspecting the bin size dependence of the jackknife error of $V_{\rm latt}$.
The number of bins is $N_1\!=\!200$ ($a_1$) or
$N_2\!=\!N_3\!=\!100$ ($a_2$ and $a_3$).
The statistical correlation is taken into account
in the fit~(\ref{WtoV}) and subsequent analyses.

\subsection{Analysis (I): Two-step analysis}
\label{sec:AnaI}
In Analysis (I), we first take the continuum limit of the lattice data in Sec~\ref{sec:CL}.
Using the extracted result, we confirm the validity of the OPE 
and compare it with other methods adopted in preceding studies in Sec.~\ref{sec:CC}.
Matching of the OPE with the lattice result to determine $\alpha_s$ 
is performed in Sec.~\ref{sec:AD}.

\subsubsection{Continuum extrapolation} 
\label{sec:CL}
We take the continuum limit of 
the dimensionless combination $X_{\rm latt}(r) \equiv r_1[V_{\rm latt}(r)-V_{\rm latt}(r_1)]$. 
Here, $r_1$ is the scale defined by $r_1^2 \frac{d V}{d r}(r_1)=1$.
We fix the $r$-independent constant by subtracting the value at $r=r_1$.\fn{
By subtracting the potential at $r=r_1$, 
we can eliminate the $r$-independent constant which exhibits a divergent behavior 
in the lattice simulations as $a \to 0$.}

To take the continuum limit $X_{\rm latt}^{\rm cont}(r)=X_{\rm latt}(r;a=0)$, we first construct sequences 
$\{X_{\rm latt}(r;a) \}_{a=a_1,a_2,a_3}$ with {\it physical distances} $r$ fixed.
We choose these reference distances $r$ as
the physical points where the coarsest lattice has original data.\fn{
Namely, $X_{\rm latt}(r;a)$ at reference distances is determined without extrapolating lattice data 
using model-assumption \eqref{inter}.} 
To obtain $X_{\rm latt}(r;a)$ for each $a$ at a reference distance $r$, 
we interpolate the lattice data, which are originally discrete,
and calculate $X_{\latt}(r;a)$ via the interpolating function. 
An extrapolation to the continuum limit of the sequence $\{ X_{\latt}(r;a) \}_{a=a_1,a_2,a_3}$
can straightforwardly be performed 
once the sequence is constructed.

To interpolate the lattice data, we use the following function form:
\be
V_{{\rm latt},d,i}^{\rm Inter.}({r})=\frac{\alpha_{d,i}}{r}+c_{0,d,i}+\sigma_{d,i}\, r+\frac{c_{1,d,i}}{r^3}+c_{2,d,i}\, r^2 \label{inter} \, ,
\ee
where $d=1,2$ and $i=1,2,3$ specify
the direction and lattice spacing, respectively.
The first three terms represent the Cornell potential,
which is consistent with the LO perturbation theory at short distances
and consistent with the string model at long distances.
If we assume this function form to be correct at the continuum limit,
correction terms can arise due to finite $a$ and $L$ effects.
The fourth term, $1/r^3$-term, is included to take into account 
the $\mathcal{O}(a^2)$ discretization error.
(Note that the potential has mass dimension one.) 
The fifth term is similarly introduced for finite $L$ effect to absorb a $1/L^3$-term.  
Furthermore, the lattice potential data are function of $\vec{r}$ rather than $r$,
since the rotational symmetry is broken.
Therefore, the coefficients can differ depending on the direction.
We interpolate the lattice data separately for each direction.
This is the reason why the subscript $d$ appears in Eq.~\eqref{inter}.

In interpolation, we use the lattice data in the range $2 a< r< L/2$.
Namely, we do not use, for instance, the data at $r=a$.
This aims at being free from serious finite $a$ and $L$ effects.
(We show in Appendix~\ref{app:includeshortest} 
that when we include the data at $r=a$,
continuum extrapolation cannot be performed reasonably.)
In fact, the function form \eqref{inter} is chosen 
assuming the hierarchy $r/a \gg 1$ and $r/L \ll 1$.

\begin{figure}[t]
\begin{center}
\includegraphics[width=9cm]{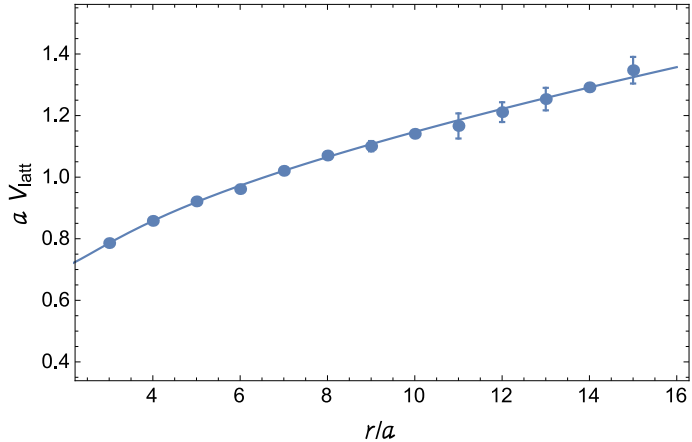}
\end{center}
\caption{Interpolation of a jackknife lattice data for $d=1$ and $i=1$.
The reduced $\chi^2$ [defined in Eq.~\eqref{chisq1}] is given by
$\chi_{\rm Inter}^2|_{d=1, i=1}/{\rm d.o.f.}=6.9/(13-5)$.}
\label{fig:inter}
\end{figure}

From the fit, we obtain an interpolating function in $a$ units.
We show an example in Fig.~\ref{fig:inter}, 
where one sees that the interpolating function can indeed fit the lattice data.
We calculate $X_{\rm latt}(r;a)$ at reference distances 
using the interpolating function in lattice units:
First, we calculate the ratio $r_1/a$ from the (slope of) interpolating function,
with which one can convert the function into $r_1$ units. 
Secondly, we read off a value of $X_{\latt}(r)$ at a reference point.

By repeating the above procedure for all the jackknife samples,
we obtain the average of $X_{\latt}(r;a)$ and its statistical error $\delta X_{\latt}(r;a_i)$ 
for each $a$ at the reference distances $r$.\fn{
More precisely, we choose the reference points as 
$r/r_1=3 \braket{a_1/r_1}, 4 \braket{a_1/r_1}, \dots , 15 \braket{a_1/r_1}$ for $d=1$, and similarly for $d=2$. 
Therefore, to calculate the average and statistical error of $X_{\latt}(r;a)$ at these distances,
we first calculate $\braket{a_1/r_1}$ by examining all the jackknife samples of $i=1$. 
Then, we read off the values of $X_{\latt}(r;a)$ at these distances for each jackknife sample.}
In our jackknife analysis, we ignore statistical fluctuation of the covariance matrix 
[$\Delta_{d,i}^{\rm latt}(r_k,r_j)$], and use [$\Delta_{d,i}^{\rm latt}(r_k,r_j)$] 
calculated with all the data for all the jackknife samples.
Our analysis using the jackknife method proceeds in the same way hereafter.

In table~\ref{tab:para1}, we summarize the fitting parameters in interpolation of Eq.~\eqref{inter}.
Generally, we have smaller statistical errors for finer lattice 
since more data are available for the interpolation.
The lattice spacings obtained via $r_1/a$ are consistent with the ones determined from the Wilson-flow scale
(where $r_1=0.311(2) ~{\rm fm}$ is assumed \cite{Bazavov:2010hj,Bazavov:2011nk,Sommer:2014mea}), 
although the former ones have much larger statistical errors.
\begin{table}
\small
\begin{center}
  \begin{tabular}{l| l| l| l| l| l| l}  \hline
 $i   ~({\rm size})$  & 
 \multicolumn{2}{|c|}{$i=1 ~(32^3\times 64$)}  &
 \multicolumn{2}{|c|}{$i=2 ~(48^3\times 96$)}  & 
 \multicolumn{2}{|c}{$i=3 ~(64^3\times 128$)}  \\
 $d$~ ($N_{i,d}$)     &
\multicolumn{2}{c|}{$d=1 ~(13)$ ~~~~~ $d=2 ~(10)$} &   
\multicolumn{2}{c|}{$d=1 ~(21)$ ~~~~~ $d=2 ~(15)$} &  
\multicolumn{2}{c}{$d=1 ~(29)$ ~~~~~ $d=2 ~(21)$}  \\ \hline
 $\chi^2/{\rm d.o.f}$         & $~~7.5/8$    & $~~2.5/5$  & $~~7.7/16$   & $~~9.6/10$     & $~~21.4/24$     &  $~~12.9/16$ \\
 $r_1/a$                          & $~~3.84(14)$    & $~~3.93(16)$       & $~~5.76(15)$     & $~~5.59(12)$      & $~~7.13(11)$      & $~~7.121(98)$\\
 $\alpha$                        & $-0.59(77)$    & $-0.31(64) $    & $-0.74(23)$    & $-0.60(20)$     & $-0.45(11)$    & $-0.577(91)$\\
 $c_0 ~[{\rm GeV}]$       & $~~2.07(66)$     & $~~1.87(56)$       & $~~3.17(24)$     & $~~2.96(23)$      & $~~3.33(14)$     & $~~3.50(12)$\\
 $\sigma~\,[{\rm GeV}^2]$  & $~~0.24(19)$     & $~~0.28(17)$       & $~~0.060(84)$   & $~~0.169(81)$    & $~~0.211(56)$   & $~~0.139(50)$\\
 $c_2~[{\rm GeV}^3]$      & $-0.004(18)$  & $-0.008(16)$    &$~~0.0223(85)$ & $~~0.0065(91)$  & $~~0.0036(68)$ & $~~0.0136(60)$ \\
 $c_1~[{\rm GeV}^{-2}]$  &  $~~0.11(33)$    & $-0.03(25)$      & $~~0.062(53)$   & $~~0.043(42)$    &$~~0.007(17)$    & $~~0.029(13)$\\
 \hline
 \end{tabular}
\end{center}
\caption{Fitting parameters in Eq.~\eqref{inter}, $r_1/a$, and the reduced $\chi^2$.
The fitting parameters are shown as dimensionful quantities (except for $\alpha$),
which are originally obtained as dimensionless parameters normalized by proper powers of $a$.
To make them dimensionful parameters, we use the lattice spacings estimated from the Wilson-flow scale.
$N_{i,d}$ is the number of the lattice data used in interpolation.
}
\label{tab:para1}
\end{table}

\begin{figure}[t!]
\begin{minipage}{0.5\hsize}
\begin{center}
\includegraphics[width=7cm]{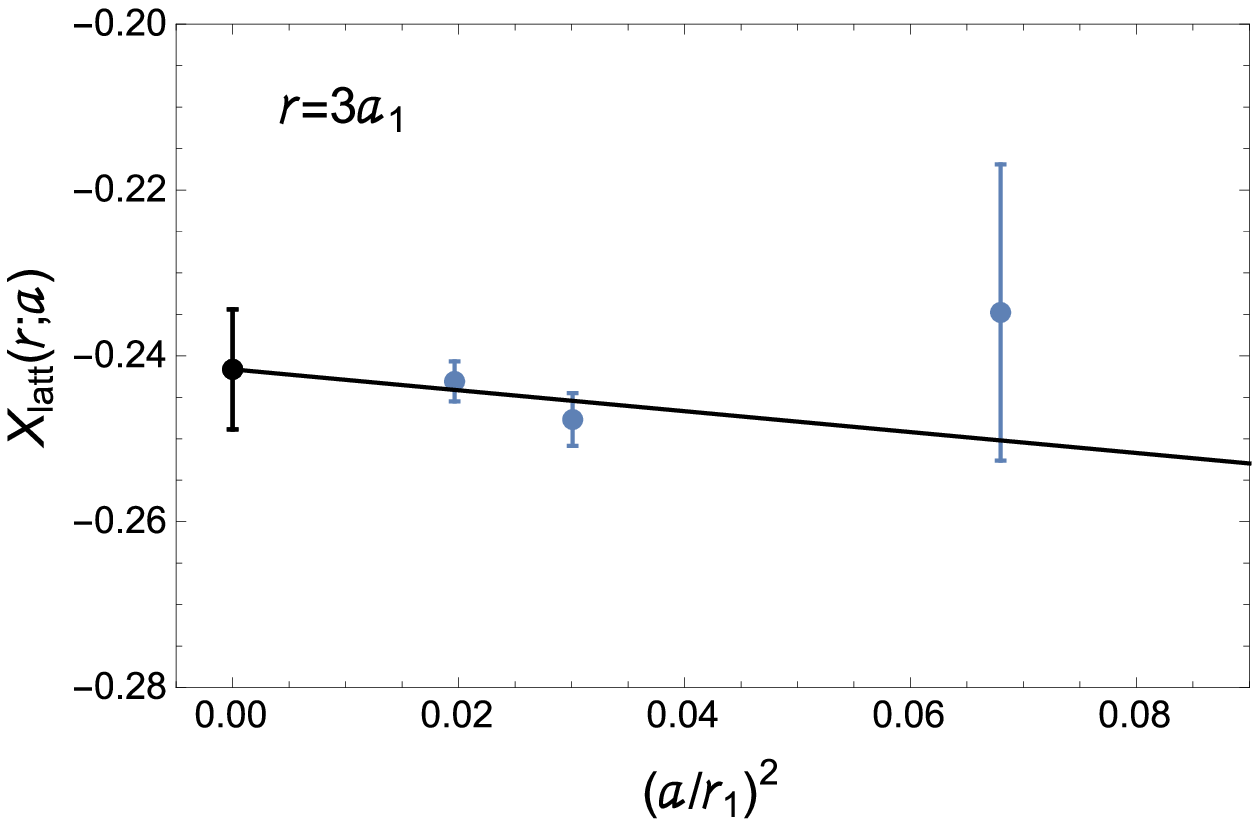}
\end{center}
\end{minipage}
\begin{minipage}{0.5\hsize}
\begin{center}
\includegraphics[width=7cm]{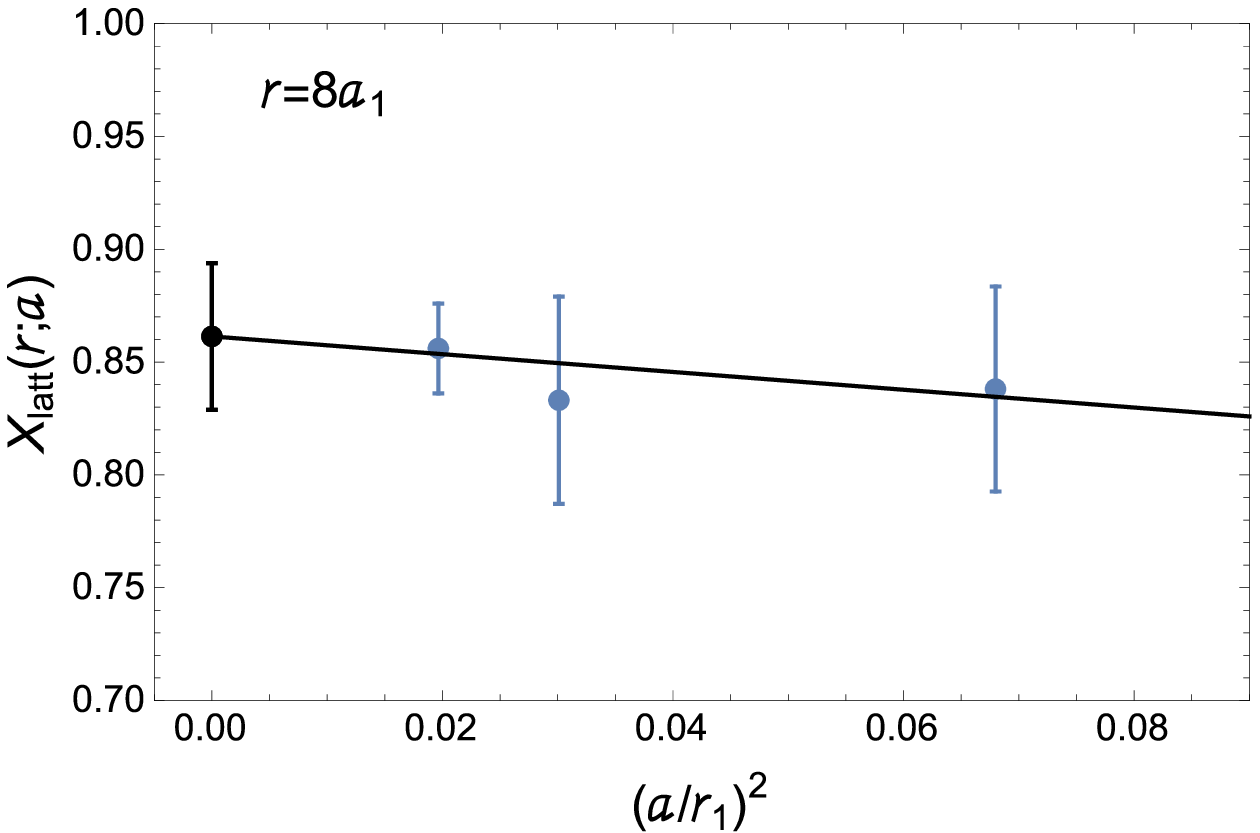}
\end{center}
\end{minipage}
\caption{$X_{\latt}(r;a)$ as functions of $(a/r_1)^2$.
We show them for $r= 3 a_1$ (left) and $r=8 a_1$ (right), 
which appear in the analysis for $d=1$, as examples.
Black lines are linear functions in $a^2$, which extrapolate the data to the continuum limit.
$\chi^2_{\rm ex}/{\rm d.o.f.}$, which is the reduced $\chi^2$ in this extrapolation, 
are $1.43$ (left) and $0.15$ (right).
The black data at $a=0$ are the extracted continuum limit with the shown statistical errors.}
\label{fig:extrapolate}
\end{figure}

\begin{figure}[t!]
\begin{minipage}{0.5\hsize}
\begin{center}
\includegraphics[width=7cm]{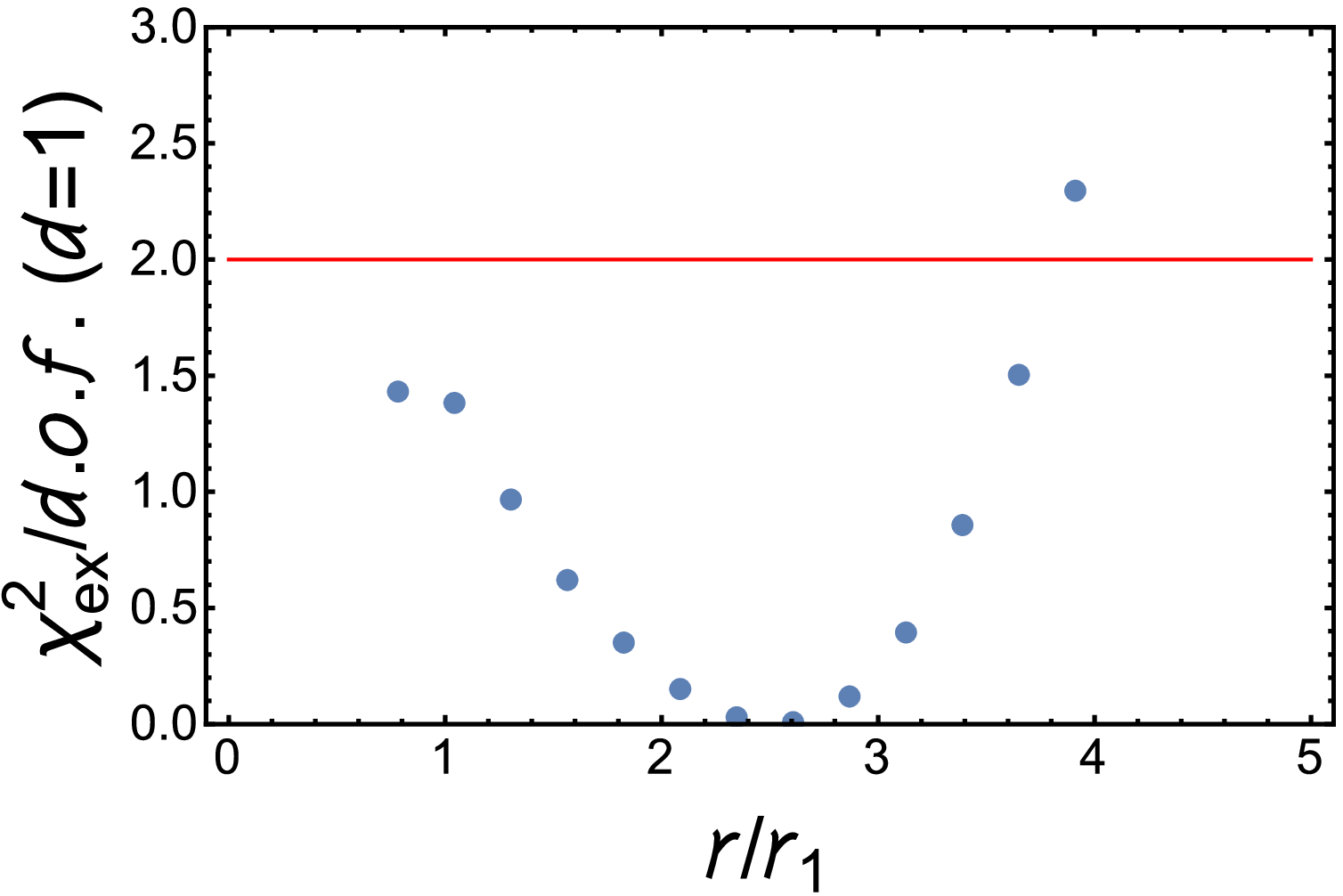}
\end{center}
\end{minipage}
\begin{minipage}{0.5\hsize}
\begin{center}
\includegraphics[width=7cm]{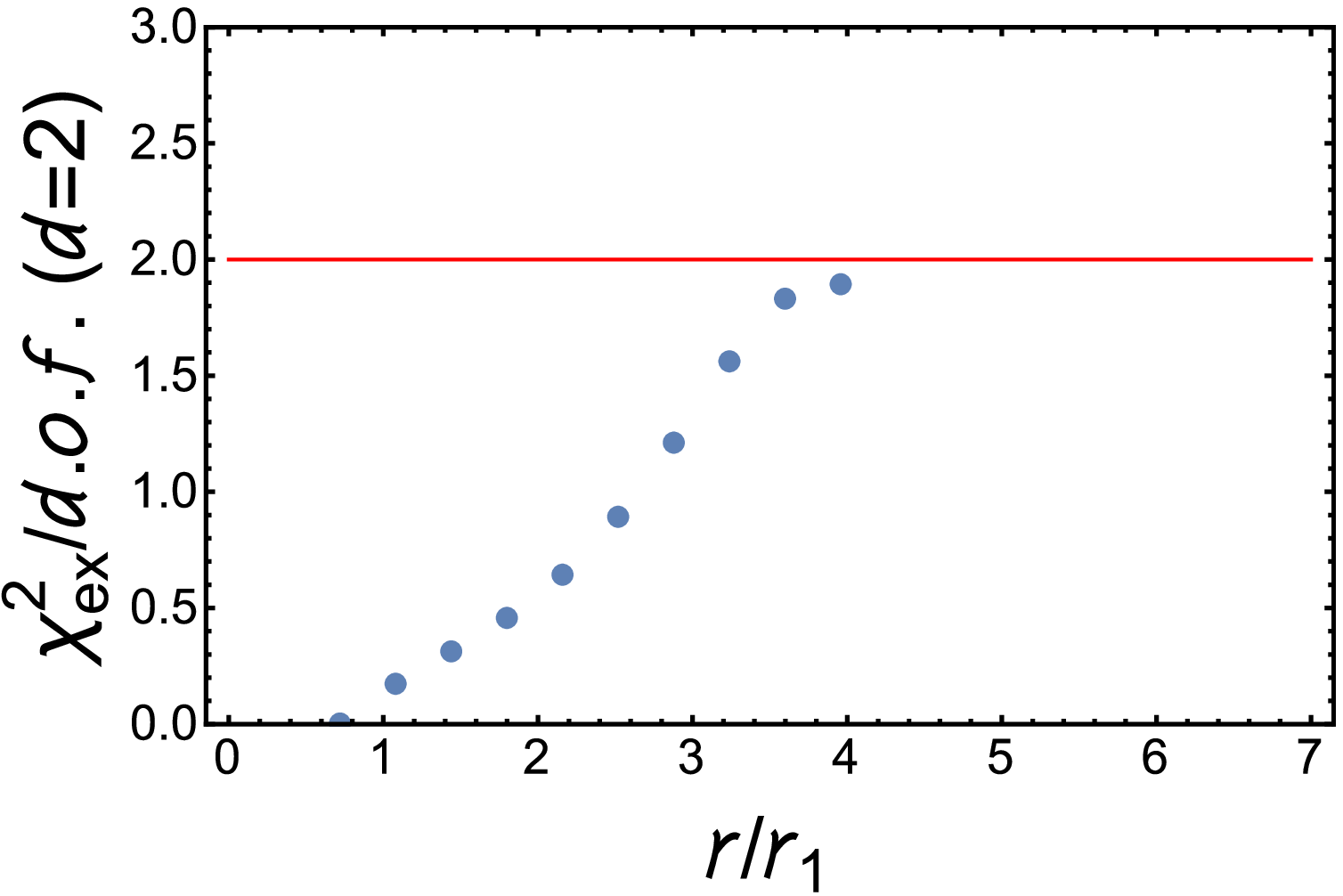}
\end{center}
\end{minipage}
\caption{The reduced $\chi^2$ in extrapolation [see Eq.~\eqref{chisqex} for definition]
for the distances where the continuum limit are taken.
As a benchmark, $\chi^2/{\rm d.o.f.}=2$ is shown by the red line.
}
\label{fig:chisq}
\end{figure}

Now we are in a position to extrapolate the sequences $\{X_{\latt}(r;a) \}_{a=a_1,a_2,a_3}$ to the continuum limit $a \to 0$.
In Fig.~\ref{fig:extrapolate}, we plot the data point of $X_{\latt}(r;a)$ as a function of $a^2/r_1^2$, 
where we choose $r=3 a_1$ and $r=8 a_1$ from the analysis for $d=1$.
We extrapolate the data by a linear fit in $a^2$,
in accord with the $\mathcal{O}(a^2)$ discretization error.
Namely, we extrapolate the lattice data at each reference distance to the continuum limit with
\be 
Y(a)=\gamma + \delta \cdot (a/r_1)^2 \, , \label{quad}
\ee
where $\gamma, \delta$ are the fitting parameters. $\gamma$ corresponds to $X_{\latt}^{\cont}$.
In Fig.~\ref{fig:extrapolate}, one can see that the data follow this function 
and are extrapolated to the continuum limit.
To see how smoothly the data are extrapolated to the continuum limit,
we show the reduced $\chi^2$ [i.e. $\chi^2_{\rm ex}/{\rm d.o.f.}$ of Eq.~\eqref{chisqex}]
at the reference distances in Fig.~\ref{fig:chisq}.
Almost all the points are extrapolated to the continuum limit smoothly
with the reduced $\chi^2$ less than 2.
Only the farthest data of $d=1$, which has $\chi^2/{\rm d.o.f.}>2$, 
is not adopted as our continuum limit result.

\begin{figure}[tbp]
\begin{center}
\includegraphics[width=11cm]{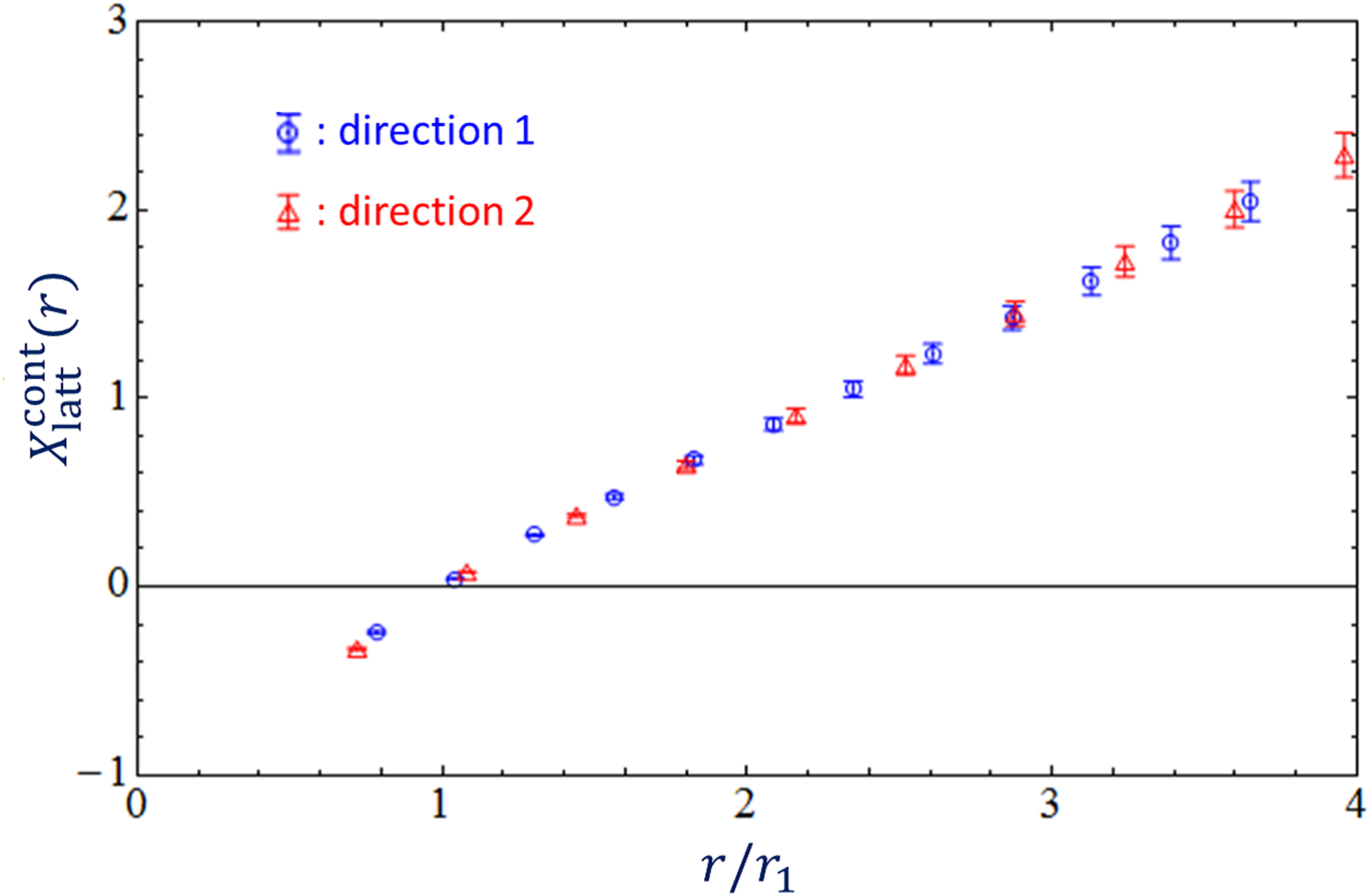}
\end{center}
\caption{Continuum limit of the lattice result, $X_{\latt}^{\rm cont}$.
Blue points originate from $d=1$ and red ones from $d=2$.}
\label{fig:contlim}
\end{figure}

In this way, we obtain the continuum limit $X_{\latt}^{\rm cont}(r)$.
It is shown in Fig.~\ref{fig:contlim}.
We also list the numerical values in table~\ref{tab:contlim}.
\begin{table}[ttbp]
\begin{minipage}{0.5\hsize}
\begin{center}
\begin{tabular}{D{.}{.}{5}D{.}{.}{7}}
\hline
\multicolumn{1}{c}{$r/r_1$}   & \multicolumn{1}{c}{$X_{\latt}^{\cont}$}     \\ \hline
0.7196    & -0.3305(93)     \\
0.7822    & -0.2416(72)     \\
1.043     & 0.04211(15)  \\
1.079    & 0.07662(33)    \\
1.304   & 0.2723(48)      \\
1.439 &  0.3775(83)     \\
1.564  &  0.478(13)        \\
1.799  &  0.645(22)        \\
1.825 &   0.672(22)        \\
2.086 &   0.861(33)         \\
2.159  &    0.908(37)  \\ \hline
\end{tabular}
\end{center}
\end{minipage}
\begin{minipage}{0.5\hsize}
\begin{center}
\begin{tabular}{D{.}{.}{4}D{.}{.}{6}}
\hline
\multicolumn{1}{c}{$r/r_1$}   & \multicolumn{1}{c}{$X_{\latt}^{\cont}$}    \\ \hline
2.347 &    1.049(43)  \\
2.519 &   1.175(52)  \\
2.607 & 1.236 (52) \\
2.868 &  1.426(63)  \\
2.878 &  1.448(66)  \\
3.129 &   1.621(75)  \\
3.238 &   1.725(81)  \\
3.389 &  1.826(88)  \\
3.598 &  2.004(97)  \\
3.650 &  2.05(10)  \\
3.958 &  2.29(12)  \\ \hline
\end{tabular}
\end{center}
\end{minipage}
\caption{Numerical results of $X_{\latt}^{\cont}(r)$.}
\label{tab:contlim}
\end{table}
The covariance matrix for $X_{\latt}^{\cont}$ 
(as well as its definition)
is presented  in Appendix~\ref{app:corrmat} 
for the first 6 points in the short distance region.\fn{
The authors can provide a larger size matrix upon request.}

%We also obtain the covariance matrix $\Delta^{\rm cont} (r_k,r_l)$ for $X_{\latt}^{\rm cont}(r)$ 
%with the jackknife method.

%We summarize the conditions for this analysis.
%\begin{itemize}
%\item Interpolate the data of $2 a < r < L/2$
%\item Interpolate the data with Eq.~\eqref{Cornell}
%\item Extract $r_1 [V(r)-V(r_1)]$
%\end{itemize}

\newpage
\subsubsection{Consistency checks and comparison with conventional methods}
\label{sec:CC}
\begin{figure}[t]
\begin{center}
\includegraphics[width=11cm]{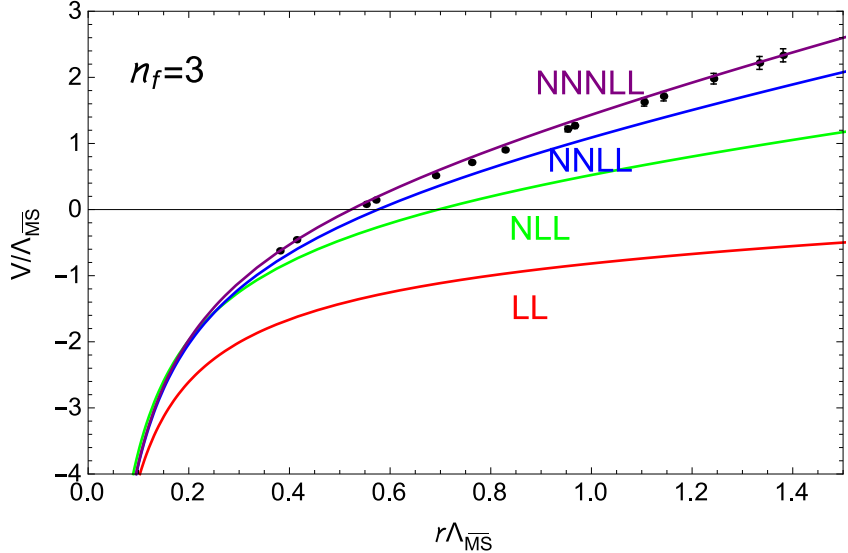}
\end{center}
\caption{Comparison of the lattice result $X_{\latt}^{\cont}$ (black dots) with 
$V_S^{\RF}$ at LL (red), NLL (green), ${\rm N^2 LL}$ (blue), and ${\rm N^3LL}$ (purple),
for the inputs $n_f=3$ and $\alpha_s(Q^2)=0.2$.
$\LMS^{\rm PDG}=336~{\rm MeV}$ is used to convert $X_{\latt}^{\cont}$ 
to $\LMS$ units. $r$-independent constant of each potential is adjusted. 
}
\label{fig:LLtoNNNLL}
\end{figure}

Before determining $\alpha_s$,
we check consistency of the OPE as given in Eq.~\eqref{RFOPE} by using the lattice data $X_{\latt}^{\cont}$.
First, we examine the perturbative part, $V_S^{\RF}$. 
We check whether $V_S^{\RF}$ has a reasonable behavior 
as we go to higher orders. 
For this purpose, we construct $V_S^{\RF}$ at ${\rm LL}$ to ${\rm N^2LL}$ in a parallel way
to Sec.~\ref{sec:BF}\fn{
The perturbative potentials at ${\rm N}^k$LL for $k=0,1,2$ do not contain IR divergences.}
and compare them with the current order prediction at ${\rm N^3}$LL.
Note that, at ${\rm N}^k \rm{LL}$, the prediction $V_S^{\RF}/\LMS^{(k+1){\text{-loop}}}$ 
is obtained as a function of $\LMS^{(k+1){\text{-loop}}} r$.
Therefore, in order to plot the $k$-th order prediction in $\LMS^{4{\text{-loop}}}$ units,
we need a conversion parameter $\LMS^{(k+1)\text{{-loop}}}/\LMS^{4{\text{-loop}}}$.
We determine these ratios by taking $\alpha_s(Q^2)=0.2$ as an input
regardless of the order of the running coupling (following Ref.~\cite{Sumino:2005cq}),
which yields $\LMS^{1{\text{-loop}}}/Q=0.0305$, $\LMS^{2{\text{-loop}}}/Q=0.0685$,
$\LMS^{3{\text{-loop}}}/Q=0.0648$, and $\LMS^{4{\text{-loop}}}/Q=0.0642$.
By regarding $Q$ as a common scale, 
we obtain the ratios $\LMS^{(k+1){\text{-loop}}}/\LMS^{4{\text{-loop}}}$ for $k=0,1,2$.
The above condition $\alpha_s(Q^2)=0.2$ assures that different order predictions 
have no large deviations at $\LMS^{4{\text{-loop}}} r \sim 0.0642$.
This is legitimate since these perturbative predictions should be accurate 
at such a high energy scale.
In Fig.~\ref{fig:LLtoNNNLL}, we plot each order prediction in $\LMS^{4{\text{-loop}}}$ units,
where the lattice result is shown as well.
The lattice result $X_{\latt}^{\cont}$ is converted to $\LMS$ units from $r_1$ units
by assuming $\LMS=\LMS^{\rm PDG} \equiv336 ~ {\rm MeV}$ 
[which corresponds to the current PDG central value of $\alpha_s(M_Z^2)$], 
and using the central value of $r_1=0.311(2) ~ {\rm fm}$.
In plotting these theoretical predictions, 
the $r$-independent constants are adjusted such that
the different order predictions have a common value at $\LMS^{4{\text{-loop}}} r=0.0642$,
and the ${\rm N^3LL}$ prediction matches the shortest distance lattice data. 
From the figure, one can see that the perturbative part, $V_S^{\RF}$,
gradually approaches the lattice result at higher order.\fn{
A similar behavior has been observed in quenched QCD in 
Fig.~17 of Ref.~\cite{Sumino:2005cq} and Fig.~13 of Ref~\cite{Sumino:2014qpa}.
Compared to the quenched case, shorter distance lattice data are absent in 
the current three-flavor lattice simulation.
Due to this, the coincidence of the lattice result and 
lower order predictions at short distances cannot be observed,
which are observed in the quenched case.
}

\begin{figure}
\begin{center}
\includegraphics[width=11cm]{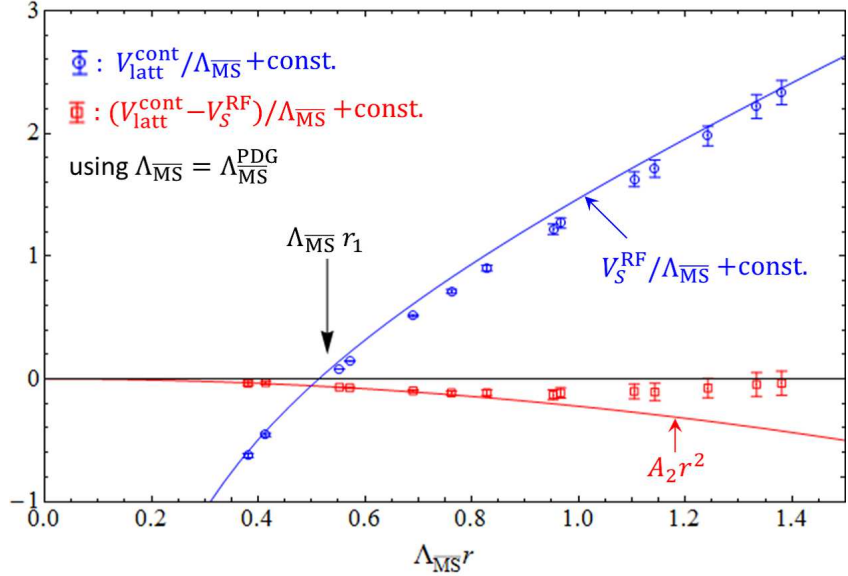}
\end{center}
\caption{Comparison of the lattice result (cont.\ limit: blue circles) and leading OPE prediction
($V_S^{\RF}/\LMS$: blue line) using $\LMS^{\rm PDG}$ and adjusting $r$-independent part.
The difference (red boxes) is fitted by const.$\times r^2$ (red line) at small $r$.}
\label{fig:consistency1}
\end{figure}

Now let us investigate a more detailed issue: 
we check if the difference between $V_S^{\RF}$ (at ${\rm N^3LL}$) and the lattice result
is $\mathcal{O}(r^2)$ as the OPE dictates. 
In Fig. \ref{fig:consistency1}, we show these two potentials in $\LMS$ units. 
The lattice potential is the same as the previous one. For the singlet potential, 
we add an $r$-independent constant so that the difference between them is zero at the origin. 
This constant is determined by a fit assuming that the difference follows 
const.$+$const.$\times r^2$.\fn{
The first six points are used in this fit.} 
We show their difference by the red boxes. 
In the difference, a linear-like behavior with an $\mathcal{O}(\LMS^2)$ coefficient,
which is observed in the lattice and perturbative potentials, vanishes. 
In fact, they can be fitted well by an $r^2$-term at short distances, as shown by the red line.\footnote{
The coefficient of the $r^2$-term (normalized by $\LMS^3$) is determined as
\be
A_2/(\LMS^{\rm PDG})^3=-0.222 \pm 0.011({\rm stat}) \, . \label{A2stat}
\ee
The reduced $\chi^2$ of this analysis is $\chi^2/$d.o.f.$=2.5/(6-2)$.}
%In the left panel of  Fig.~\ref{fig:consistency}, we show the lattice result (black) 
%and $V_S^{\RF}$ at ${\rm N^3LL}$ (purple) in $\LMS$ units.
%(The lattice potential is the same as Fig.~\ref{fig:LLtoNNNLL},
% whereas the constant of $V_S^{\RF}$ is re-adjusted.)
%In the right panel, one sees that the difference between them 
%can be well fitted by a const.+const.$\times r^2$ function at short distances.\fn{
%The constant does not matter in examining the OPE structure since 
%it can always be removed by a shift.}
%The solid and dotted lines, obtained from different fitting ranges (see the caption of the figure),
%have quite similar behaviors,
%which supports a clear $r^2$ behavior in the examined range.
From this figure, the OPE turns out to be valid up to $\LMS r \lesssim 0.8$,
corresponding to $r \lesssim 0.5 ~ {\rm fm}$ or $r^{-1} \gtrsim 0.5 ~ {\rm GeV}$.
We remark that this curve is almost unchanged even if we adopt the first 3 points in the fit, 
although we use the first 6 points in drawing the figure.

\begin{figure}[tbp]
\begin{minipage}{0.5\hsize}
\begin{center}
\includegraphics[width=7.3cm]{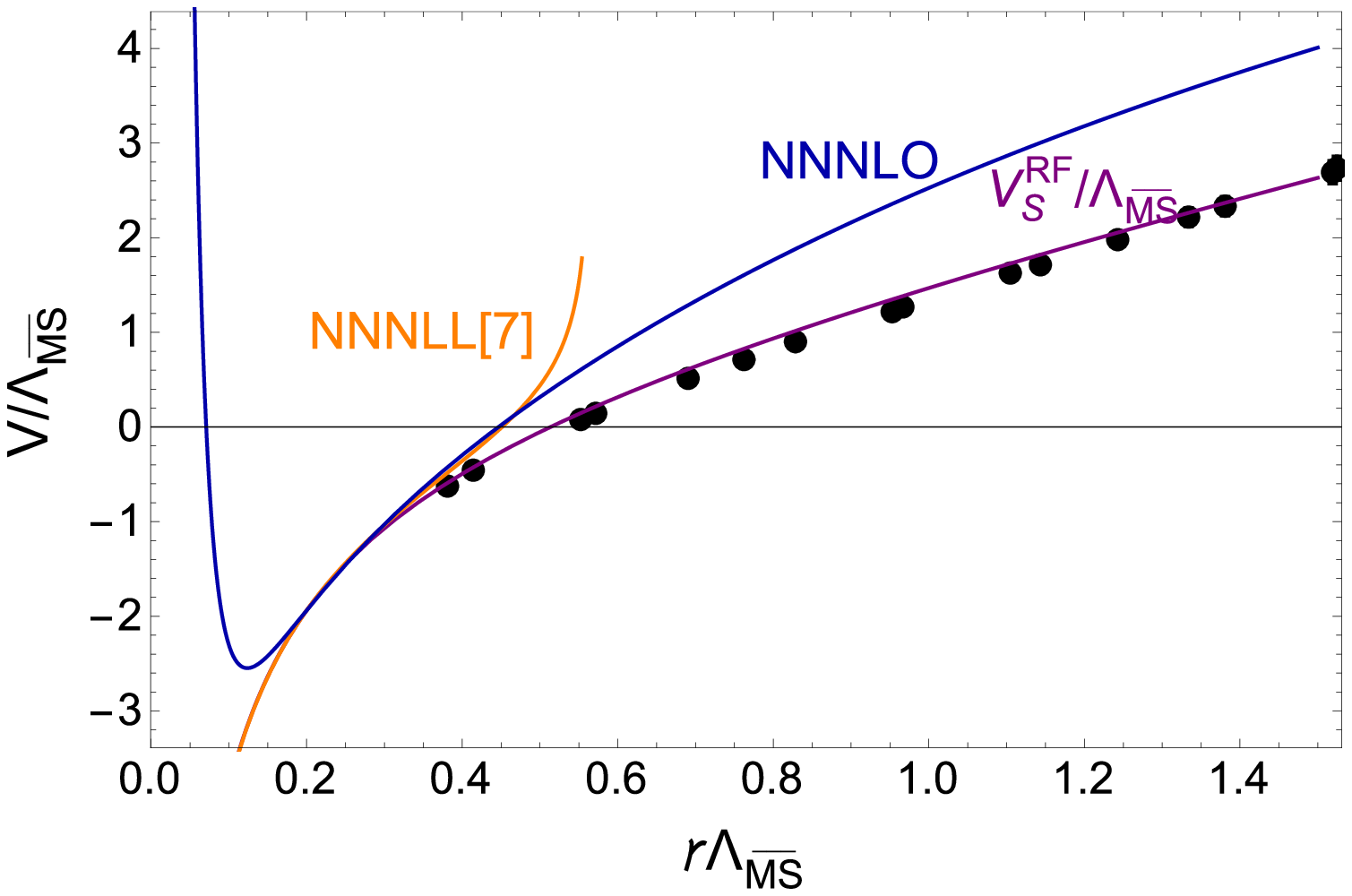}
\end{center}
\end{minipage}
\begin{minipage}{0.5\hsize}
\begin{center}
\includegraphics[width=7.3cm]{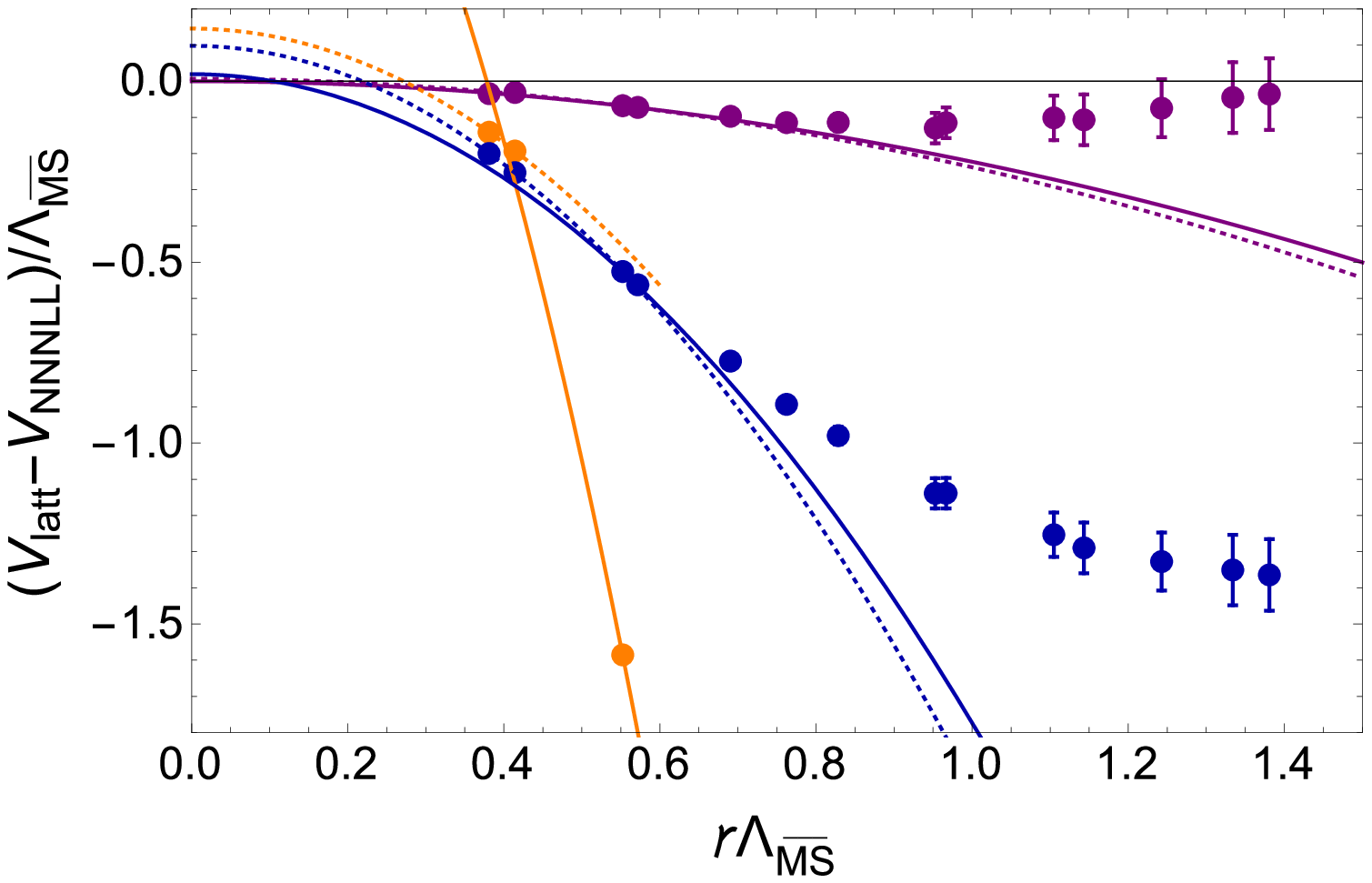}
\end{center}
\end{minipage}
\caption{(Left)
Static potentials obtained from lattice (black),
$V_S^{\RF}/\LMS$ (purple),
the ${\rm N^3}$LL prediction of Ref.~\cite{Bazavov:2014soa} (orange),
and the fixed order ${\rm N^3}$LO prediction with $\LMS/\mu=0.4$ (dark blue).
(Right) Differences between the lattice result and the theoretical predictions.
Purple data represent the difference from $V_S^{\RF}$, 
the orange ones from the ${\rm N^3}$LL prediction of Ref.~\cite{Bazavov:2014soa},
and the dark blue ones from the ${\rm N^3}$LO prediction.
The curves in this figure are const.+const.$\times r^2$ functions 
determined by fits.
The purple solid line is determined with the first six points [$\chi^2/{\rm d.o.f.}=2.5/(6-2)$],
and the purple dotted one with the first three points [$\chi^2/{\rm d.o.f.}=0.19/(3-2)$].
The orange solid line is determined with the first three points [$\chi^2/{\rm d.o.f.}=85/(3-2)$],
and the orange dotted one with the first two points [${\rm d.o.f.}=0$].
The dark blue solid line is determined with the first six points [$\chi^2/{\rm d.o.f.}=141/(6-2)$],
and the dark blue dotted one with the first three points [$\chi^2/{\rm d.o.f.}=0.002/(3-2)$]. }
\label{fig:consistency2}
\end{figure}

To clarify the impact of the above result, in Fig.~\ref{fig:consistency2}
we compare the validity range of theoretical prediction with the methods adopted 
in the preceding analyses using the static potential.
We first consider the case adopting the ${\rm N^3}$LL prediction used in the main analysis of Ref.~\cite{Bazavov:2014soa}, 
instead of $V_S^{\RF}$.
The prediction in Ref.~\cite{Bazavov:2014soa} has the $u=3/2$ renormalon and 
the unphyiscal singularity at $\LMS r \simeq 0.56$ 
unlike $V_S^{\RF}$,
although it is free from the $u=1/2$ renormalon.\fn{
The prediction of Ref.~\cite{Bazavov:2014soa} is obtained as follows.
First, the fixed order perturbative prediction for the QCD force is considered,
which is free from the $u=1/2$ renormalon.
Then, the RG improved potential is obtained by integrating the RG improved force with respect to $r$.
Here, the RG improved force has a singularity due to the running coupling
for the reason explained in footnote~\ref{fn:unphysical}.
Hence, the integration cannot be performed in the region containing this singularity.} 
Due to this singularity, the prediction cannot be obtained at $\LMS r \gtrsim 0.56$, 
and it starts to be distorted around this region
as seen from the left panel of Fig.~\ref{fig:consistency2} (orange line).
In the right panel, 
the difference from our continuum lattice result is shown (orange points).
One cannot observe a const.+const.$\times r^2$ behavior
even at $\LMS r \lesssim 0.6$.\fn{
The determination of Ref.~\cite{Bazavov:2014soa} is performed
at the high energy scale $\LMS r \lesssim 0.3$
after carefully examining the perturbative regime.}

Secondly, we consider the fixed order perturbative prediction of $V_S$ at ${\rm N^3}$LO.
It is free from the $u=1/2$ renormalon 
(once a value at some distance is subtracted) and from the unphysical singularity,
while it has the $u=3/2$ renormalon.
Since it is a fixed-order potential, 
the prediction is reliable only around the region $r \sim \mu^{-1}$.
We choose $\mu$ as $\LMS/\mu=0.4$, 
where $\mu^{-1}$ is close to the smallest $r$ among the lattice data points in the continuum limit. 
This also fixes the value of $\alpha_s(\mu^2)$ as $\alpha_s(\mu^2)=0.59$.
In the right panel of Fig.~\ref{fig:consistency2},
the difference from the lattice data is shown (dark blue points).
We can observe the OPE structure up to a certain distance region:
the first three points ($\LMS r \lesssim 0.55$) can be fitted reasonably
by a const.+const.$\times r^2$ function,
while the first six points ($\LMS r \lesssim 0.8$) cannot be.
However, we note that the result of the analysis is sensitive to a choice of the renormalization scale. 
If we make the renormalization scale twice ($\LMS/\mu=0.2$),
the range that the OPE is applicable and the coefficient of an $r^2$-term 
vary considerably. 
(We cannot take the scale $1/2$, since the running coupling constant
diverges above this scale.)
This indicates that the OPE structure is not held stable
against the higher order correction.
In contrast, if we perform a parallel analysis using $V_S^{\RF}$,\fn{
Namely, we vary $\mu=q$ to $q/2$ or $2 q$ in $\alpha_V(q^2)$
in obtaining the renormalon free part $V_S^{\RF}$.}
we always confirm that the OPE is valid up to $\LMS r \lesssim 0.8$
and the variation of the coefficient of an $r^2$-term is milder.
Namely, the OPE structure is stably observed.
This allows us to treat the nonperturbative effect (coefficient of an $r^2$-term)
in a more reasonable and reliable way.

The above arguments show that our theoretical calculation 
indeed allows us to use a range up to larger $r$ than previous studies.
We confirmed that the OPE structure is observed up to $\LMS r \lesssim 0.8$.
This is achieved thanks to a stable and reliable prediction of $V_S^{\RF}$ 
at short to relatively long distances.
This feature originates from the RG improvement, the absence of the unphysical singularity,
and the $u=3/2$ renormalon subtraction.
The latter two features result from 
subtraction of IR contributions [see Eq.~\eqref{VS}] in constructing $V_S^{\RF}$.
This subtraction removes instability caused by IR dynamics.

\noindent
{\large {\it Discussion on $r^2$ behavior}}:
We provide a supplementary explanation on at which level the OPE structure is confirmed in this study.
The OPE of pNRQCD predicts that the difference between lattice result and $V_S^{\rm RF}$ is order $r^2$,
and that a coefficient of a linear term in $r$ is zero if it is considered.
We investigate size of the coefficient, expected to be zero, 
by including a term $A_1 r$ in addition to $A_2 r^2$ for fitting the difference.
We obtain 
$A_1/\LMS^2=-0.33\pm 0.23 (\rm{stat})^{+0.25}_{-0.22}(\LMS)\pm 0.32 (h.o.)=-0.33^{+0.47}_{-0.45}$, 
where the statistical and systematic errors are combined in quadrature.
Here, we consider only the dominant systematic errors.
The error associated with $\LMS$ is estimated by varying $\LMS$ within the current PDG error $\LMS=336 \pm 17$ MeV;
the other stems from higher order uncertainty of $V_S^{\rm RF}$,
which is estimated by shifting $V_S^{\rm RF} \to V_S^{\rm RF} \pm \delta V_S^{\rm RF}$.
One can see that $A_1$ is consistent with zero. In addition to this result,
a fit with an $r^2$-term alone (assuming $A_1=0$) can be reasonably performed as shown above. 
These facts suggest correctness of the OPE.

At this stage, however, $A_1/\LMS^2$ of nearly order one is still allowed.
Hence, we refrain from making a stronger statement on confirmation of the OPE structure 
before $A_1/\LMS^2$ is constrained to be much smaller than unity.\fn{
Our result for $A_2$, the coefficient of the $r^2$-term, is consistent with zero as well
when systematic errors are considered.
However, in fact, this value is dependent on a scheme to factorize a $\mu_f$-independent part \cite{Mishima:2016vna}.
There always exists a scheme to render $A_2$ non-zero.
Therefore, validity of the OPE, which predicts $\mathcal{O}(r^2)$ difference, is exclusively
exhibited by smallness of $A_1$. ($A_1$ is independent of a scheme choice.)
 }
 
Nevertheless, it is worth making a comment on an example of the hypothesis which conflicts with the OPE.
One may find the literature where a nonperturbative linear potential 
with the coefficient of the string tension is considered {\it at short distances} $\LMS r \lesssim 0.8$.
This possibility is excluded more than at 8 $\sigma$ level from our estimate of $A_1$.\fn{
Here, we assume the string tension to be $\sigma_s/\LMS^2 \sim 3.8$.}

\subsubsection{$\alpha_s$ determination: Matching between OPE and lattice result}
\label{sec:AD}
We now determine $\alpha_s(M_Z^2)$ by matching the lattice result with the OPE.
Our determination of $\alpha_s$ reduces to the problem to find an appropriate $x=\LMS \, r_1$
such that the OPE agrees with the lattice result.
Once $x$ is determined, we obtain $\LMS$ through the value $r_1=0.311(2) ~ {\rm fm}$.
Then, we obtain $\alpha_s(M_Z^2)$ by solving the RG equation for $\alpha_s(\mu^2)$.
%where $\LMS$ appears as the integration constant.  
%(The relation between $\LMS$ and $\alpha_s(\mu^2)$ is presented in Appendix.~\ref{app:Lambda}.)

We compare the lattice and theoretical potentials in $\LMS$ units
by converting the lattice result to $\LMS$ units with $x$:
\be
\tilde{X}^{\rm cont}_{\latt}(r) = x^{-1} X_{\latt}^{\cont} \, .
\ee
The OPE prediction is given by
\be
v_{\rm OPE}(r)=\LMS^{-1} V_{\rm OPE}(r)
=\LMS^{-1} [V_S(r)+A_0+A_2  r^2 ]\, , \label{vope}
\ee
where $A_0$ and $A_2$ are the fitting parameters.
In the matching, we choose the lattice data points satisfying $\LMS^{\rm PDG} r<0.8$ 
in order for the OPE to be valid.
Hence, the first six points of Fig.~\ref{fig:contlim} are used.
The covariance matrix required in this analysis is presented in Appendix~\ref{app:corrmat}.

The results are summarized in table~\ref{tab:para1-2}.
\begin{table}[b]
\begin{center}
\begin{tabular}{c|c} \hline
$x$           &  $0.496 (24)$ \\
$A_0/\LMS$   &  $0.580 (44)$   \\
$A_2/\LMS^3$   &  $0.04 (22)$ \\ \hline \hline
$\chi^2_{\rm match}/$d.o.f.  & $0.4/(6-3)$
\end{tabular}
\end{center}
\caption{Fitting parameters in Analysis (I). Values inside parentheses denote statistical errors.
See Eq.~\eqref{chisqmatch} for definition of $\chi^2_{\rm match}$.
}
\label{tab:para1-2}
\end{table}
From the result of $x$ in this table, we obtain
\be
\LMS=315 \pm 15 ({\rm stat}) ~ {\rm MeV} \, , \label{Lambda1}
\ee
using $r_1=0.311 \, {\rm fm}$.

The obtained 3-flavor $\Lambda_{\overline {\rm MS}}$ of Eq.~\eqref{Lambda1} gives the 5-flavor coupling as
\be
\alpha_s(M_Z^2)=0.1166^{+0.0010}_{-0.0011}({\rm stat}) \, . \label{alpha1stat}
\ee
This value is obtained as follows. 
First, we calculate $\alpha_s(\mu^2)|_{n_f=3}$ below the charm $\MSb$ mass $\mu<\overline{m}_c=1.3~{\rm GeV}$, 
from the obtained $\LMS$ using Eq.~\eqref{Lambdadef} in Appendix~\ref{app:Lambda}.
Secondly, we obtain the 4-flavor coupling at the charm $\MSb$ mass (which we take as a matching scale)
$\mu=\overline{m}_c$, using the 3-loop matching equation \cite{Chetyrkin:1997sg}.
Then, we obtain $\alpha_s(\mu^2)$ for $\overline{m}_c < \mu < \overline{m}_b=4.2~{\rm GeV}$
by solving the RG equation for $n_f=4$.
Similarly, we obtain the 5-flavor coupling at the bottom $\MSb$ mass by the matching equation.
Then, we obtain the coupling at the $Z$ boson mass $M_Z=91.187~{\rm GeV}$, 
$\alpha_s(M_Z^2)$, by solving the RG equation for $n_f=5$.
We solve the RG equation for $\alpha_s(\mu^2)$ numerically; see footnote~\ref{fn:running}.

For convenience, we summarize the conditions used in our main analysis, 
with which we determine the central value of $\alpha_s(M_Z^2)$.
\begin{itemize}
\item{Controlling finite $a$ and $L$ effects: Lattice data in the range $2 a <r< L/2$ are used in interpolation}
\item{Interpolating function: Cornell type potential [Eq.~\eqref{inter} }]
\item{Lattice result extrapolated to $a=0$: $X(r)=r_1[V_{\latt}(r)-V_{\latt}(r_1)]$}
\item{Singlet potential: $V_S^{\RF}(r)$ defined by Eq.~\eqref{VSRF}, which has ${\rm N^3 LL}$ accuracy}
\item{Regularization of US divergence: Prescription I [Eq.~\eqref{regI}] }
\item{Matching range (Used lattice data in the continuum limit): $\LMS^{\rm PDG} r <0.8$}
\item{Conversion of $x$ to $\LMS$: Central value of $r_1=0.311 (2) \, {\rm fm}$}
\end{itemize}

We now estimate systematic errors of our determination.
For this purpose, we perform re-analyses by changing the conditions as follows
and examine variations of determined $\alpha_s(M_Z^2)$. 

\begin{itemize}
\item{{\it Finite $a$ effects}: We use the lattice data of $a < r<L/2$ in interpolation such that 
the shorter distance points  $r \gtrsim a$ are included, although we still omit the data at $r=a$.\fn{
For the case including the data at $r=a$, see Appendix~\ref{app:includeshortest}.}}
\item{{\it Interpolating function}: The Cornell potential has a defect that it does not have a logarithmic correction
in the Coulomb part at short distances
where it should be $1/(r \log(r \LMS))$ 
rather than $1/r$.
Such a logarithmic correction follows from the one-loop $\beta$ function.
We replace the Coulomb part by the one consistent with the one-loop $\beta$ function:
\be
{V'}_{\latt}^{\rm Inter.}(r)=V_C^{{\text{large-}}\beta_0}(r \LMS^{1{\text{-loop}}})+c_0+\sigma r+\frac{c_1}{r^3}+c_2 r^2 \, , \label{Vinter2}
\ee
where $V_C^{{\text{large-}}\beta_0}(r)$ is the Coulomb-like potential calculated in the large-$\beta_0$ approximation
according to the method of Ref.~\cite{Sumino:2005cq} or Appendix~\ref{app:VSRF}. 
Its asymptotic form is given by\fn{
In interpolating the lattice unit potential with the above fitting form,
we introduce $y= \LMS^{1{\text{-loop}}} a$ as the fitting parameter 
in order to convert $V_C^{{\text{large-}}\beta_0}/\LMS^{1{\text{-loop}}}$
to $a$ units as
\be
a {V'}_{\latt}^{\rm Inter.}(r)=y [V_C^{{\text{large-}}\beta_0}/\LMS^{1{\text{-loop}}}](y r/a )+\dots \, .
\ee
}
\be
[V_C^{{\text{large-}}\beta_0}/\LMS^{1{\text{-loop}}}](\rho= r \LMS^{1{\text{-loop}}}) \to
\begin{cases}
& -C_F \frac{2 \pi}{\beta_0} \frac{1}{\rho \log(1/\rho)}   ~~~~(r \LMS  \ll 1)  \\
& -C_F \frac{4 \pi}{\beta_0 \rho}    ~~~~~~~~~~~~~~~~~(r \LMS  \gg 1) \, .
\end{cases}
\ee
%It has the correct logarithmic dependence.
}
\item{{\it Subtraction point}: We take the continuum limit of
\be
r_1 [V_{\latt}(r)-V_{\latt}(0.8 r_1)] \, ,
\ee
where the subtraction point is changed.}
\item{{\it Higher order uncertainty}: We replace $V_S^{\RF}$ in matching as
\be
V_S^{\RF} + t \delta V_S^{\RF}
\ee
with $t=-1$ or $1$ in order to estimate higher order uncertainty; see Eq.~\eqref{deltaV} for $\delta V_S^{\RF}$.}
\item{{\it US regularization}: We adopt the regularization prescription II, given by Eq.~\eqref{regII}.
We choose $\mu_{\rm US}$ as $3 \LMS$ and $4 \LMS$.}
\item{{\it Matching range}: We vary the range of the lattice result used in the matching as
\be
\LMS^{\rm PDG} r < 0.7 ~\text{or}~ 0.9 
\ee
to examine the stability of the OPE truncated at $\mathcal{O}(r^2)$.}

\item{{$r_1$}: We vary $r_1$ in the range $r_1=0.311 \pm 0.002 ~ {\rm fm}$.}
\end{itemize}
The estimated systematic errors are summarized in table~\ref{tab:sys1}.
\begin{table}[t!]
\small
\begin{center}
\begin{tabular}{c||ccccccc}
\hline
                                & ~~finite $a$~~            & interpol.\ fn.     & ~subt.\ point  &  ~~~~h.o.~~ ~                          &    ~~~US~~~                      &~range~ & $r_1$        \\ \hline
Obtained value         &    $- 4$           &   $+ 4$                      &  $- 8$              &  $^{+14~(t=-1)}_{-12~(t=1)}$   & $^{+1~ (3\LMS)}_{-0 ~ (4 \LMS)}$  &  $^{+5~(0.7)}_{-8~(0.9)}$ &  $\pm 1$        \\ \hline 
Assigned error       &   $\pm 4$         &   $ \pm 4$                  &  $\pm 8$        & $^{+14}_{-12}$                       & $\pm 1$                   &  $^{+5}_{-8}$      &  $\pm 1$ \\ \hline
\end{tabular}
\end{center}
\caption{Estimates of systematic errors in Analysis (I) from variations of the central value of $\alpha_s(M_Z^2)$ 
in units of $10^{-4}$ when varying the analysis conditions.
In the upper row, variations are shown. (Detailed conditions are shown inside brackets).
Assigned systematic errors are shown in the lower row. 
}
\label{tab:sys1}
\end{table}

By taking the root-sum-square of the errors, we obtain
\be
\alpha_s(M_Z^2)=0.1166 ^{+0.0010}_{-0.0011}({\rm stat})^{+0.0018}_{-0.0017}({\rm sys}) 
\ee
from Analysis (I).

\subsection{Analysis (II): Global fit}
\label{sec:AnaII}

In Analysis (I), an interpolating function is assumed  
in order to take the continuum limit of the potential, although the exact functional form is unknown. 
This is a short-coming from the viewpoint of first principles.
In Analysis (II), we perform a first-principle determination,
without using such a model-like interpolating function. 
This is achieved by a global fit
in which the continuum extrapolation and the matching with a theoretical calculation
are performed at once.

This analysis is based on the idea that the OPE prediction should be correct
at short distances and coincide with the lattice data 
once the discretization errors are removed.
Then, the OPE is matched with the modified lattice data 
which can be regarded as the result in the continuum limit:
\be
V_{{\rm latt}}^{\rm cont}(r)=
V_{{\rm latt},d,i}(r)
-\kappa_{d,i} \lt(\frac{1}{r}- \lt[\frac{1}{r} \rt]_{d,i}  \rt)
+f_d \frac{a_i^2}{r^3} - c_{0,d,i}\, . \label{Vcont}
\ee
Discretization errors contained in the original lattice data $V_{{\latt},d,i}$
are removed by the second and third terms (depending on $i$ and $d$), 
and the last term adjusts the $r$-independent constant;
$\lt[ \frac{1}{r} \rt]$ is the LO result in the lattice perturbation theory,
which deviates from a smooth $1/r$-function due to finite $a$ and $L$ effects.
Hence, the second term removes the discretization error at the tree-level.
Note that the tree-level potential is given by a one-gluon exchanging diagram
and is order $\alpha_s$. 
Here, $\kappa$ is regarded as an effective coupling of lattice perturbation theory,
and is treated as a fitting parameter.
The third term extrapolates the data to the continuum limit 
by removing the remaining error of order $\alpha_s^2 a^2$.
In Eq.~\eqref{Vcont}, we do not include a term related to finite $L$ effects
because in Analysis (I) the finite $L$ effects, shown by the size of $c_2$,
turn out to be small (see table~\ref{tab:para1}).
On the other hand, the term $f_d a_i^2/r^3$,
which is also small in Analysis (I) (see $c_1$ in table~\ref{tab:para1}),
is kept just in case because Analysis (II) uses shorter distance data. 

We perform matching by converting lattice and theoretical potentials to ${\rm GeV}$ units.
Lattice data are converted to these units using $a$'s estimated by the Wilson-flow scale.
The theoretical potential is converted with $z=\LMS [\rm GeV]$,
which is unknown in advance and thus is treated as a fitting parameter.
Therefore, an OPE prediction used here is given by
\be
V_{\rm OPE}(r)=z [V_S/\LMS](z r)+A_2 r^2 \, . \label{OPEII}
\ee
Since an $r$-independent constant is already 
included in Eq.~\eqref{Vcont}, it is not included here.

In matching, we adopt the lattice data in the range $r \LMS^{\rm PDG}<0.6$.
Here, we choose a shorter distance region than in Analysis (I)
since we have more available data points.
It serves to reduce our dominant error, given by higher order perturbative uncertainty.
In this analysis, we do not omit short distance data at $r \sim a$, 
and in particular we include the data at $r=a$.
(Note that the continuum extrapolation cannot be taken reasonably if we
include the data at $r=a$ in Analysis (I), as discussed in Appendix~\ref{app:includeshortest}.)
Thus, we take into account the tree-level correction,
which is powerful to remove the discretization error at short distances (where perturbation theory works)
and does not need the hierarchy $r \gg a$.\fn{
In Analysis (I), the tree-level correction is not considered.
It is because we try to examine the validity range of the OPE, and thus
need the continuum limit result in a wide distance region, where the tree-level correction is not generally valid.
}
The number of the $i$-th lattice data used in the matching is $7, 10, 13$ points for $i=1,2,3$, respectively.

In this analysis, we determine 16 parameters in total:
$\LMS$, $A_2$, six tree-level correction parameters $\kappa$'s,
two $f$'s, and six $r$-independent constants $c_0$'s.
Due to the nature of this global fit, the lattice result in the continuum limit is determined
such that it matches with the OPE prediction.
In this respect, the continuum extrapolation is not taken within lattice simulation,
but it is constrained by the OPE prediction.\fn{
We do not interpolate each lattice data (for each $d$ and $i$).
The continuous function appearing in this analysis is only the OPE prediction,
and each lattice data is modified to agree with this function according to Eq.~\eqref{Vcont}.}
Thus, the lattice data in the continuum limit in Analyses (I) and (II)
have qualitatively different meanings.

In this global fit, we obtain
\be
\LMS=334 \pm  10 ({\rm stat}) \, {\rm MeV} \, . \label{Lambda2}
\ee
We summarize the other parameters in this global fit in table~\ref{tab:para2}.
The reduced $\chi^2$ of this fit is 
$\chi^2_{\rm GF}/{\rm d.o.f.} = 8.7/(30-16)$ [see Eq.~\eqref{chiGF} for definition of $\chi^2_{\rm GF}$],
showing the validity of the analysis.
$A_2$ is consistent with our previous estimate Eq.~\eqref{A2stat}, 
which is obtained while assuming $\LMS=\LMS^{\rm PDG}$. (It is also consistent with Analysis (I).)
$f$'s are consistent with zero, which suggests that the discretization error is quite small after 
the tree-level correction is taken into account. 
\begin{table}
\small
\begin{center}
\begin{tabular}{l| l| l| l| l| l| l} \hline
 $i   ~({\rm size})$  & 
 \multicolumn{2}{c|}{$i=1 ~(32^3\times 64$)}  &
 \multicolumn{2}{c|}{$i=2 ~(48^3\times 96$)}  & 
 \multicolumn{2}{c}{$i=3 ~(64^3\times 128$)}  \\
 $d$~ ($N_{i,d}$)     &
\multicolumn{2}{c|}{$d=1~(4) $ ~~~~~ $d=2~(3) $} &   
\multicolumn{2}{c|}{$d=1~(6) $ ~~~~~ $d=2~(4) $} &  
\multicolumn{2}{c}{$d=1~(8) $ ~~~~~ $d=2~(5)$}  \\ \hline
$\kappa$ &~~$0.19(15)$ & ~~$-0.26(85)$ & ~~$0.27(12)$ & ~~$-0.53(88)$ & ~~$0.27(11)$ & ~~$-0.57(91)$\\
$c_0~[{\rm GeV}]$  & ~~$2.245(11)$  & ~~$2.300(87)$  & ~~$3.012(11)$  & ~~$3.099(89)$  & ~~$3.546(10)$  & ~~$3.631(86)$\\ \hline
$\chi^2$ &  \multicolumn{6}{c}{ ~~~~$\chi^2/{\rm d.o.f.}=8.7/(30-16)$  ~~~~~(global fit)} \hspace{2cm} \\
$f_d$  &  \multicolumn{6}{c}{ \hspace{1.5cm} $f_1=0.0004(18),~~~f_2=-0.025(32)$ ~~~(common to all $i$)} \\
$A_2$ &  \multicolumn{6}{c}{ $A_2 =-0.0091(54) ~{\rm GeV^3}$ ~~~(common to all $i,d$)} \\
\hline
\end{tabular}
\end{center}
\caption{Fitting parameters in Analysis (II). Only statistic errors are shown.
$N_{i,d}$ expresses the number of data points for direction $d$ of the $i$-th lattice.}
\label{tab:para2}
\end{table}

\begin{figure}[t!]
\begin{center}
\includegraphics[width=10cm]{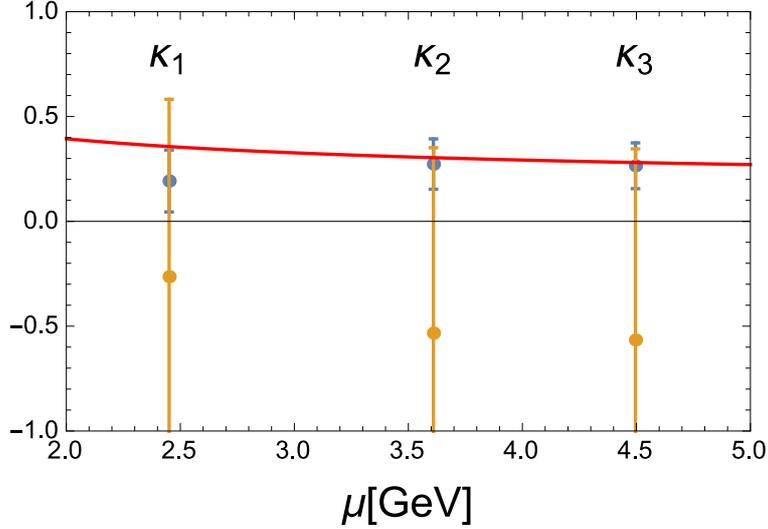}
\end{center}
\caption{Determined values of $\kappa$. Blue (orange) data 
represent $\kappa_{i,d=1}$ ($\kappa_{i,d=2}$).
Red curve represents the running of 
$C_F \alpha_s(\mu^2)$ assuming $\LMS=\LMS^{\rm PDG}$ and $n_f=3$.
We plot $\kappa_{d,i}$ at $\mu=a_i^{-1}$ for comparison.}
\label{fig:alphas}
\end{figure}

To check if the tree-level correction works in a reasonable way,
we show the determined values of $\kappa$'s in Fig.~\ref{fig:alphas}.  
In this figure, we compare $\kappa_{d,i}$ with its naively expected value, $C_F \alpha_s(\mu^2)$,
while taking the renormalization scale as $\mu=a_i^{-1}$.
Note that $C_F=4/3$ is multiplied since the LO result in the continuum theory 
is $V_{\rm QCD}(r)|_{\rm tree}=-C_F \alpha_s/r$.
In plotting the running coupling, we assume $\LMS=\LMS^{\rm PDG}$ and $n_f=3$.
The determined $\kappa$'s are consistent with the naively expected values 
within the statistical errors,
which supports validity of our analysis.
Large statistical errors for $\kappa_{d=2,i}$ stem from 
the small number of data for $d=2$.

\begin{figure}[t]
\begin{center}
\includegraphics[width=13cm]{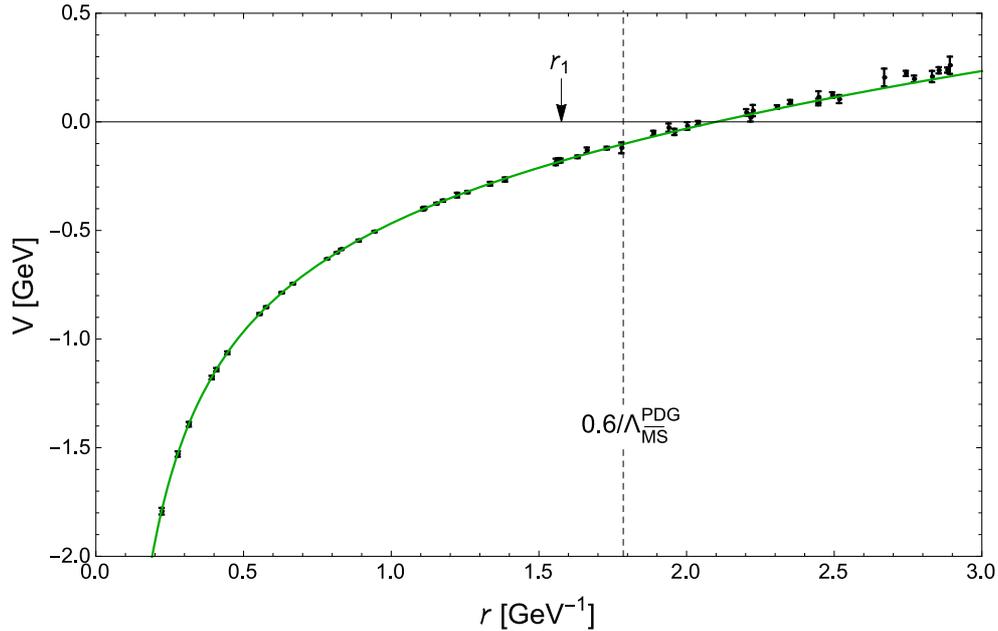}
\end{center}
\caption{Lattice result in the continuum limit (black points) and the OPE calculation (green)
determined simultaneously by the fit in Analysis (II).
The distance region used in this fit $r \LMS^{\rm PDG}<0.6$ 
is shown by the dotted line.
For reference, $r=r_1$ is also shown.}
\label{fig:Globalfit}
\end{figure}

We show the lattice result in the continuum limit [Eq.~\eqref{Vcont}] and 
the OPE prediction [Eq.~\eqref{OPEII}] which are determined by the fit 
in Fig.~\ref{fig:Globalfit}.
From the figure, one can see that the analysis is performed reasonably,
and that the OPE calculation and the lattice result are mutually consistent 
in the examined region.

The obtained $\LMS$ in Eq.~\eqref{Lambda2} gives
\be
\alpha_s(M_Z^2)=0.1179 \pm 0.0007 ({\rm stat}) \, . \label{alpha2stat}
\ee
The procedure to obtain $\alpha_s(M_Z^2)$ is the same as for Analysis (I).

For convenience, we summarize the conditions used in our main analysis, 
with which we determine the central value of $\alpha_s(M_Z^2)$.
\begin{itemize}
\item{Controlling finite $a$ effects: The data at $r \geq a$ are used combined with the tree-level correction.}
\item{Singlet potential: $V_S^{\RF}(r)$ defined by Eq.~\eqref{VSRF}, which has ${\rm N^3 LL}$ accuracy}
\item{Regularization of US divergence: Prescription I [Eq.~\eqref{regI}] }
\item{Quark masses: We use the lattice data obtained with unphysical quark mass 
inputs and $V_S^{\RF}$ in the massless quark approximation.}
\item{Matching range: $\LMS^{\rm PDG} r <0.6$}
\end{itemize}

Now we estimate systematic errors of our determination.
We perform the following re-analyses.
Since this analysis will give our final result,
some additional aspects are studied in comparison to Analysis (I).
\begin{itemize}
\item{{\it Finite $a$ effects}: We use the lattice data at $r \geq 2 a$.
In this case, we omit the tree-level correction by setting $\kappa$'s to zero.
This is because the role of the tree-level correction is similar to that of the $a^2/r^3$-term 
under the current hierarchy $a/r \leq 1/2$, where the tree-level correction is well approximated 
in expansion in $a/r$.\fn{
If we include both $\kappa$'s and $f$'s, the fit is destabilized due to a flat direction 
caused by this degeneracy.
We adopt the $a^2/r^3$-term rather than the tree-level correction since
the tree-level correction becomes less reliable 
when the matching range shifts to lower energy region. \label{fn:treevsa2}}
}
\item{{\it Higher order uncertainty}: We replace $V_S^{\RF}$ in matching as
\be
V_S^{\RF} + t \delta V_S^{\RF}
\ee
with $t=-1$ or $1$ in order to estimate higher order uncertainty; see Eq.~\eqref{deltaV} for $\delta V_S^{\RF}$.}
\item{{\it US regularization}: We adopt the regularization method II, given by Eq.~\eqref{regII}.
We have chosen $\mu_{\rm US}$ as $3 \LMS$ and $4 \LMS$.}
\item{{\it Mass effects}: Lattice data are obtained with the unphysical mass inputs.
We include an estimation of this mass difference effect as a systematic error,
since we do not know the true correction.
We estimate the lattice data on the physical point as
\be
V_{{\latt}, d, i}(r; m^{{\latt},i}) \to 
V_{{\latt}, d, i}(r; \overline{m})=V_{{\latt}, d, i}(r; m^{{\latt},i})+[V_{{\rm pt},i}(r; \overline{m})-V_{{\rm pt},i}(r; m^{{\latt},i})] \, , \label{masscorr}
\ee
where $\overline{m}$ is the $\MSb$ masses for the light quarks ($u,d,s$);
$V_{\rm pt}$ is the finite mass correction 
evaluated in perturbative QCD at ${\rm N^2LO}$ \cite{Hoang:2000fm, Melles:2000dq, Recksiegel:2001xq}.
More precisely, it is a function of $\{r, m , \mu \}$ of the form 
\be
V_{\rm pt}(r;m)=c_1(r, m) \alpha_s^2+c_2 (r, m , \mu) \alpha_s^3 \, ,
\ee 
which vanishes in the limit $m \to 0$.
In the above estimation, we take the renormalization scale as $\mu=a_i^{-1}$ 
and choose $\alpha_s$ as 0.27, 0.23, 0.21 for $i=1,2,3$, respectively,
so that it is close to $\alpha_s(\mu^2=a_i^{-2})$.
For the $\MSb$ mass values of the light quarks, we use 
$\overline{m}_u=2.2 ~ {\rm MeV}, \overline{m}_d=4.7 ~ {\rm MeV}, \overline{m}_s=96 ~ {\rm MeV}$.
To model a nonperturbative effect, 
we also substitute a constituent quark mass of $300~{\rm MeV}$ for $\overline{m}$ in Eq.~\eqref{masscorr}
as an additional test (while the other parameters are kept fixed).
Furthermore, since $V_S^{\rm RF}$ is obtained by treating the light quarks as massless,
the finite mass effects are also added to $V_S^{\rm RF}$ as
\be
V_S^{\rm RF} \to V_S^{\rm RF}+V_{{\rm pt}}(r; \overline{m}) \, .
\ee 
For this $V_{\rm pt}$, we take $\mu=3 ~ {\rm GeV}$ and $\alpha_s=0.25$.
In this way, we estimate both theoretical prediction and lattice result at the physical point.
}
\item{{\it Matching range}: We vary the range of the lattice result used in the matching as
\be
\LMS^{\rm PDG} r < 0.5 ~\text{or}~ 0.8 
\ee
to examine the stability of the OPE truncated at $\mathcal{O}(r^2)$.}
\item{{\it Factorization scheme}: In extracting the renormalon free part $V_S^{\rm RF}$,
we rewrite the integrand of $V_S$ by a complex function; see \eqref{rewrite} in Appendix~\ref{app:VSRF}.
In general, there can be other choices for this function, and in this regard, we have chosen a certain scheme.
A different scheme practically causes an $\mathcal{O}(r^3)$ difference in the OPE prediction truncated at $\mathcal{O}(r^2)$;
see Ref.~\cite{Mishima:2016vna} for details.\fn{
In Ref.~\cite{Mishima:2016vna}, it is shown that the current choice is natural from the viewpoint of analyticity.}
To see an effect of this scheme dependence, we add an $A_3 r^3$-term in the fit
so that this scheme dependence is absorbed.
(Note that, in order to determine coefficients up to higher orders in $r$, 
a wider fitting range is required.
We choose the range in this analysis as $\LMS^{\rm PDG} r<0.8$,
where $A_2$ and $A_3$ are stable against variation of the range.\fn{
This range is chosen after studying the stability for various ranges.})
\item{{\it Lattice spacing}: The lattice spacing $a$, used to convert $r$ and $V_{\rm latt}$
into physical units, has an error as shown in table~\ref{tab:param},  
and has an additional error of 1.7 \% due to the uncertainty of the physical value of the Wilson-flow scale \cite{Borsanyi:2012zs}.
For the former one, the error is etimated by the largest deviation 
detected from a set of six data, $\{ \{a_1 \pm \delta a_1, a_2, a_3 \}, \{a_1, a_2 \pm \delta a_2, a_3 \}, \{a_1, a_2, a_3 \pm \delta a_3 \} \}$, where $\delta a_i$ denotes the error shown in table~\ref{tab:param}.  
The error associated with the latter is estimated by shifting all the $a$'s 
simultaneously by its uncertainty.
By combining these two errors in $\alpha_s(M_Z^2)$ in quadrature, 
the uncertainty from the lattice spacing is estimated.}
}
\end{itemize}
The estimated systematic errors are summarized in table~\ref{tab:sys2}.
Some error sources included in Analysis (I) are absent 
thanks to the first-principle nature of this analysis.
\begin{table}
\scriptsize
\begin{center}
\begin{tabular}{c||ccccccc}
\hline
                        & ~~finite $a$~~&  ~~~~h.o.~~ ~                                     &     ~~~US~~~                           &  Mass                        & ~range~     & fact. scheme   & latt. spacing     \\ \hline
Obtained value &   $- 2$      &  $^{+12~(t=-1)}_{-10~(t=1)}$  & $^{+2~(3 \LMS)}_{+0~(4 \LMS)}$   &  $-0 (^{\rm \MSb mass}_{\rm Constituent~mass})$  &   $^{-3~(0.5)}_{-4~(0.8)}$     &  $+ 3$  & $\pm 4$ \\ \hline 
Assigned error  &  $\pm 2$  &      $^{+12}_{-10}$               &      $\pm 2$                 & $ \pm 0$           &  $\pm4$      &  $\pm 3$  & $\pm 4$ \\ \hline
\end{tabular}
\end{center}
\caption{Estimates of systematic errors in Analysis (II) from variations of the central value of $\alpha_s(M_Z^2)$ 
in units of $10^{-4}$ when varying the analysis conditions.
In the upper row, variations are shown. (Detailed conditions are shown inside brackets).
Mass effects are negligibly small in both cases.
Assigned systematic errors are shown in the lower row. 
}
\label{tab:sys2}
\end{table}
In addition, most of the systematic errors are reduced compared to Analysis (I).
In particular, the higher order uncertainty is smaller 
since a shorter distance region is used; see Fig.~\ref{fig:theoerror}.
The mass effects turn out to be negligibly small even if we consider the constituent quark mass.
This is because we are probing a sufficiently short-distance region.
(Additional analyses on systematic errors are given in Appendix~\ref{app:further}.)

As a result of Analysis (II), we obtain
\be
\alpha_s(M_Z^2)=0.1179 \pm 0.0007 ({\rm stat}) ^{+0.0014}_{-0.0012} (\rm sys) \, . 
\ee

\subsection{Summary of results}
We have performed two determinations of $\alpha_s$.
In Analysis (I), which is a preparatory analysis, we first took the continuum limit of the lattice data,
and then we matched the result with the OPE prediction.
Although this analysis partially relies on a model-like assumption,
we explicitly showed that (a) the continuum extrapolation of the lattice data
can be taken smoothly, and 
that (b) the OPE combined with our renormalon subtraction is indeed consistent; see Fig.~\ref{fig:consistency1}.   
We obtained $\LMS=315\pm15 ({\rm stat})^{+26}_{-25} ({\rm sys}) =315^{+30}_{-29} \, {\rm MeV}$
and $\alpha_s(M_Z^2)=0.1166^{+0.0010}_{-0.0011}({\rm stat})^{+0.0018}_{-0.0017}({\rm sys})=0.1166 ^{+0.0021}_{-0.0020}$.
The total errors are obtained by combining the statistic and systematic errors in quadrature.

In Analysis (II), we performed a global fit, where theoretical constraints are fully used.
Analysis (II) is superior to Analysis (I) in the sense that
it is a first-principle analysis and that our dominant error, higher order uncertainty, is reduced thanks to 
the use of short distance range. 
This gives our final result:
\be
\begin{cases}
&\LMS=334 \pm 10 ({\rm stat}) ^{+21}_{-18}({\rm sys}) \, {\rm MeV}=334^{+23}_{-21} \, {\rm MeV}  \, , \\
&  \alpha_s(M_Z^2)=0.1179 \pm 0.0007 ({\rm stat}) ^{+0.0014}_{-0.0012} (\rm sys)=0.1179 ^{+0.0015}_{-0.0014} \, .
\end{cases}
\ee

\begin{figure}[htbp]
\begin{center}
\includegraphics[width=7cm]{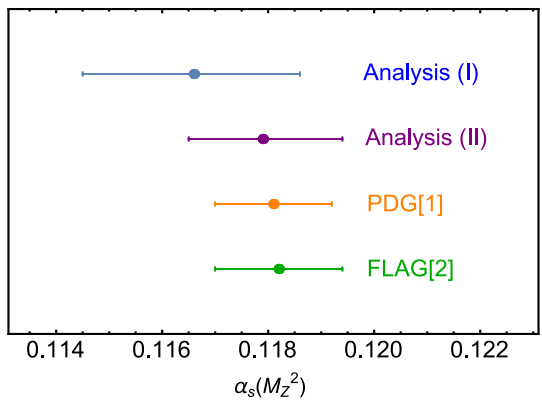}
\end{center}
\caption{Comparison of various $\alpha_s$ determinations.}
\label{fig:comp}
\end{figure}

One can see that both analyses give consistent values.
Our results of $\alpha_s(M_Z^2)$ are compared with the current PDG and FLAG results in Fig.~\ref{fig:comp},
where one can see that our results are also consistent with them.

\section{Conclusions and discussion}
\label{sec:4}

We determined the strong coupling constant $\alpha_s$ from the static QCD potential by
matching a lattice result with a new OPE calculation
where renormalons are subtracted from the leading Wilson coefficient.  
We subtract both $u=1/2$ and $u=3/2$ renormalons from the Wilson coefficient.
In particular, we confirmed the following features regarding the renormalon subtraction.
%The specific features  can be listed as follows.
%We have demonstrated that our calculation has a wider validity range than the ordinary perturbation theory.
%Furthermore, we have confirmed that the OPE prediction agrees with a good quality lattice result at $\LQ r \lesssim 0.8$.
%(See Figs.~\ref{fig:consistency1} and \ref{fig:consistency2}.)
%In particular, a leading nonperturbative effect of $r^2$, indicated from OPE,
%was clearly observed.
\begin{enumerate}
\item Theoretically the cancellation of the $u=3/2$ renormalon 
against the nonperturbative term is checked at the LL order. Furthermore, 
logarithmic contributions at IR region in the Fourier integral, which cause factorial divergence, 
are subtracted at the NNNLL level.

\item To check that ignoring renormalons contained in
$\alpha_V(q)$ is harmless at the current level of the analysis, we confirmed that the
renormalon-free Wilson coefficient $V_S^{\RF}(r)$ approaches the
lattice data as we raise the order: LL, NLL, NNLL, NNNLL (Fig.~\ref{fig:LLtoNNNLL}).

\item 
As a result of the renormalon subtraction, convergence and stability against scale
variation are improved as compared to the conventional
methods.
The difference between the Wilson coefficient $V_S^{\RF}(r)$
and the lattice data can be fitted with $r^2$ consistently with the prediction of the OPE.  
This $r^2$ behavior is observed up to $\LMS r \sim 0.8$ ($r \sim 0.4$  fm).
(Figs.~\ref{fig:consistency1},\ref{fig:consistency2})
\end{enumerate}
Based on these confirmations, we adopt the OPE framework
where a power correction term of order $r^2$ is added to the renormalon-free Wilson coefficient. 

In our $\alpha_s$ determination, the matching range 
is taken as $\LMS r \lesssim 0.6$ based on the above observation.
This range is significantly wider than preceding determinations using the static QCD potential, where
typically $\LMS r \lesssim 0.3$ has been used. 
This enables us to use the data not only at $r \sim a$ but also at $r \gg a$,
where lattice simulation is considered to be accurate.
We performed a reasonable fit in this wide region,
which leads to a reliable determination.
Our final result is $\alpha_s(M_Z^2)=0.1179 ^{+0.0015}_{-0.0014}$.
This result is obtained by a global fit [Analysis (II)] 
and is consistent with our another analysis [Analysis (I)],
where we examined intermediate processes step by step. 
The reasonable value of $\alpha_s$ with respect to today's other determinations again indicates the validity of our analysis. 
We also confirmed that although the energy region extends to lower energy side than conventional determinations using perturbative calculation,
varying the matching range does not induce significant systematic errors.

Dominant error of our determination comes from systematic errors,
in particular from the higher order perturbative uncertainty of the leading Wilson coefficient.
We emphasize that a finer lattice simulation will straightforwardly reduce this error,
since we can adopt a shorter distance range in the fit, 
where the uncertainty becomes smaller.\fn{
Reduction of the higher order uncertainty can be estimated as follows.
The relative perturbative accuracy in our formulation at ${\rm N}^3$LL is order $\alpha_s(\mu^2)^4$,
not affected by renormalon uncertainties. Here, $\mu$ is the typical scale used 
in the $\alpha_s$ determination.
For instance, suppose that currently $\mu \sim 7 \LMS$, and suppose that
$\mu$ can be raised by a factor 2 (corresponding to twice finer lattices).
Then the perturbative error would reduce, which is multiplied by 
$[\alpha_s((2 \mu)^2)/\alpha_s(\mu^2)]^4 \sim 0.3$--0.4.
}$^{,}$\fn{
We remark that if the coarsest lattice becomes finer while the finest lattice spacing is kept fixed,
it serves to reduce the error.
This is because our current range $\LMS r < 0.6$ is chosen 
so that the number of data points from the coarsest lattice is sufficient.}

We believe that our analyses are useful not only in determining $\alpha_s$ 
but also in promoting understanding on the OPE structure and lattice discretization errors.
As stated, this is a first numerical observation that
the difference between the Wilson coefficient and the lattice result is consistent 
with $\mathcal{O}(r^2)$ behavior at $\LMS r  \lesssim 0.8$ in accordance with the OPE structure.
We also give a constraint on the linear term in $r$ in the difference, which should be zero in the OPE.
(See discussion in Sec.~\ref{sec:CC}.)
%We have demonstrated that our formulation enlarges the validity range 
%of the theoretical prediction to lower energy as compared to conventional methods. 
Concerning the lattice discretization error, we clarified that
(i) the data at $r=a$ indeed has a serious finite $a$ effect (Appendix~\ref{app:includeshortest}),
and (ii) once the tree-level correction is considered combined with the OPE calculation,
the finite $a$ effect can be largely removed with reasonable values of lattice effective couplings. 

\vspace{2mm}

\noindent
{\bf Acknowledgements} The authors are grateful to the JLQCD collaboration
for providing the lattice data.
They thank G. Mishima for collaboration at an early stage of this study
and also thank S. Aoki, S. Hashimoto, T. Onogi, and S. Sasaki for fruitful discussion.
The works of Y.K. and Y.S. are supported in part by Grant-in-Aid for scientific research
(Nos. 26400255 and 17K05404) from MEXT, Japan.

\newpage
\appendix

\section{Coefficients of perturbative calculation}
\label{app:coeff}
The coefficients $a_n$ of Eq.~\eqref{pertexp2} are given by
\begin{align}
&a_0=1 \, , \non
&a_1=\frac{31}{3}-\frac{10}{9} n_f \, , \non 
&a_2=\frac{4343}{18}+36 \pi^2+66 \zeta_3-\frac{9 \pi^4}{4}-\lt(\frac{1229}{27}+\frac{52}{3} \zeta_3 \rt) n_f+\frac{100}{81} n_f^2 \, ,  \non 
&a_3=a_3^{(0)}+a_3^{(1)} n_f+a_3^{(2)} n_f^2+a_3^{(3)} n_f^3 \, ,
\end{align}
with
\begin{align}
&a_3^{(0)}=\frac{385645}{108}+\pi^2 \lt[\frac{893}{3}+816 \alpha_4+(1844-1302 \zeta_3) \log{2}+295 \zeta_3 \rt]+5256 \zeta_3   \non  
&~~~~~~+\pi^4 \lt(-\frac{227}{20}+115 \log{2}+35 \log^2{2} \rt)-\frac{17343}{2} \zeta_5 -\frac{1643 \pi^6}{168}-\frac{3861 \zeta_3^2}{2}+3888 s_6  \, , \non
&a_3^{(1)}=-\frac{452213}{324}+\pi^2 \lt[\frac{274}{27}-\frac{409}{9} \zeta_3-144 \alpha_4+\lt(-\frac{8}{3}-28 \zeta_3 \rt) \log{2} \rt]-\frac{26630 \zeta_3}{27}  \non
&~~~~~~~~+\pi^4 \lt(-\frac{293}{18}-\frac{35}{18} \log{2}+\frac{17}{6} \log^2{2} \rt)
+\frac{30097}{36} \zeta_5+\frac{1931}{1260} \pi^6+\frac{513}{4} \zeta_3^2-216 s_6 \, , \non
&a_3^{(2)}=\frac{93631}{972}+\frac{16 \pi^4}{45}+\frac{412 \zeta_3}{9} \, ,  \non
&a_3^{(3)}=-\lt(\frac{10}{9} \rt)^3 \, .
\end{align}
Here, $\alpha_4$ and $s_6$ are given by
\be
\alpha_4={\rm Li}_4(1/2)+\frac{(-\log{2})^4}{4!}=0.527097... \, ,
\ee
\be
s_6=\zeta(-5,-1)+\zeta(6)=0.987441... \, .
\ee
The above analytic expression for $a_3$ has been obtained in Ref.~\cite{Lee:2016cgz}.

\section{Formulation to extract $V_S^{\RF}(r)$ from $V_S(r)$}
\label{app:VSRF}
\begin{figure}[htbp] 
\begin{minipage}{1\hsize}
\begin{center}
\includegraphics[width=6cm]{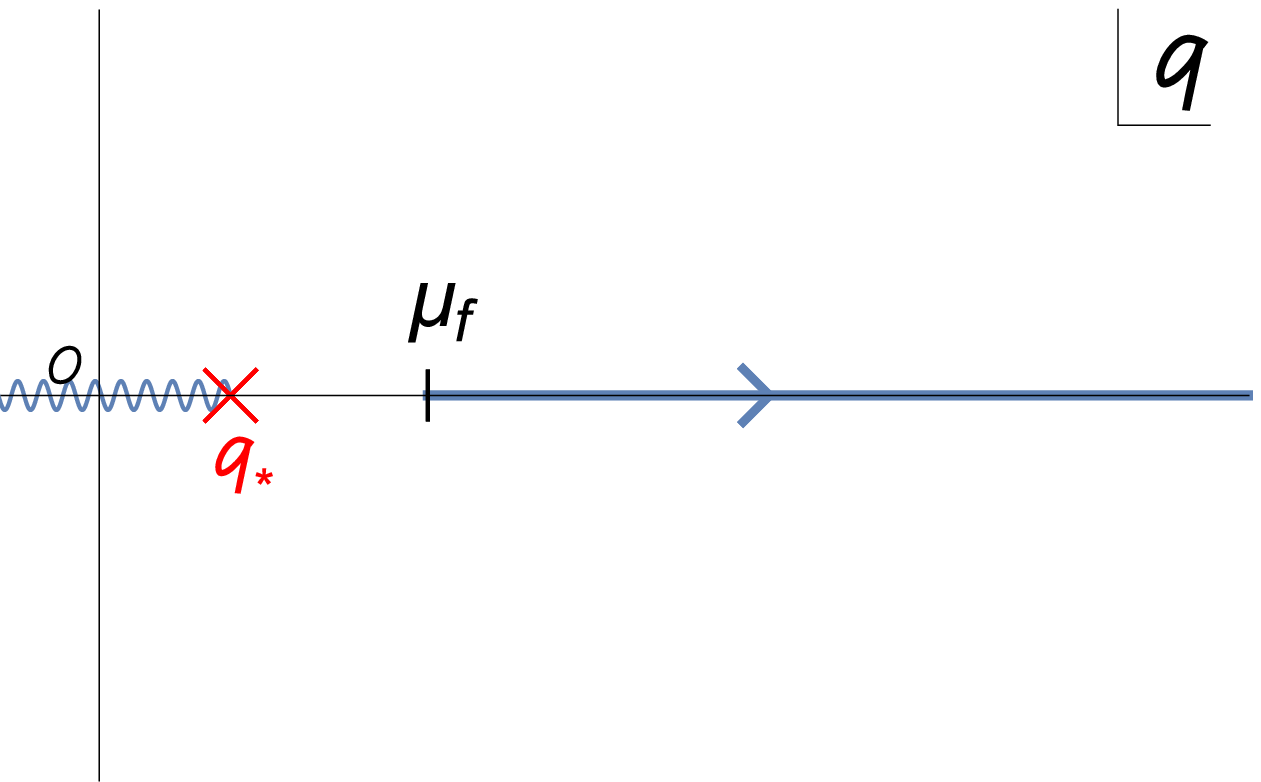}
\end{center}
\end{minipage}
\begin{minipage}{0.5\hsize}
\begin{center}
\includegraphics[width=5cm]{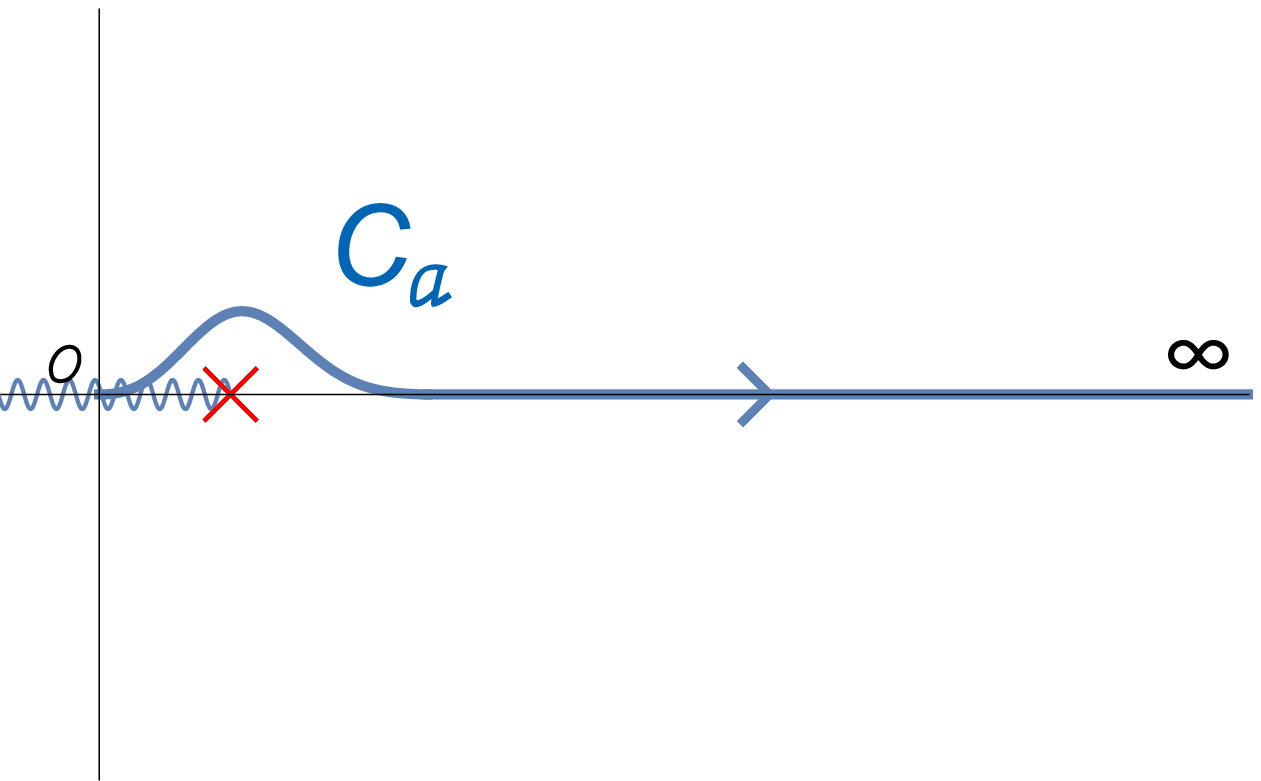}
\end{center}
\end{minipage}
\begin{minipage}{0.5\hsize}
\begin{center}
\includegraphics[width=5cm]{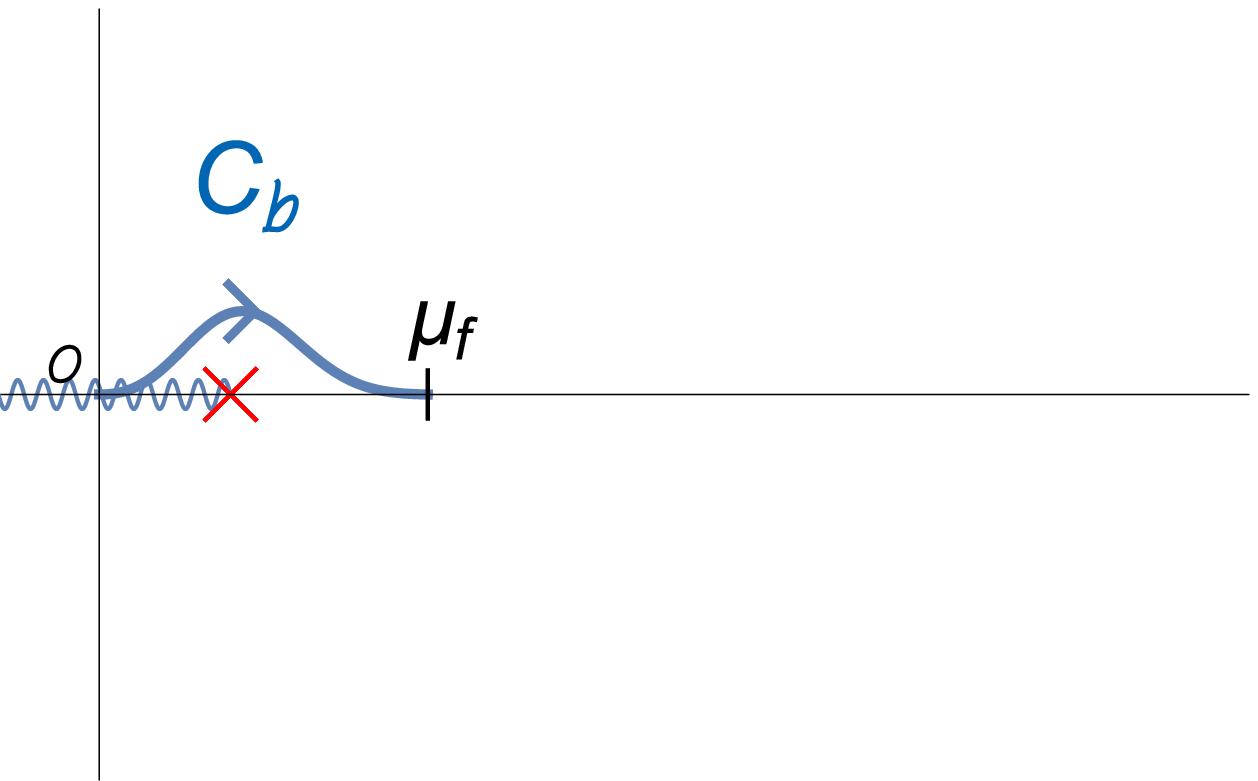}
\end{center}
\end{minipage}
\caption{Contour $C_a$ and $C_b$ in the complex $q$-plane.
$q_*$ shows the singular point of $\alpha_V(q^2)$.}
\label{fig:complex1}
\end{figure}

We explain the formula to extract $V_S^{\RF}(r)$ from Eq.~\eqref{VS}.
We reduce Eq.~\eqref{VS} to the one-dimensional integral representation:
\be
V_S(r;\mf)=-\frac{2 C_F}{\pi r} \int_{\mf}^{\infty} \frac{d q}{q} \sin(q r) \alpha_V(q^2) \, , \label{start1}
\ee
with $q=|\vec{q}|$. We rewrite the integral as
\begin{align}
V_S(r;\mf)
&=-\frac{2 C_F}{\pi r} {\rm Im} \int_{\mf}^{\infty} \frac{d q}{q} e^{i q r} \alpha_V(q^2)  \non
&=-\frac{2 C_F}{\pi r} {\rm Im}\lt(  \int_{C_a}-\int_{C_b}  \rt) \frac{d q}{q} e^{i q r} \alpha_V(q^2) \, . \label{rewrite}
\end{align}
The contours $C_a$ and $C_b$ are displayed in Fig.~\ref{fig:complex1}.
The integral along $C_a$ is clearly independent of $\mf$.
Although the integral along $C_b$ looks $\mf$ dependent,
it contains a $\mf$-independent part.
We evaluate this integral as
\begin{align}
&\frac{2 C_F}{\pi r} {\rm Im} \int_{C_b}  \frac{d q}{q} e^{i q r} \alpha_V(q^2) \non
&=\frac{2 C_F}{\pi r} {\rm Im} \int_{C_b} \frac{d q}{q} \lt[ 1+i q r -\frac{1}{2} (qr)^2-\frac{i}{6} (qr)^3+\dots \rt] \alpha_V(q^2) \, ,
\end{align}
since $|q r| < \mf r \ll 1$.
In expansion of the exponential factor, the real and pure imaginary coefficients appear in turn.

\begin{figure}[h!]
\begin{center}
\includegraphics[width=5cm]{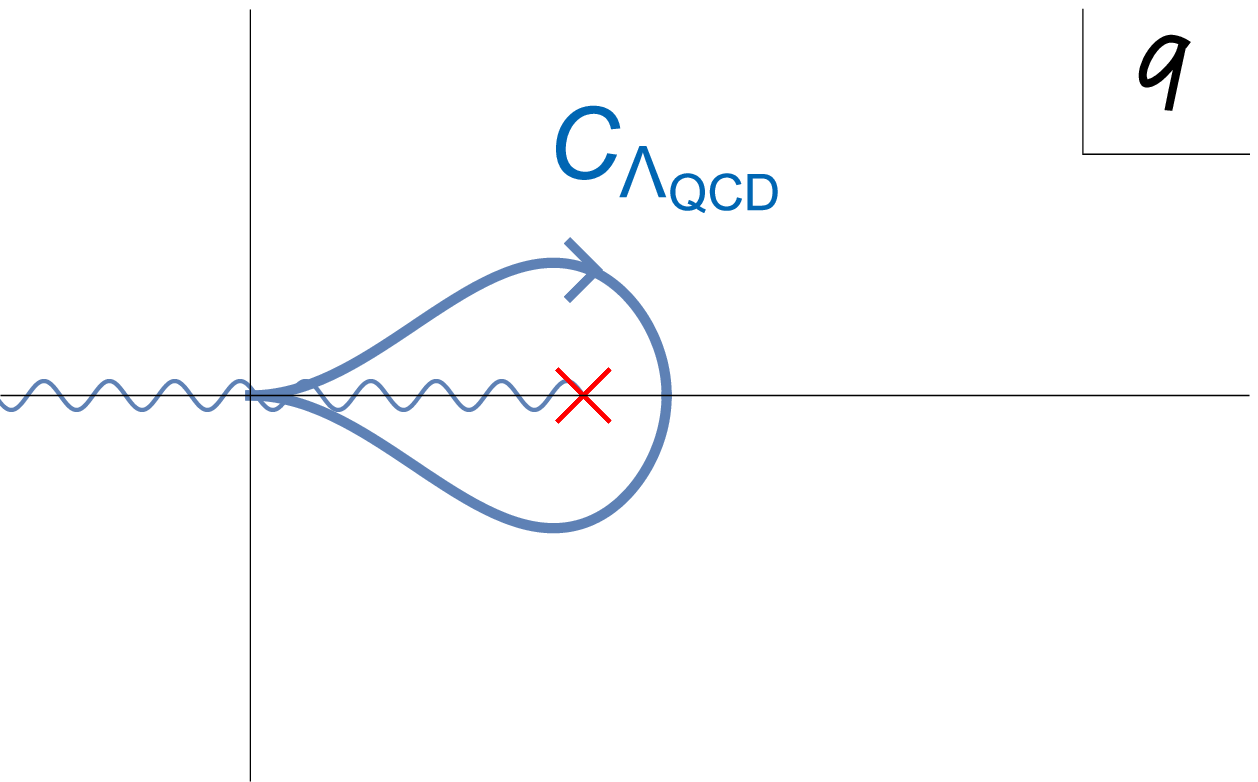}
\end{center}
\caption{Contour $C_{\LQ}$.}
\label{fig:complex2}
\end{figure}

The terms with real coefficients satisfy the relation $\{ f(z) \}^*=f(z^*)$.
Owing to this, these parts can be calculated as
\begin{align}
&\frac{2 C_F}{\pi r} {\rm Im} \int_{C_b} \frac{d q}{q} \lt[ 1-\frac{1}{2} (qr)^2 \rt] \alpha_V(q^2) \non
%&=\frac{2 C_F}{\pi r} \frac{1}{2 i} \lt( \int_{C_b}-\int_{C_b^*} \rt) \frac{d q}{q} \lt[ 1-\frac{1}{2} (qr)^2 \rt] \alpha_V(q^2) \non
&=\frac{2 C_F}{\pi r} \frac{1}{2 i} \int_{C_{\LQ}} \frac{d q}{q} \lt[ 1-\frac{1}{2} (qr)^2 \rt] \alpha_V(q^2)  \non
&=\frac{1}{r} \mathcal{C}_{-1}+\mathcal{C}_1 r
\end{align}
with
\be
\mathcal{C}_{-1}=2 C_F \frac{1}{2 \pi i} \int_{C_{\LQ}} \frac{d q}{q} \alpha_V(q^2) \, ,
\ee
\be
\mathcal{C}_1=-C_F \frac{1}{2 \pi i} \int_{C_{\LQ}} \frac{d q}{q} q^2 \alpha_V(q^2) \, ,
\ee
where $C_{\LQ}$ is shown in Fig.~\ref{fig:complex2}.
The coefficients $\mathcal{C}_{-1}$ and $\mathcal{C}_1$ are 
$\mf$ independent and real. 
Numerical evaluation of these coefficients is sufficient for our purpose. 
[$\mathcal{C}_1$ is given by Eq.~\eqref{linear}.]
We remark that the analytical results up to ${\rm N}^2$LL can be found in Ref.~\cite{Sumino:2005cq}.

On the other hand, the terms with imaginary coefficients do {\it not} satisfy the relation $\{ f(z) \}^*=f(z^*)$,
and the above deformation cannot be applied.
Therefore, we have
\begin{align}
&\frac{2 C_F}{\pi r} {\rm Im} \int_{C_b} \frac{d q}{q} \lt[i q r-\frac{i}{6} (qr)^3\rt] \alpha_V(q^2) \non
&=\mathcal{C}_{0}(\mf)+\mathcal{C}_2(\mf) r^2 \, ,
\end{align}
where $\mf$ dependence remains.

Based on the above argument, we can construct a $\mf$-independent quantity $V_S^{\RF}$.
Note that a $\mf$-independent part is also given by the integral along $C_a$. 
Then by collecting all the $\mf$-independent part, we obtain
\be
V_S^{\rm RF}(r)=V_C(r)+\mathcal{C}_1 r
\ee
with
\begin{align}
V_C(r)
&=-\frac{1}{r} \lt[\frac{2 C_F}{\pi} {\rm Im} \int_{C_a} \frac{d q}{q} e^{i q r} \alpha_V(q^2) -\mathcal{C}_{-1} \rt] \non
&=-\frac{1}{r} \lt[\frac{2 C_F}{\pi}  \int_{0}^{\infty} \frac{d q}{q} e^{- q r} {\rm Im} \, \alpha_V(-q^2+i 0) -\mathcal{C}_{-1} \rt] \, . \label{VC}
\end{align}
In the last line, we rotate the contour $C_a$ to the line along $e^{i \pi/2} q$ with real positive $q$.

Once the $\mf$-dependent part of $V_S(r;\mf)$ is considered as well, 
one obtains the decomposition shown in Eq.~\eqref{decom}.

\section{Definition of $\LMS$}
\label{app:Lambda}
The definition of the scale $\Lambda$ in the ${\overline {\rm MS}}$ scheme, $\LMS$,
is given by 
\begin{align}
&\log{ \lt(\frac{\mu^2}{\LMS^2}\rt)} 
=\frac{4 \pi}{\alpha_s \bz}+\frac{\beta_1}{\bz^2} \log{\lt(\frac{\bz \alpha_s}{4 \pi} \rt)} 
+ \int_0^{\alpha_s} dx \lt(\frac{1}{\beta(x)}+ \frac{4 \pi}{\bz x^2}-\frac{\beta_1}{\bz^2 x}  \rt) \, , \label{Lambdadef}
\end{align}
where $\alpha_s$ represents the coupling at the renormalization scale $\mu$.
We approximate the $\beta$ function at four-loop as in Eq.~\eqref{running},
which gives the definition of $\LMS^{4{\text{-loop}}}$, used extensively in this paper.

\section{$\chi^2$ and covariance matrix}
\label{app:corrmat}
We present definitions of $\chi^2$ and covariance matrices used in our analyses, 
which may be useful especially for non-expert readers.

\noindent
{\bf Interpolation [Analysis (I)]} 
We define $\chi^2$ in the interpolation with a covariance matrix as\fn{
$\chi^2$ is a dimensionless quantity. 
Accordingly, each quantity appearing in Eq.~\eqref{chisq1} can be made
dimensionless.
In practice, we normalize all the quantities with $a$.
}  
\be
\chi_{\rm Inter}^2(\alpha,c_0,\sigma,c_1,c_2)|_{d,i}
=\sum_{k,l} [V_{{\latt},d,i}(r_k)-V_{{\latt}, d, i}^{\rm Inter.}(r_k)] {\Delta^{\latt}_{d,i} (r_k, r_l)}^{-1} 
 [V_{{\latt}, d, i}(r_l)-V_{{\latt}, d, i}^{\rm Inter.}(r_l)] \, , \label{chisq1}
\ee
where $V_{{\latt}, d, i}^{\rm Inter.}(r)$ is defined in Eq.~\eqref{inter} and
$k, l$ run over the lattice points under consideration.
The covariance matrix $\Delta^{\latt} (r_k , r_l)$ is calculated as
\be
\Delta^{\latt}_i (r_k , r_l)=(N_i-1) \braket{(V_{{\latt}, i}(r_k)-\braket{V_{{\latt}, i}(r_k)}) \cdot (V_{{\latt}, i}(r_l)-\braket{V_{{\latt}, i}(r_l)}) } \label{corrmat1}
\ee
in the jackknife method, where $N_i$ is the number of bins for the $i$-th lattice simulation; 
see table~\ref{tab:param}.
%In the above, the average $\braket{\dots}$ is taken in the jackknife method,
%where for instance $\braket{A B}=\frac{1}{N_s} \sum_{i=1}^{N_s} A_i B_i$ follows.
If the subscript $d$ is shown, it expresses a covariance matrix 
among the potentials $V_{{\latt}, i, d}$.\fn{
Although in Analysis (I) we treat the data separately according to each direction,
we will use them simultaneously in Analysis (II).
This is the reason why we suppress the subscript $d$ in Eq.~\eqref{corrmat1}.}

\noindent
{\bf Continuum extrapolation [Analysis (I)]} 
$\chi^2$ in the extrapolation to the continuum limit is defined as
\be
\chi^2_{\rm ex}(\gamma,\delta;r)=\sum_{i=1,2,3} \lt(\frac{X_{\latt}(r;a_i)-Y(a_i)}{\delta X_{\latt}(r;a_i)} \rt)^2 \label{chisqex} \, ,
\ee
where $Y(a)$ is defined by Eq.~\eqref{quad}.

The covariance matrix for $X_{\latt}^{\cont}$ is calculated as
\be
 \Delta^{\cont}(r_i, r_j) =(N_{\rm tot}-1) \braket{(X_{{\latt}}^{\cont}(r_i)-\braket{X_{{\latt}}^{\cont}(r_i)}) \cdot (X_{{\latt}}^{\cont}(r_j)-\braket{X_{{\latt}}^{\cont}(r_j)}) }   \,  . \label{Deltacont}
\ee
Note that in the continuum extrapolation, 
the jackknife samples with the size $N_{\rm tot}=\sum_{i=1}^3 N_i=400$ are generated 
since we have three independent lattice measurements.
We present the numerical result of $\Delta^{\cont}$ in table~\ref{tab:corrmatX}.
\begin{table}[h!]
\small
\begin{center}
\begin{tabular}{c|cccccc} 
             & 0.7196                    & 0.7822                      & 1.043                       & 1.079                     & 1.304                      & 1.439 \\ \hline
0.7196  & $8.61\times10^{-5}$& $-1.47\times10^{-6}$ &$-8.72\times10^{-8}$ &$1.79\times10^{-6}$&$-5.14\times10^{-6}$&$2.36\times10^{-5}$ \\
0.7822  &                                &$5.24\times10^{-5}$  &$9.45\times10^{-7}$ &$-1.76\times10^{-7}$&$2.25\times10^{-5}$&$-4.33\times10^{-6}$ \\  
1.043   &                                 &                                 &$2.18\times10^{-8}$ &$-3.54\times10^{-9}$&$6.34\times10^{-7}$&$-7.32\times10^{-8}$ \\
1.079   &                                 &                                 &                                &$1.07\times10^{-7}$&$-1.25\times10^{-7}$&$2.40\times10^{-6}$ \\
1.304   &                                 &                                 &                                &                               &$2.27\times10^{-5}$ &$-1.65\times10^{-6}$\\
1.439   &                                 &                                 &                                &                               &                                &$6.90\times10^{-5}$ \\ \hline
\end{tabular}
\end{center}
\caption{Covariance matrix for $X_{\latt}^{\cont}$, $\Delta^{\cont}(r_i, r_j)$.
The first row is $r_i/r_1$ and the first column is $r_j/r_1$.
The $(i,j)$ component is the numerical value of $\Delta^{\cont} (r_i, r_j)$.
Note that $\Delta^{\cont }(r_i,r_j)$ is a symmetric matrix,
and hence, we only show the elements of the upper triangular part. }
\label{tab:corrmatX}
\end{table}

\noindent
{\bf Matching [Analysis (I)]} 
We define $\chi^2$ in the matching of Analysis (I) as
\be
\chi_{\rm match}^2(x,A_0,A_2)
=\sum_{i,j} [\tilde{X}_{\latt}(r_i)-v_{\rm OPE}(r_i)] 
\tilde{\Delta}^{\rm cont} (r_i, r_j)^{-1}
[\tilde{X}_{\latt}(r_j)-v_{\rm OPE}(r_j)] \, , \label{chisqmatch}
\ee
where $v_{\rm OPE}$ is given in Eq.~\eqref{vope}
and $\tilde{\Delta}^{\rm cont}$ is the covariance matrix 
for $\tilde{X}_{\latt}^{\rm cont}(r)$:
\begin{align}
&\tilde{\Delta}^{\cont} (r_i , r_j)=x^{-2}  \Delta^{\cont}(r_i , r_j) \, .
\end{align}
See Eq.~\eqref{Deltacont} and table~\ref{tab:corrmatX} for $\Delta^{\cont}$.

\noindent
{\bf Global fit [Analysis (II)]}
We define $\chi^2$ in the global fit in Analysis (II) as
%\fn{Dimension of each quantity is $V_{\latt}^{\cont} [{\rm GeV}], V_{\rm OPE} [{\rm GeV}]$ 
%or $\Delta^{\latt}(r_i, r_j)^{-1} [{\rm GeV}^{-2}]$. 
%}
\be
\chi^2_{\rm GF}(z=\LMS, A_2, \kappa, f, c_0)
=\sum_{i,j} 
[V_{\latt}^{\cont}(r_i)-V_{\rm OPE}(r_i)] \Delta^{\latt}(r_i ,r_j)^{-1}  [V_{\latt}^{\cont}(r_j)-V_{\rm OPE}(r_j)] \, . \label{chiGF}
\ee
Here, the covariance matrix consists of three matrices of dimension $7, 10,$ and $13$ 
in a block diagonal form:
\be
\Delta^{\latt}=\lt(
\begin{array}{ccc} 
\Delta^{\latt}_1 & O & O \\
O & \Delta^{\latt}_2 & O \\
O & O & \Delta^{\latt}_3
\end{array} 
\rt) \, ,
\ee
where the definition of each matrix is given by Eq.~\eqref{corrmat1}.

\section{Case including data at $r=a$ in Analysis (I)}
\label{app:includeshortest}

In Analysis (I), we do not use the data at $r=a$ in interpolating lattice data in our analyses,
in order to suppress serious finite $a$ effects.
Here, let us see what happens if we include this shortest point.
\begin{figure}[h!]
\begin{minipage}{0.5\hsize}
\begin{center}
\includegraphics[width=7cm]{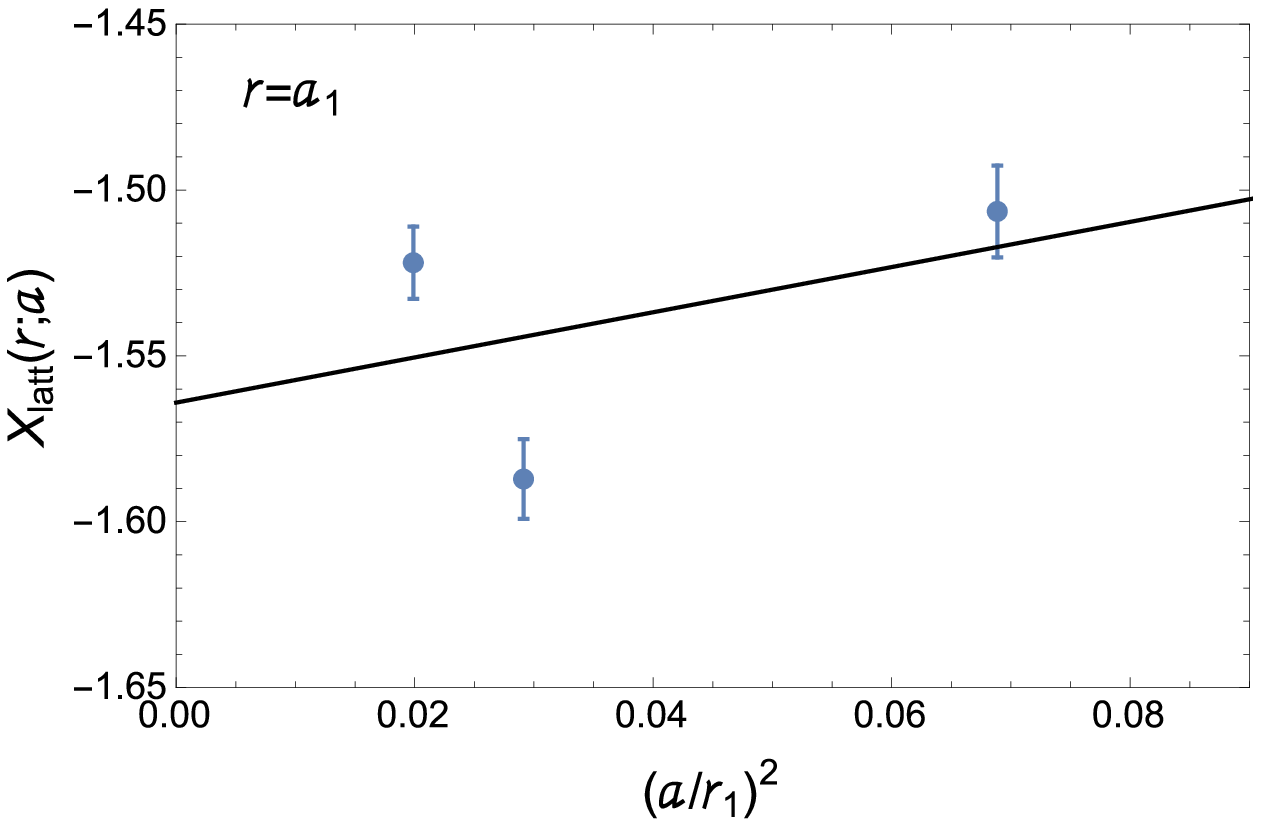}
\end{center}
\end{minipage}
\begin{minipage}{0.5\hsize}
\begin{center}
\includegraphics[width=7cm]{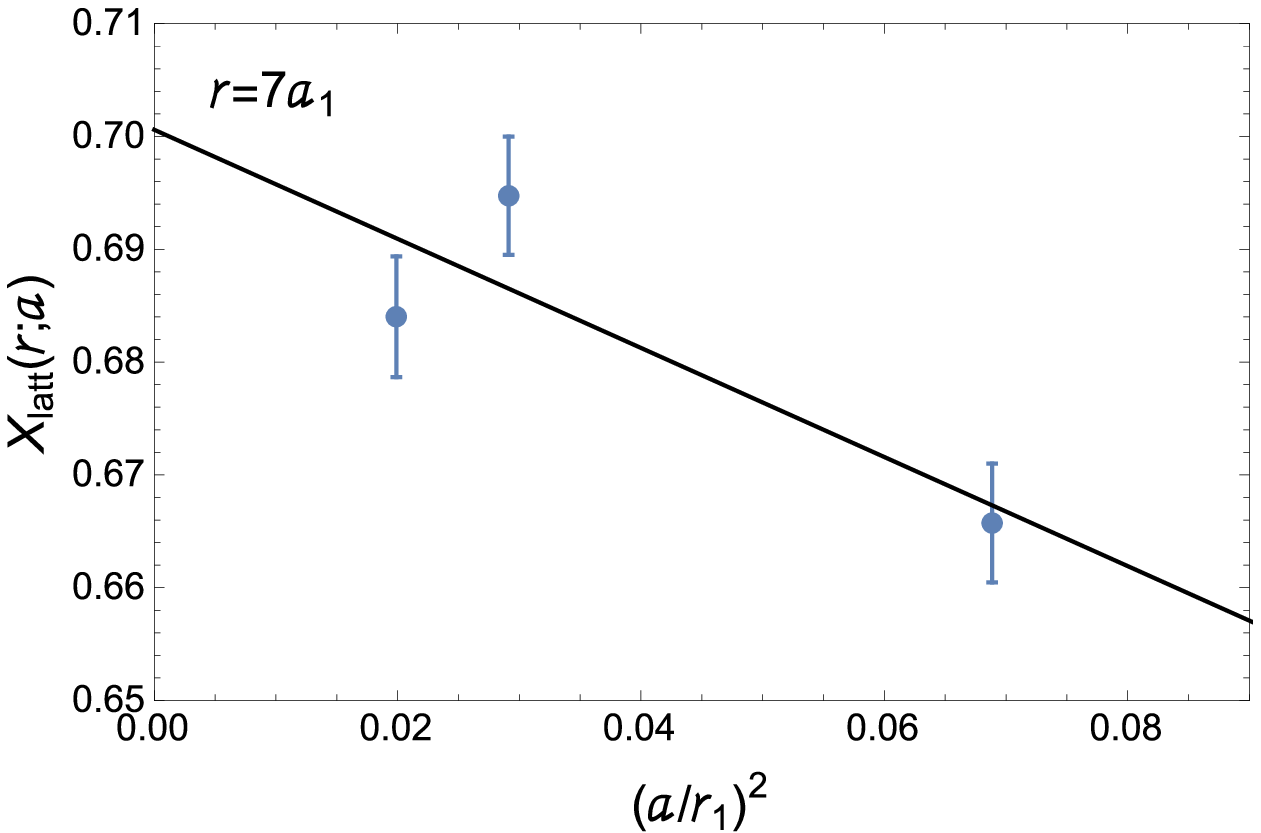}
\end{center}
\end{minipage}
\caption{$X_{\latt}(r;a)$ as functions of $(a/r_1)^2$ when we include $r=a$ in interpolation.
We show them for $r= a_1$ (left) and $r=7 a_1$ (right), which are reference distances for $d=1$.
Black lines are linear functions in $a^2$ determined by fits.
$\chi^2_{\rm ex}/{\rm d.o.f.}$, which is the reduced $\chi^2$ in this extrapolation, 
are $20$ (left) and $4.3$ (right).}
\label{fig:includeshortest}
\end{figure}
We interpolate lattice data including the ones at $r=a$,
and obtain $X_{\latt}(r;a)$ in the same way.
In Fig.~\ref{fig:includeshortest}, we plot the data points of $X_{\latt}(r;a)$ 
taking the horizontal axis as $(a/r_1)^2$. 
One can see that they do not obey linear behaviors in $a^2$.
We remark that even the data for $r=7 a_1$, where the finite $a$ effect is considered 
to be well suppressed, cannot smoothly be extrapolated to the continuum limit.  
It shows that the data at $r=a$, which has a small statistical error,
dominantly contributes to determining the interpolating function,
and thus, the interpolating function is seriously distorted.
In Fig.~\ref{fig:chisqincludeshortest}, we show $\chi^2_{\rm ex}/{\rm d.o.f.}$ in this case, 
corresponding to Fig.~\ref{fig:chisq}.
\begin{figure}[t!]
\begin{minipage}{0.5\hsize}
\begin{center}
\includegraphics[width=7cm]{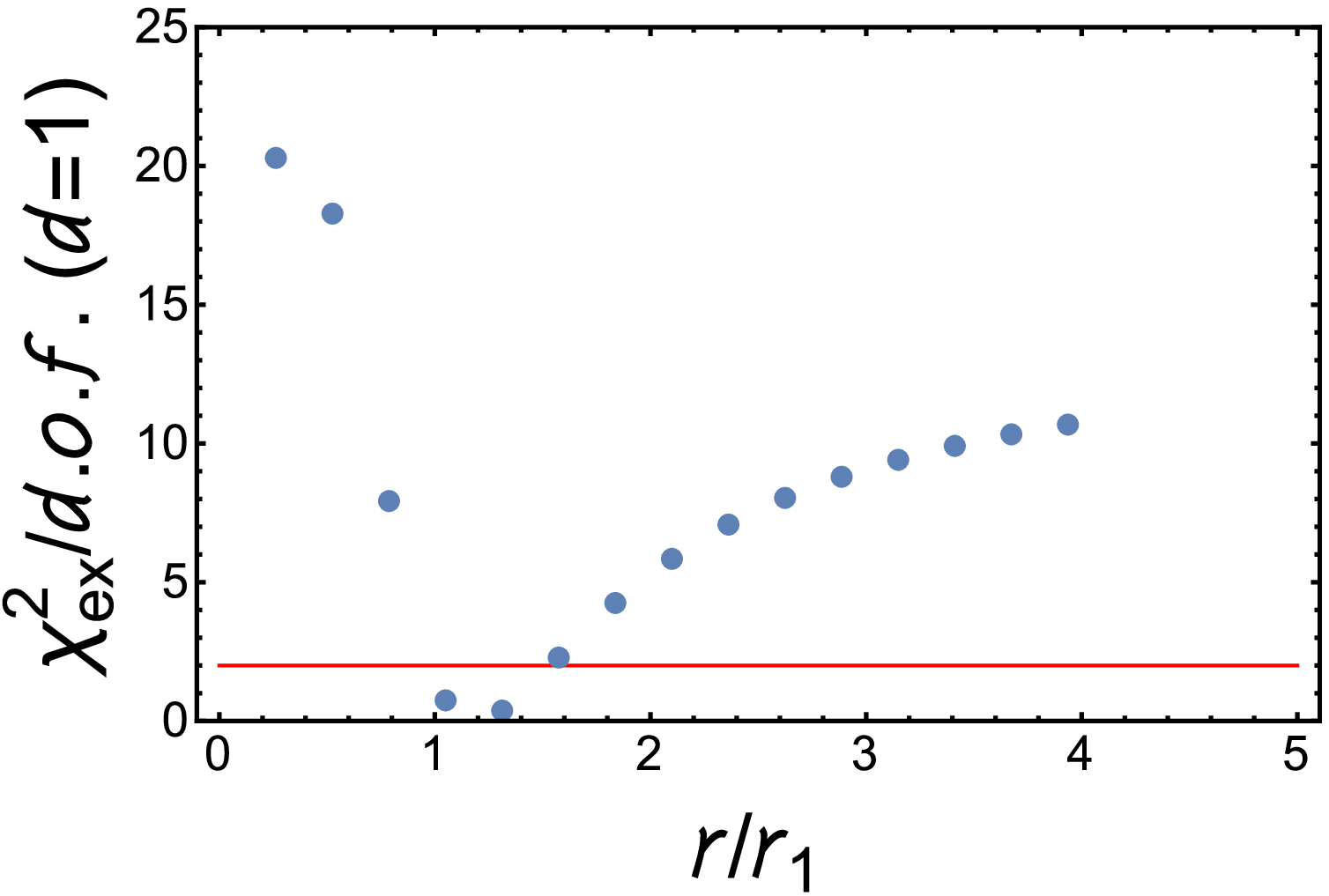}
\end{center}
\end{minipage}
\begin{minipage}{0.5\hsize}
\begin{center}
\includegraphics[width=7cm]{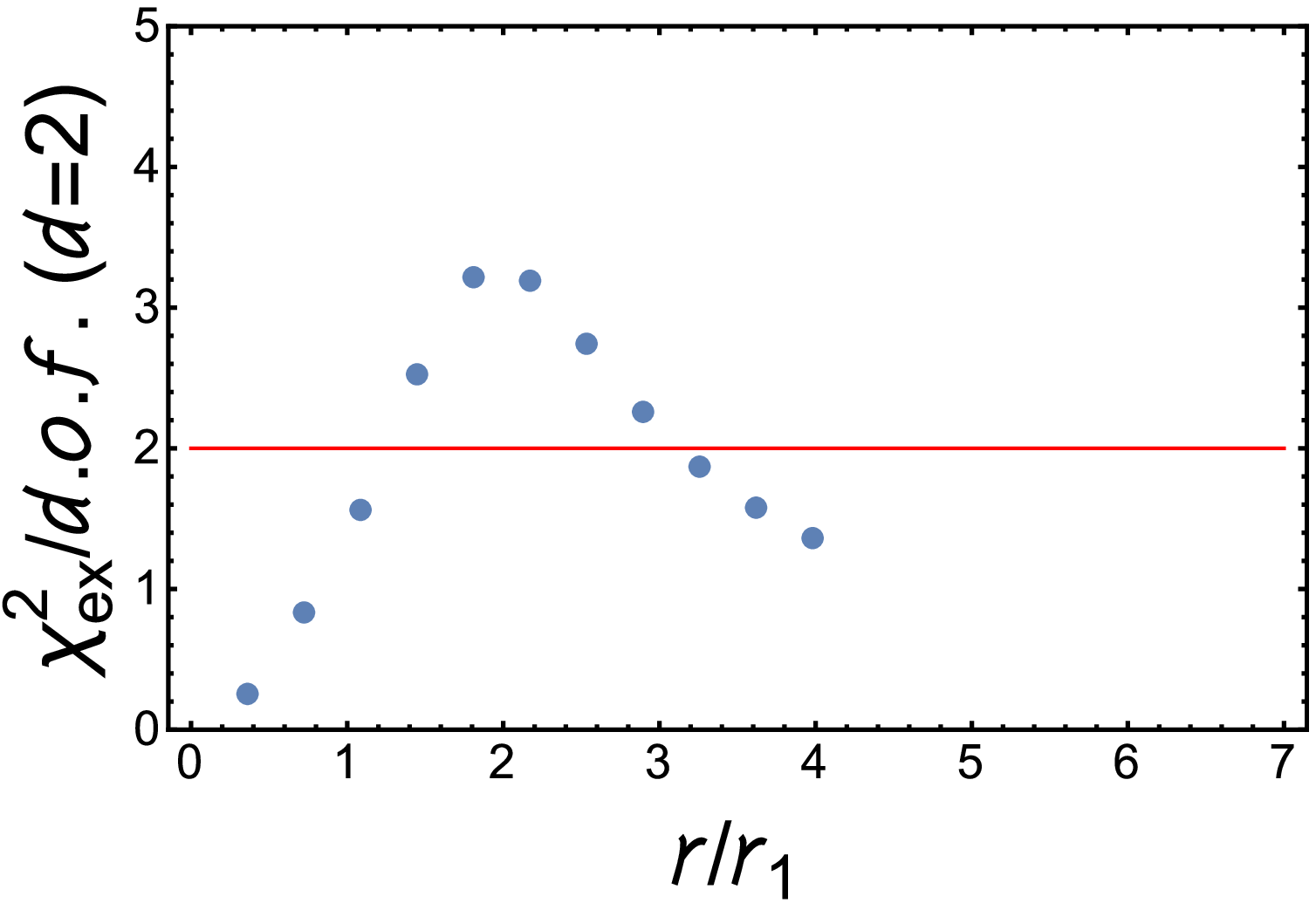}
\end{center}
\end{minipage}
\caption{The reduced $\chi^2_{\rm ex}$ in extrapolations.
$\chi^2_{\rm ex}/{\rm d.o.f.}=2$ is shown by red lines as a benchmark.}
\label{fig:chisqincludeshortest}
\end{figure}
Extrapolations to the continuum limit do not work for $d=1$. 
For $d=2$, where the shortest point is located at $r=\sqrt{2} a$,
extrapolations work better than for $d=1$.
We conclude that the data at $r=a$ has serious discretization error,
and we should be cautious about treating it.

\section{Additional analyses on systematic errors}
\label{app:further}
In this appendix, 
we provide supplemental analyses to check validity of our error analyses from additional aspects. 

\noindent
{\bf ${\boldmath \mathcal{O}(a^4)}$ effect} \\
In Sec.~\ref{sec:alphas}, we considered the leading discretization error, the quadratic effect in $a$, of the lattice data.
Here, we estimate the possible effect coming from the neglected $\mathcal{O}(a^4)$ error.\fn{
Due to chiral symmetry, an $\mathcal{O}(a^3)$ error is prohibited.}

In both Analyses (I) and (II), if we perform fits including $\mathcal{O}(a^4)$ terms 
it turns out that the fits have little sensitivities to these terms, 
given the current lattice data.
As a result, the fits become fairly unstable,
leading to fairly uncertain results for $\alpha_s(M_Z^2)$ [even though
they are consistent with Eqs. \eqref{alpha1stat} and \eqref{alpha2stat} within
estimated (large) errors].

Instead we can confirm that our analysis is stable against possible $\mathcal{O}(a^4)$ effects in
the following way in the case of Analysis (II).
We add an $\mathcal{O}(a^4)$-term to Eq.~\eqref{Vcont} as $g_d a_i^4/r^5$ while fixing $g_d$.
In this analysis, to properly consider the expansion in $a/r$ up to NLO, 
we omit the data at $r < 2 a$ because 
this expansion is not legitimate when the data at $r=a$ is included, as clarified in Appendix~\ref{app:includeshortest}.
In this case, the tree-level correction is not necessary and is omitted (see footnote~\ref{fn:treevsa2}).
%However, both $f_d$ and $g_d$ cannot be treated as fitting parameters
%because the fit is destabilized
To assume a reasonable size of $g_d$, we refer to the size of $f_d$, 
the coefficients of the $\mathcal{O}(a^2)$ error,
determined from the data at $r \geq 2 a$. They read $f_1=0.04, f_2=-0.008$.
Then, we assume $g_1=g_2=0.04 t$ and vary $t=-1$ to $+1$. 
The other parameters (such as $f_d$ and $\LMS$) are treated as fitting parameters.
The largest variation of $\alpha_s$ caused by the $\mathcal{O}(a^4)$-term is obtained as
$\Delta \alpha_s(M_Z^2)=-0.0003$.
This is comparable to the assigned error in Table~\ref{tab:sys2} in Analysis (II).
This result indicates that our error analysis concerning finite $a$ effects is reasonable
even if we take into account the neglected higher order discretization errors.
\\

\begin{figure}[h]
\begin{center}
\includegraphics[width=11cm]{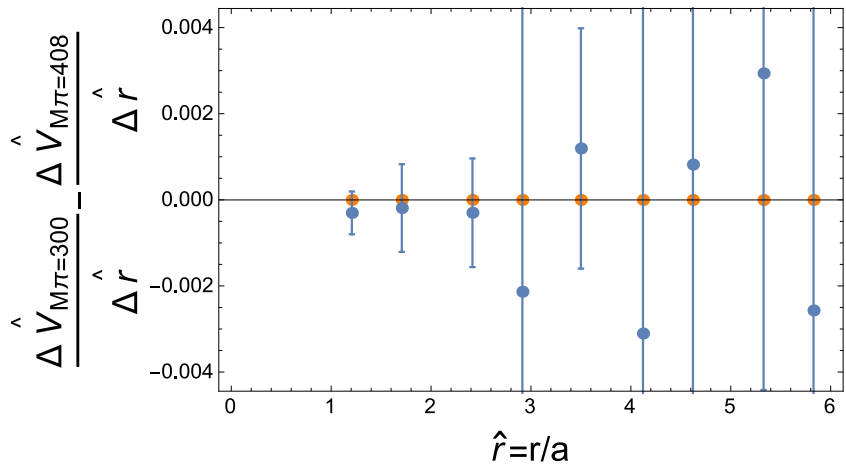}
\end{center}
\caption{Difference of the (approximate) slopes of the potentials in the different pion masses ($M_{\pi}=300$~MeV and $408$~MeV). 
The potential $V$ and distance $r$ are normalized by lattice spacing $a$.
The blue data show the lattice results with statistical errors. The orange points are the perturbative estimate.}
\label{fig:MassEffect}
\end{figure}

\noindent
{\bf Mass correction}\\
In Sec.~\ref{sec:AnaII}, the effect of the mass deviation in the lattice simulations was estimated based on perturbation theory,
where it was found negligibly small.
We support this result by directly comparing lattice results with different pion masses.
We analyze the lattice data with $M_{\pi}=300$ and $408~{\rm MeV}$ for the lattice spacing $a_2$ \cite{JLQCD:future}.
(Here, we neglect finite $a$ effects.)
Since the slope of the potential affects $\alpha_s$, 
we examine the {\it{difference}} of the slopes.
The slopes are approximately obtained from the difference of the potentials at the nearest neighbor.
In Fig.~\ref{fig:MassEffect}, the difference of the (approximate) slopes is shown,
where it is consistent with zero.\fn{
The slope itself (before taking the difference) is about 0.1 in the same units.}
(In Analysis (II), we use the first 10 points.) 
This result is consistent with our error estimate that the mass effect is negligibly small.

\noindent
{\bf Possible logarithmic correction to $r^2$-term}\\
We have treated the nonperturbative effect as $\delta E^{\rm RF}_{\rm US}(r)=A_2 r^2$. 
There is a possibility that this $r^2$-term is modified by logarithmic corrections,
which may stem from higher order computations of Wilson coefficients.
Here, we examine how large such a logarithmic correction affects our $\alpha_s$ determination.

In Analysis (II), we assume $\delta E_{\rm US}^{\rm RF}(r)$ as 
\be
\delta E_{\rm US}^{\rm RF}(r)=A_2 \lt[ 1+ t \log{\lt(r \cdot 1~{\rm GeV} \rt)} \rt] r^2 \label{logcorr}
\ee
with $t=-0.3,-0.1, 0.1,0.3$.
Then, we obtain the result in Table~\ref{tab:logcorr}.
(We also use 1.5~GeV and 0.5~GeV instead of 1~GeV as a scale in the logarithm.
We find that the results hardly change.)
\begin{table}
\begin{center}
\begin{tabular}{c|cccc} \hline
$t$  & $-0.3$ & $-0.1$ & 0.1 & 0.3  \\ \hline
$A_2$~[GeV$^3$] &  $-0.01$ & $-0.01$ & $-0.008$ &  $-0.007$  \\
$\Delta \alpha_s(M_Z^2) \times 10^4$  & 2 & 1 & $-$0 & $-1$  \\ \hline
\end{tabular}
\caption{Values of $A_2$ and variation of $\alpha_s(M_Z^2)$ when the logarithmic correction of Eq.~\eqref{logcorr} is considered.}
\label{tab:logcorr}
\end{center}
\end{table} 
One sees that this uncertainty dose not induce a dominant systematic error, and thus, 
is not included in our final result.

\bibliographystyle{utphys}
\bibliography{Bibalphas}

\providecommand{\href}[2]{#2}\begingroup\raggedright\begin{thebibliography}{10}

\bibitem{37ce3e5843594be4beddf3c7540d08bc}
C.~Patrignani, {\em et~al.}, [Particle Data Group], ``Review of particle
  physics,'' \href{http://dx.doi.org/10.1088/1674-1137/40/10/100001}{{\em
  Chinese Physics C} {\bfseries 40} no.~10, (10, 2016) }.

\bibitem{Aoki:2016frl}
S.~Aoki {\em et~al.}, ``{Review of lattice results concerning low-energy
  particle physics},''
  \href{http://dx.doi.org/10.1140/epjc/s10052-016-4509-7}{{\em Eur. Phys. J.}
  {\bfseries C77} no.~2, (2017) 112},
\href{http://arxiv.org/abs/1607.00299}{{\ttfamily arXiv:1607.00299 [hep-lat]}}.
%%CITATION = ARXIV:1607.00299;%%.

\bibitem{Maltman:2008bx}
K.~Maltman, D.~Leinweber, P.~Moran, and A.~Sternbeck, ``{The Realistic Lattice
  Determination of alpha(s)(M(Z)) Revisited},''
  \href{http://dx.doi.org/10.1103/PhysRevD.78.114504}{{\em Phys. Rev.}
  {\bfseries D78} (2008) 114504},
\href{http://arxiv.org/abs/0807.2020}{{\ttfamily arXiv:0807.2020 [hep-lat]}}.
%%CITATION = ARXIV:0807.2020;%%.

\bibitem{Aoki:2009tf}
 S.~Aoki {\em et~al.}, (PACS-CS Collaboration), ``{Precise determination of the
  strong coupling constant in N(f) = 2+1 lattice QCD with the Schrodinger
  functional scheme},''
  \href{http://dx.doi.org/10.1088/1126-6708/2009/10/053}{{\em JHEP} {\bfseries
  10} (2009) 053},
\href{http://arxiv.org/abs/0906.3906}{{\ttfamily arXiv:0906.3906 [hep-lat]}}.
%%CITATION = ARXIV:0906.3906;%%.

\bibitem{McNeile:2010ji}
C.~McNeile, C.~T.~H. Davies, E.~Follana, K.~Hornbostel, and G.~P. Lepage,
  ``{High-Precision c and b Masses, and QCD Coupling from Current-Current
  Correlators in Lattice and Continuum QCD},''
  \href{http://dx.doi.org/10.1103/PhysRevD.82.034512}{{\em Phys. Rev.}
  {\bfseries D82} (2010) 034512},
\href{http://arxiv.org/abs/1004.4285}{{\ttfamily arXiv:1004.4285 [hep-lat]}}.
%%CITATION = ARXIV:1004.4285;%%.

\bibitem{Chakraborty:2014aca}
B.~Chakraborty, C.~T.~H. Davies, B.~Galloway, P.~Knecht, J.~Koponen, G.~C.
  Donald, R.~J. Dowdall, G.~P. Lepage, and C.~McNeile, ``{High-precision quark
  masses and QCD coupling from $n_f=4$ lattice QCD},''
  \href{http://dx.doi.org/10.1103/PhysRevD.91.054508}{{\em Phys. Rev.}
  {\bfseries D91} no.~5, (2015) 054508},
\href{http://arxiv.org/abs/1408.4169}{{\ttfamily arXiv:1408.4169 [hep-lat]}}.
%%CITATION = ARXIV:1408.4169;%%.

\bibitem{Bazavov:2014soa}
A.~Bazavov, N.~Brambilla, X.~Garcia~i Tormo, P.~Petreczky, J.~Soto, and
  A.~Vairo, ``{Determination of $\alpha_s$ from the QCD static energy: An
  update},'' \href{http://dx.doi.org/10.1103/PhysRevD.90.074038}{{\em Phys.
  Rev.} {\bfseries D90} no.~7, (2014) 074038},
\href{http://arxiv.org/abs/1407.8437}{{\ttfamily arXiv:1407.8437 [hep-ph]}}.
%%CITATION = ARXIV:1407.8437;%%.

\bibitem{Luscher:1991wu}
M.~Luscher, P.~Weisz, and U.~Wolff, ``{A Numerical method to compute the
  running coupling in asymptotically free theories},''
\href{http://dx.doi.org/10.1016/0550-3213(91)90298-C}{{\em Nucl. Phys.}
  {\bfseries B359} (1991) 221--243}.
%%CITATION = NUPHA,B359,221;%%.

\bibitem{Luscher:1992an}
M.~Luscher, R.~Narayanan, P.~Weisz, and U.~Wolff, ``{The Schrodinger
  functional: A Renormalizable probe for nonAbelian gauge theories},''
  \href{http://dx.doi.org/10.1016/0550-3213(92)90466-O}{{\em Nucl. Phys.}
  {\bfseries B384} (1992) 168--228},
\href{http://arxiv.org/abs/hep-lat/9207009}{{\ttfamily arXiv:hep-lat/9207009
  [hep-lat]}}.
%%CITATION = HEP-LAT/9207009;%%.

\bibitem{Luscher:1993gh}
M.~Luscher, R.~Sommer, P.~Weisz, and U.~Wolff, ``{A Precise determination of
  the running coupling in the SU(3) Yang-Mills theory},''
  \href{http://dx.doi.org/10.1016/0550-3213(94)90629-7}{{\em Nucl. Phys.}
  {\bfseries B413} (1994) 481--502},
\href{http://arxiv.org/abs/hep-lat/9309005}{{\ttfamily arXiv:hep-lat/9309005
  [hep-lat]}}.
%%CITATION = HEP-LAT/9309005;%%.

\bibitem{Sint:1993un}
S.~Sint, ``{On the Schrodinger functional in QCD},''
  \href{http://dx.doi.org/10.1016/0550-3213(94)90228-3}{{\em Nucl. Phys.}
  {\bfseries B421} (1994) 135--158},
\href{http://arxiv.org/abs/hep-lat/9312079}{{\ttfamily arXiv:hep-lat/9312079
  [hep-lat]}}.
%%CITATION = HEP-LAT/9312079;%%.

\bibitem{Bruno:2017gxd}
 M.~Bruno, M.~Dalla~Brida, P.~Fritzsch, T.~Korzec, A.~Ramos,
  S.~Schaefer, H.~Simma, S.~Sint, and R.~Sommer, (ALPHA Collaboration),``{QCD Coupling from a
  Nonperturbative Determination of the Three-Flavor $\Lambda$ Parameter},''
  \href{http://dx.doi.org/10.1103/PhysRevLett.119.102001}{{\em Phys. Rev.
  Lett.} {\bfseries 119} no.~10, (2017) 102001},
\href{http://arxiv.org/abs/1706.03821}{{\ttfamily arXiv:1706.03821 [hep-lat]}}.
%%CITATION = ARXIV:1706.03821;%%.

\bibitem{Beneke:1998ui}
M.~Beneke, ``{Renormalons},''
  \href{http://dx.doi.org/10.1016/S0370-1573(98)00130-6}{{\em Phys. Rept.}
  {\bfseries 317} (1999) 1--142},
\href{http://arxiv.org/abs/hep-ph/9807443}{{\ttfamily arXiv:hep-ph/9807443
  [hep-ph]}}.
%%CITATION = HEP-PH/9807443;%%.

\bibitem{Mueller:1984vh}
A.~H. Mueller, ``{On the Structure of Infrared Renormalons in Physical
  Processes at High-Energies},''
\href{http://dx.doi.org/10.1016/0550-3213(85)90485-7}{{\em Nucl. Phys.}
  {\bfseries B250} (1985) 327--350}.
%%CITATION = NUPHA,B250,327;%%.

\bibitem{Sumino:2005cq}
Y.~Sumino, ``{Static QCD Potential at $r<\Lambda_\mathrm{QCD}^{-1}$:
  Perturbative Expansion and Operator-Product Expansion},''
  \href{http://dx.doi.org/10.1103/PhysRevD.76.114009}{{\em Phys. Rev.}
  {\bfseries D76} (2007) 114009},
\href{http://arxiv.org/abs/hep-ph/0505034}{{\ttfamily arXiv:hep-ph/0505034
  [hep-ph]}}.
%%CITATION = HEP-PH/0505034;%%.

\bibitem{Mishima:2016vna}
G.~Mishima, Y.~Sumino, and H.~Takaura, ``{Subtracting infrared renormalons from
  Wilson coefficients: Uniqueness and power dependences on $\Lambda$QCD},''
  \href{http://dx.doi.org/10.1103/PhysRevD.95.114016}{{\em Phys. Rev.}
  {\bfseries D95} no.~11, (2017) 114016},
\href{http://arxiv.org/abs/1612.08711}{{\ttfamily arXiv:1612.08711 [hep-ph]}}.
%%CITATION = ARXIV:1612.08711;%%.

\bibitem{Beneke:1998rk}
M.~Beneke, ``{A Quark Mass Definition Adequate for Threshold Problems},''
  \href{http://dx.doi.org/10.1016/S0370-2693(98)00741-2}{{\em Phys. Lett.}
  {\bfseries B434} (1998) 115--125},
\href{http://arxiv.org/abs/hep-ph/9804241}{{\ttfamily arXiv:hep-ph/9804241
  [hep-ph]}}.
%%CITATION = HEP-PH/9804241;%%.

\bibitem{Hoang:1998nz}
A.~H. Hoang, M.~C. Smith, T.~Stelzer, and S.~Willenbrock, ``{Quarkonia and the
  pole mass},'' \href{http://dx.doi.org/10.1103/PhysRevD.59.114014}{{\em Phys.
  Rev.} {\bfseries D59} (1999) 114014},
\href{http://arxiv.org/abs/hep-ph/9804227}{{\ttfamily arXiv:hep-ph/9804227
  [hep-ph]}}.
%%CITATION = HEP-PH/9804227;%%.

\bibitem{Brambilla:1999xf}
N.~Brambilla, A.~Pineda, J.~Soto, and A.~Vairo, ``{Potential NRQCD: an
  Effective Theory for Heavy Quarkonium},''
  \href{http://dx.doi.org/10.1016/S0550-3213(99)00693-8}{{\em Nucl. Phys.}
  {\bfseries B566} (2000) 275},
\href{http://arxiv.org/abs/hep-ph/9907240}{{\ttfamily arXiv:hep-ph/9907240
  [hep-ph]}}.
%%CITATION = HEP-PH/9907240;%%.

\bibitem{Smirnov:2008pn}
A.~V. Smirnov, V.~A. Smirnov, and M.~Steinhauser, ``{Fermionic contributions to
  the three-loop static potential},''
  \href{http://dx.doi.org/10.1016/j.physletb.2008.08.070}{{\em Phys. Lett.}
  {\bfseries B668} (2008) 293--298},
\href{http://arxiv.org/abs/0809.1927}{{\ttfamily arXiv:0809.1927 [hep-ph]}}.
%%CITATION = ARXIV:0809.1927;%%.

\bibitem{Anzai:2009tm}
C.~Anzai, Y.~Kiyo, and Y.~Sumino, ``{Static QCD Potential at Three-Loop
  Order},'' \href{http://dx.doi.org/10.1103/PhysRevLett.104.112003}{{\em Phys.
  Rev. Lett.} {\bfseries 104} (2010) 112003},
\href{http://arxiv.org/abs/0911.4335}{{\ttfamily arXiv:0911.4335 [hep-ph]}}.
%%CITATION = ARXIV:0911.4335;%%.

\bibitem{Smirnov:2009fh}
A.~V. Smirnov, V.~A. Smirnov, and M.~Steinhauser, ``{Three-Loop Static
  Potential},'' \href{http://dx.doi.org/10.1103/PhysRevLett.104.112002}{{\em
  Phys. Rev. Lett.} {\bfseries 104} (2010) 112002},
\href{http://arxiv.org/abs/0911.4742}{{\ttfamily arXiv:0911.4742 [hep-ph]}}.
%%CITATION = ARXIV:0911.4742;%%.

\bibitem{Lee:2016cgz}
R.~N. Lee, A.~V. Smirnov, V.~A. Smirnov, and M.~Steinhauser, ``{Analytic
  Three-Loop Static Potential},''
  \href{http://dx.doi.org/10.1103/PhysRevD.94.054029}{{\em Phys. Rev.}
  {\bfseries D94} no.~5, (2016) 054029},
\href{http://arxiv.org/abs/1608.02603}{{\ttfamily arXiv:1608.02603 [hep-ph]}}.
%%CITATION = ARXIV:1608.02603;%%.

\bibitem{Kaneko:2013jla}
 T.~Kaneko, S.~Aoki, G.~Cossu, H.~Fukaya, S.~Hashimoto, and
  J.~Noaki,  (JLQCD Collaboration),``{Large-scale simulations with chiral symmetry},'' {\em PoS}
  {\bfseries LATTICE2013} (2014) 125,
\href{http://arxiv.org/abs/1311.6941}{{\ttfamily arXiv:1311.6941 [hep-lat]}}.
%%CITATION = ARXIV:1311.6941;%%.

\bibitem{JLQCD:future}
JLQCD Collaboration, in preparation

\bibitem{Karbstein:2018mzo}
F.~Karbstein, M.~Wagner, and M.~Weber, ``{Determination of
  $\Lambda_{\overline{\textrm{MS}}}^{(n_f=2)}$ and analytic parameterization of
  the static quark-antiquark potential},''
\href{http://arxiv.org/abs/1804.10909}{{\ttfamily arXiv:1804.10909 [hep-ph]}}.
%%CITATION = ARXIV:1804.10909;%%.

\bibitem{Takaura:2018lpw}
H.~Takaura, T.~Kaneko, Y.~Kiyo, and Y.~Sumino, ``{Determination of $\alpha_s$
  from static QCD potential with renormalon subtraction},''
  \href{http://dx.doi.org/10.1016/j.physletb.2018.12.060}{{\em Phys. Lett.}
  {\bfseries B789} (2019) 598--602},
\href{http://arxiv.org/abs/1808.01632}{{\ttfamily arXiv:1808.01632 [hep-ph]}}.
%%CITATION = ARXIV:1808.01632;%%.

\bibitem{Takaura:2017lwd}
H.~Takaura, ``{Renormalon free part of an ultrasoft correction to the static
  QCD potential},''
  \href{http://dx.doi.org/10.1016/j.physletb.2018.07.014}{{\em Phys. Lett.}
  {\bfseries B783} (2018) 350--356},
\href{http://arxiv.org/abs/1712.05435}{{\ttfamily arXiv:1712.05435 [hep-ph]}}.
%%CITATION = ARXIV:1712.05435;%%.

\bibitem{Appelquist:1977tw}
T.~Appelquist, M.~Dine, and I.~J. Muzinich, ``{The Static Potential in Quantum
  Chromodynamics},''
\href{http://dx.doi.org/10.1016/0370-2693(77)90651-7}{{\em Phys. Lett.}
  {\bfseries 69B} (1977) 231--236}.
%%CITATION = PHLTA,69B,231;%%.

\bibitem{Brambilla:1999qa}
N.~Brambilla, A.~Pineda, J.~Soto, and A.~Vairo, ``{The Infrared behavior of the
  static potential in perturbative QCD},''
  \href{http://dx.doi.org/10.1103/PhysRevD.60.091502}{{\em Phys. Rev.}
  {\bfseries D60} (1999) 091502},
\href{http://arxiv.org/abs/hep-ph/9903355}{{\ttfamily arXiv:hep-ph/9903355
  [hep-ph]}}.
%%CITATION = HEP-PH/9903355;%%.

\bibitem{Weisz:1982zw}
P.~Weisz, ``{Continuum Limit Improved Lattice Action for Pure Yang-Mills
  Theory. 1.},''
\href{http://dx.doi.org/10.1016/0550-3213(83)90595-3}{{\em Nucl. Phys.}
  {\bfseries B212} (1983) 1--17}.
%%CITATION = NUPHA,B212,1;%%.

\bibitem{Brower:2012vk}
R.~C. Brower, H.~Neff, and K.~Orginos
  \href{http://dx.doi.org/10.1016/j.cpc.2017.01.024}{{\em Comput. Phys.
  Commun.} {\bfseries 220} (2017) 1--19},
\href{http://arxiv.org/abs/1206.5214}{{\ttfamily arXiv:1206.5214 [hep-lat]}}.
%%CITATION = ARXIV:1206.5214;%%.

\bibitem{Luscher:2010iy}
``{Properties and uses of the Wilson flow in lattice QCD},''
  \href{http://dx.doi.org/10.1007/JHEP08(2010)071,
  10.1007/JHEP03(2014)092}{{\em JHEP} {\bfseries 08} (2010) 071},
  \href{http://arxiv.org/abs/1006.4518}{{\ttfamily arXiv:1006.4518 [hep-lat]}}.
[Erratum: JHEP03,092(2014)].
%%CITATION = ARXIV:1006.4518;%%.

\bibitem{Tomii:2016xiv}
M.~Tomii, G.~Cossu, B.~Fahy, H.~Fukaya, S.~Hashimoto,
  T.~Kaneko, and J.~Noaki,  (JLQCD Collaboration), ``{Renormalization of domain-wall bilinear operators
  with short-distance current correlators},''
  \href{http://dx.doi.org/10.1103/PhysRevD.94.054504}{{\em Phys. Rev.}
  {\bfseries D94} no.~5, (2016) 054504},
\href{http://arxiv.org/abs/1604.08702}{{\ttfamily arXiv:1604.08702 [hep-lat]}}.
%%CITATION = ARXIV:1604.08702;%%.

\bibitem{Bali:1992ab}
G.~S. Bali and K.~Schilling, ``{Static quark - anti-quark potential: Scaling
  behavior and finite size effects in SU(3) lattice gauge theory},''
\href{http://dx.doi.org/10.1103/PhysRevD.46.2636}{{\em Phys. Rev.} {\bfseries
  D46} (1992) 2636--2646}.
%%CITATION = PHRVA,D46,2636;%%.

\bibitem{Bazavov:2010hj}
 A.~Bazavov {\em et~al.}, (MILC Collaboration),``{Results for light pseudoscalar
  mesons},'' {\em PoS} {\bfseries LATTICE2010} (2010) 074,
\href{http://arxiv.org/abs/1012.0868}{{\ttfamily arXiv:1012.0868 [hep-lat]}}.
%%CITATION = ARXIV:1012.0868;%%.

\bibitem{Bazavov:2011nk}
A.~Bazavov {\em et~al.}, ``{The chiral and deconfinement aspects of the QCD
  transition},'' \href{http://dx.doi.org/10.1103/PhysRevD.85.054503}{{\em Phys.
  Rev.} {\bfseries D85} (2012) 054503},
\href{http://arxiv.org/abs/1111.1710}{{\ttfamily arXiv:1111.1710 [hep-lat]}}.
%%CITATION = ARXIV:1111.1710;%%.

\bibitem{Sommer:2014mea}
R.~Sommer, ``{Scale setting in lattice QCD},'' {\em PoS} {\bfseries
  LATTICE2013} (2014) 015,
\href{http://arxiv.org/abs/1401.3270}{{\ttfamily arXiv:1401.3270 [hep-lat]}}.
%%CITATION = ARXIV:1401.3270;%%.

\bibitem{Sumino:2014qpa}
Y.~Sumino, {\it Lecture Note}, ``{Understanding Interquark Force and Quark Masses in Perturbative
  QCD},''
\newblock 2014.
\newblock \href{http://arxiv.org/abs/1411.7853}{{\ttfamily arXiv:1411.7853
  [hep-ph]}}.
\newblock
\url{http://inspirehep.net/record/1331450/files/arXiv:1411.7853.pdf}.
\newblock
%%CITATION = ARXIV:1411.7853;%%.

\bibitem{Chetyrkin:1997sg}
K.~G. Chetyrkin, B.~A. Kniehl, and M.~Steinhauser, ``{Strong coupling constant
  with flavor thresholds at four loops in the MS scheme},''
  \href{http://dx.doi.org/10.1103/PhysRevLett.79.2184}{{\em Phys. Rev. Lett.}
  {\bfseries 79} (1997) 2184--2187},
\href{http://arxiv.org/abs/hep-ph/9706430}{{\ttfamily arXiv:hep-ph/9706430
  [hep-ph]}}.
%%CITATION = HEP-PH/9706430;%%.

\bibitem{Hoang:2000fm}
A.~H. Hoang, ``{Bottom quark mass from Upsilon mesons: Charm mass effects},''
\href{http://arxiv.org/abs/hep-ph/0008102}{{\ttfamily arXiv:hep-ph/0008102
  [hep-ph]}}.
%%CITATION = HEP-PH/0008102;%%.

\bibitem{Melles:2000dq}
M.~Melles, ``{The Static QCD potential in coordinate space with quark masses
  through two loops},''
  \href{http://dx.doi.org/10.1103/PhysRevD.62.074019}{{\em Phys. Rev.}
  {\bfseries D62} (2000) 074019},
\href{http://arxiv.org/abs/hep-ph/0001295}{{\ttfamily arXiv:hep-ph/0001295
  [hep-ph]}}.
%%CITATION = HEP-PH/0001295;%%.

\bibitem{Recksiegel:2001xq}
S.~Recksiegel and Y.~Sumino, ``{Perturbative QCD potential, renormalon
  cancellation and phenomenological potentials},''
  \href{http://dx.doi.org/10.1103/PhysRevD.65.054018}{{\em Phys. Rev.}
  {\bfseries D65} (2002) 054018},
\href{http://arxiv.org/abs/hep-ph/0109122}{{\ttfamily arXiv:hep-ph/0109122
  [hep-ph]}}.
%%CITATION = HEP-PH/0109122;%%.

\bibitem{Borsanyi:2012zs}
S.~Borsanyi {\em et~al.}, ``{High-precision scale setting in lattice QCD},''
  \href{http://dx.doi.org/10.1007/JHEP09(2012)010}{{\em JHEP} {\bfseries 09}
  (2012) 010},
\href{http://arxiv.org/abs/1203.4469}{{\ttfamily arXiv:1203.4469 [hep-lat]}}.
%%CITATION = ARXIV:1203.4469;%%.


\end{thebibliography}\endgroup

\end{document}